%% file: ZZPaper.tex
\pdfoutput=1
\newcommand*{\ATLASLATEXPATH}{latex/}
\documentclass[cernpreprint,texlive=2011,txfonts,UKenglish]{latex/atlasdoc} 
\usepackage[subcaption=true]{\ATLASLATEXPATH atlaspackage}
\usepackage{latex/atlascover}
\usepackage{dcolumn}
\usepackage{multirow}
\newcolumntype{d}[1]{D{.}{.}{#1}}
\usepackage{\ATLASLATEXPATH atlasbiblatex}
\usepackage{\ATLASLATEXPATH atlascontribute}
\usepackage{cleveref} 
\usepackage{harpoon}

\newcommand{\tightoverset}[2]{\mathop{#2}\limits^{\vbox to -.5ex{\kern+0.75ex\hbox{$#1$}\vss}}}
\usepackage{tikz}

\usepackage{feynmf}
\graphicspath{{logos/}{figures/}}

\usepackage{atlasphysics} 

\input{ZZCommands}

\input{Results.DB}

\input{ZZPaper-metadata}
\hypersetup{pdftitle={ATLAS draft},pdfauthor={The ATLAS Collaboration}}

\begin{document}

\maketitle

\tableofcontents

\newpage
%-------------------------------------------------------------------------------
\input{Introduction}
%-------------------------------------------------------------------------------
\input{ATLASDetector}

%-------------------------------------------------------------------------------
\input{PhaseSpace}
%-------------------------------------------------------------------------------
\input{SMPredictions}

%-------------------------------------------------------------------------------
\input{SimulationSamples}

%-------------------------------------------------------------------------------
\section{Data samples, reconstruction of leptons, jets, and \met\ and event selections}
\label{sec:selection}
\input{ObjectSelectionAndEvtReco}
\input{EventSelection}
%-------------------------------------------------------------------------------
\input{BackgroundEstimate}
%-------------------------------------------------------------------------------
\section{Event yields}
\label{sec:yields}
\input{EventYields.tex}
%-------------------------------------------------------------------------------
\input{SignalAcceptance}

%-------------------------------------------------------------------------------
\input{SystematicUncertainties}
%-------------------------------------------------------------------------------
\section{Cross-section measurements}
\label{sec:results}
%-------------------------------------------------------------------------------
\subsection{Cross-section extraction}
\label{sec:xsec}
\input{CrossSectionExtraction}

\subsection{Differential cross sections}
\label{sec:diffxsec}
\input{DifferentialCrossSections}

%------------------------------------------------------------------------------
\input{aTGCs}

%-------------------------------------------------------------------------------
\section{Conclusion}
\label{sec:conclusion}
%-------------------------------------------------------------------------------

A measurement of the \ZZ\ production cross section in LHC  $pp$ collisions at $\sqrt{s}
= 8 \TeV$ is presented, using 
data corresponding to an integrated luminosity of
\ZZLuminosityFbOneDp~\invfb ~collected by the ATLAS detector in
2012. Fiducial cross sections are measured for every final state in
the \zzlmlplmlpprimed\ and \zzllvv\ $(\ell = e, \mu)$ decay channels
and the results are compatible with the SM expected cross
sections. The combined total \ZZ\ production cross section
is measured to be:
\begin{center}
$\sigma_{pp \to ZZ}^{\mathrm{total}}$ = \ZZTotalCrossSectionOneDp\ \pb
\end{center}
The result is consistent with the SM prediction: 
\begin{center}
$\sigma_{pp \to ZZ}^{\mathrm{total}}$    = 6.6$^{+0.7}_{-0.6}$ \pb
\end{center}
which includes predictions from QCD at NLO for the $q\bar{q}$ process corrected for virtual NLO EW effects and 
predictions from LO gluon--gluon fusion.  

Differential cross sections in the total phase space in the
\zzlmlplmlpprimed\ channel are derived for the transverse momentum
of the leading $Z$ boson, the number of jets, the azimuthal angle
between the two leptons originating from the leading $Z$ boson and the difference in rapidity
between the two $Z$ bosons of the $ZZ$ system. In
the \zzllvv\ channel, the differential cross sections are measured in
the fiducial phase space for the transverse momentum of the $Z$ boson, the azimuthal angle
between the two leptons originating from the $Z$  and the
transverse mass of the \ZZ\ system.

The event yields as a function of the  \pT\  of the leading \Z ~boson
for the \zzlmlplmlpprimed\ and \zzllvv\ event selections are used to
derive 95$\%$ confidence intervals for anomalous neutral triple gauge
boson couplings. These limits are more stringent than the previous ATLAS results by approximately a factor of four.

%-------------------------------------------------------------------------------
\section*{Acknowledgements}
%-------------------------------------------------------------------------------

\input{acknowledgements/Acknowledgements.tex}

%-------------------------------------------------------------------------------
\clearpage
\appendix

\printbibliography

\newpage \input{atlas_authlist}
%-------------------------------------------------------------------------------
\clearpage
\appendix

\end{document}

%% file: ZZCommands.tex
\def\Z{\ensuremath{Z}}

\def\ggtwovv{\texttt{gg2VV}\xspace}

\def\ggVV{\ggtwovv}
\def\gg2ZZjimmy{\textsc{gg2ZZ+Herwig+Jimmy}\xspace}
\def\powhegboxgg2VV{\textsc{PowhegBox}+gg2\textsc{VV}\xspace}

\def\mcatnlo{MC@NLO\xspace}

\usepackage{xspace}

\def\madgraph{\textsc{MadGraph}\xspace}
\def\powhegbox{\textsc{PowhegBox}\xspace}
\def\pythia{\textsc{Pythia}\xspace}

\def\alpgen{\textsc{Alpgen}\xspace}

\def\sherpa{\textsc{Sherpa}\xspace}
\def\herwig{\textsc{Herwig}\xspace}
\def\Jimmy{\textsc{Jimmy}\xspace}

\def\ptsp1{\ensuremath{p_T} }

\def\MET{\ensuremath{E_{\mathrm{T}}^{\mathrm{miss}}}}%
\def\met{\ensuremath{E_{\mathrm{T}}^{\mathrm{miss}}}}%
\def\metvec{\ensuremath{\vec{E}_{\mathrm{T}}^{\mathrm{~miss}}}}%
\def\Zpt{\ensuremath{p_{\mathrm{T}}^{Z}}}%
\def\qqZZ{\ensuremath{q\bar{q}\to ZZ}}%

\def\ns{\ifmmode {\mathrm{\ ns}}\else
                   \textrm{ns}\fi}%
\def\pb{\mbox{pb}}%  picobarns.
\def\fb{\mbox{fb}}%  femtobarns.
\def\invfb{{\mbox{fb}$^{-1}$}}%  inverse femtobarns.
\def\ifb{{\mbox{fb}$^{-1}$}}%  inverse femtobarns.

\def\ll{\ensuremath{\ell^{-} \ell^{+}}}%
\def\lnu{\ensuremath{\ell \nu}}%

\def\qqbar{\ensuremath{q\bar{q}}}

\newcommand{\zzllll}{\ensuremath{\ZZ\to\ell^{-}\ell^{+}\ell^{-}\ell^{+}}}

\newcommand{\zzlmlplmlpprimed}{\ensuremath{\ZZ\to\ell^{-}\ell^{+}\ell^{\prime\, -}\ell^{\prime\, +}}}
\newcommand{\zzeeee}{\ensuremath{\ZZ\to e^{-}e^{+}e^{-}e^{+}}}
\newcommand{\zzeemm}{\ensuremath{\ZZ\to e^{-}e^{+}\mu^{-}\mu^{+}}}
\newcommand{\zzmmmm}{\ensuremath{\ZZ\to \mu^{-}\mu^{+}\mu^{-}\mu^{+}}}

\newcommand{\zzllvv}{\ensuremath{\ZZ\to \ell^{-}\ell^{+}\nu\bar{\nu}}}
\newcommand{\zzeevv}{\ensuremath{\ZZ\to e^{-}e^{+}\nu\bar{\nu}}}
\newcommand{\zzmmvv}{\ensuremath{\ZZ\to \mu^{-}\mu^{+}\nu\bar{\nu}}}

\newcommand{\eeee}{\ensuremath{e^{-}e^{+}e^{-}e^{+}}}

\newcommand{\mumumumu}{\ensuremath{\mu^{-}\mu^{+}\mu^{-}\mu^{+}}}
\newcommand{\mmmm}{\ensuremath{\mu^{-}\mu^{+}\mu^{-}\mu^{+}}}
\newcommand{\eemm}{\ensuremath{e^{-}e^{+}\mu^{-}\mu^{+}}}

\newcommand{\eevv}{\ensuremath{e^{-}e^{+}\nu\bar{\nu}}}
\newcommand{\mumuvv}{\ensuremath{\mu^{-}\mu^{+}\nu\bar{\nu}}}
\newcommand{\mmvv}{\mumuvv}
\newcommand{\llvv}{\ensuremath{\ell^{-}\ell^{+}\nu\bar{\nu}}}

\newcommand{\llll}{\ensuremath{\ell^{-}\ell^{+}\ell^{\prime\, -}\ell^{\prime\, +}}}
\newcommand{\ZZ}{\ensuremath{ZZ}}

%% file: Results.DB.tex
\def\ZZllllCzzMMMM{0.846}
\def\ZZllllCzzEEEE{0.495}
\def\ZZllllCzzEEMM{0.643}

\def\ZZllllCzzSystMMMM{0.034}
\def\ZZllllCzzSystEEEE{0.023}
\def\ZZllllCzzSystEEMM{0.021}

\def\ZZllllAzzMMMM{0.645}
\def\ZZllllAzzEEEE{0.817}
\def\ZZllllAzzEEMM{0.725}

\def\ZZllllAzzSystMMMM{0.020}
\def\ZZllllAzzSystEEEE{0.017}
\def\ZZllllAzzSystEEMM{0.017}

\newcommand{\ZZCzzEEEE}{\ZZllllCzzEEEE\ensuremath{\pm}\ZZllllCzzSystEEEE}
\newcommand{\ZZCzzMMMM}{\ZZllllCzzMMMM\ensuremath{\pm}\ZZllllCzzSystMMMM}
\newcommand{\ZZCzzEEMM}{\ZZllllCzzEEMM\ensuremath{\pm}\ZZllllCzzSystEEMM}
\newcommand{\ZZAzzEEEE}{\ZZllllAzzEEEE\ensuremath{\pm}\ZZllllAzzSystEEEE}
\newcommand{\ZZAzzMMMM}{\ZZllllAzzMMMM\ensuremath{\pm}\ZZllllAzzSystMMMM}
\newcommand{\ZZAzzEEMM}{\ZZllllAzzEEMM\ensuremath{\pm}\ZZllllAzzSystEEMM}

\def\ZZllvvAzzEENN{0.0413}
\def\ZZllvvAzzMMNN{0.0400}
\def\ZZllvvAzzTotalEENN{0.0022}
\def\ZZllvvAzzTotalMMNN{0.0019}
\def\ZZllvvCzzEENN{0.678}
\def\ZZllvvCzzMMNN{0.752}
\def\ZZllvvCzzTotalEENN{0.039}
\def\ZZllvvCzzTotalMMNN{0.048}

\newcommand{\ZZAzzEENN}{\ZZllvvAzzEENN\ensuremath{\pm}\ZZllvvAzzTotalEENN}
\newcommand{\ZZAzzMMNN}{\ZZllvvAzzMMNN\ensuremath{\pm}\ZZllvvAzzTotalMMNN}
\newcommand{\ZZCzzEENN}{\ZZllvvCzzEENN\ensuremath{\pm}\ZZllvvCzzTotalEENN}
\newcommand{\ZZCzzMMNN}{\ZZllvvCzzMMNN\ensuremath{\pm}\ZZllvvCzzTotalMMNN}

\def\ZZfidEEEEXsecTOneDp{5.9}%pb
\def\ZZfidEEEEXsecStatOnlyErrUpOneDp{0.8}%pb
\def\ZZfidEEEEXsecAllSystNoLumiErrUpOneDp{0.4}%pb
\def\ZZfidEEEEXsecLUMIErrUpOneDp{0.1}%pb
\def\ZZfidEEMMXsecTOneDp{12.4}%pb
\def\ZZfidEEMMXsecStatOnlyErrUpOneDp{1.0}%pb
\def\ZZfidEEMMXsecAllSystNoLumiErrUpOneDp{0.6}%pb
\def\ZZfidEEMMXsecAllSystNoLumiErrDwOneDp{-0.5}%pb
\def\ZZfidEEMMXsecLUMIErrUpOneDp{0.3}%pb
\def\ZZfidEEMMXsecLUMIErrDwOneDp{-0.2}%pb

\def\ZZfidMMMMXsecTOneDp{4.9}%pb
\def\ZZfidMMMMXsecStatOnlyErrUpOneDp{0.6}%pb
\def\ZZfidMMMMXsecStatOnlyErrDwOneDp{-0.5}%pb
\def\ZZfidMMMMXsecAllSystNoLumiErrUpOneDp{0.3}%pb
\def\ZZfidMMMMXsecAllSystNoLumiErrDwOneDp{-0.2}%pb
\def\ZZfidMMMMXsecLUMIErrUpOneDp{0.1}%pb
\newcommand{\ZZTotalCrossSectionTOneDp}{7.3}%pb
\newcommand{\ZZTotalCrossSectionStatOnlyUpOneDp}{0.4}%pb
\newcommand{\ZZTotalCrossSectionAllSystNoLumiUpOneDp}{0.3}%pb
\newcommand{\ZZTotalCrossSectionLUMIUpOneDp}{0.2}%pb
\newcommand{\ZZTotalCrossSectionLUMIDwOneDp}{-0.1}%pb

\newcommand{\ZZTotalCrossSectionOneDp}{\ZZTotalCrossSectionTOneDp\ $\pm$ \ZZTotalCrossSectionStatOnlyUpOneDp\ (stat) $\pm$ \ZZTotalCrossSectionAllSystNoLumiUpOneDp\ (syst) $^{+\ZZTotalCrossSectionLUMIUpOneDp}_{\ZZTotalCrossSectionLUMIDwOneDp}$ (lumi)}%pb
\def\ZZfidXsecEENuNuOneDp{5.0}%pb
\def\ZZfidXsecStatErrUpEENuNuOneDp{0.8}%pb
\def\ZZfidXsecStatErrDwEENuNuOneDp{-0.7}%pb
\def\ZZfidXsecSysErrUpEENuNuOneDp{0.5}%pb
\def\ZZfidXsecSysErrDwEENuNuOneDp{-0.4}%pb
\def\ZZfidXsecLumiErrUpEENuNuOneDp{0.1}%pb
\def\ZZfidXsecMMNuNuOneDp{4.7}%pb
\def\ZZfidXsecStatErrUpMMNuNuOneDp{0.7}%pb
\def\ZZfidXsecSysErrUpMMNuNuOneDp{0.5}%pb
\def\ZZfidXsecSysErrDwMMNuNuOneDp{-0.4}%pb
\def\ZZfidXsecLumiErrUpMMNuNuOneDp{0.1}%pb
\def\ZZLuminosityFbOneDp{20.3}

\def\CMSLuminosityFbOneDp{19.6}

\def\ZZTheoryTotalCrossSectionOneDpCentral{6.6}

\def\ZZTheoryTotalCrossSectionOneDpSystUp{0.7}
\def\ZZTheoryTotalCrossSectionOneDpSystDown{0.6}

\newcommand{\ZZTheoryTotalCrossSectionOneDp}{\ensuremath{\ZZTheoryTotalCrossSectionOneDpCentral^{+\ZZTheoryTotalCrossSectionOneDpSystUp}_{-\ZZTheoryTotalCrossSectionOneDpSystDown}\textrm{ pb}}}

%% file: ZZPaper-metadata.tex
\AtlasTitle{Measurement of the \ZZ\ production cross section in proton--proton
collisions at $\mathbf{\sqrt{s} = 8 \TeV}$ using the
$\zzlmlplmlpprimed$ and $\zzllvv$ decay channels with the ATLAS detector}
\author{The ATLAS Collaboration}
\AtlasRefCode{STMD-2014-16}
\AtlasNote{ATL-COM-PHYS-2015-1582}
\PreprintIdNumber{CERN-EP-2016-194}
\AtlasJournal{JHEP}
\AtlasAbstract{
A  measurement of the \ZZ\ production cross section in the \llll\ and \llvv\ 
channels $(\ell = e, \mu)$ in proton--proton collisions at $\sqrt{s}
= 8 \TeV$ at the Large Hadron Collider at CERN, using
data corresponding to an integrated luminosity of \ZZLuminosityFbOneDp~\invfb ~collected by
the ATLAS experiment in 2012 is presented.
The fiducial cross sections for \zzlmlplmlpprimed\
and \zzllvv\ are measured in selected phase-space regions. The
total cross section for \ZZ\ events produced with both \Z\ bosons
in the mass range $66$ to 116 \GeV\ is measured from the combination
of the two channels to
be \ZZTotalCrossSectionOneDp\ \pb, which is consistent with the Standard Model
prediction of \ZZTheoryTotalCrossSectionOneDp. The differential cross
sections in bins of various kinematic variables are
presented. The differential event yield as a function of the
transverse momentum of the leading \Z\ boson is used to set limits on
anomalous neutral triple gauge boson couplings in \ZZ\
production. 
}
\AtlasJournalRef{JHEP 01 (2017) 099}
\AtlasDOI{10.1007/JHEP01(2017)099}

%% file: Introduction.tex
\section{Introduction}
\label{sec:intro}
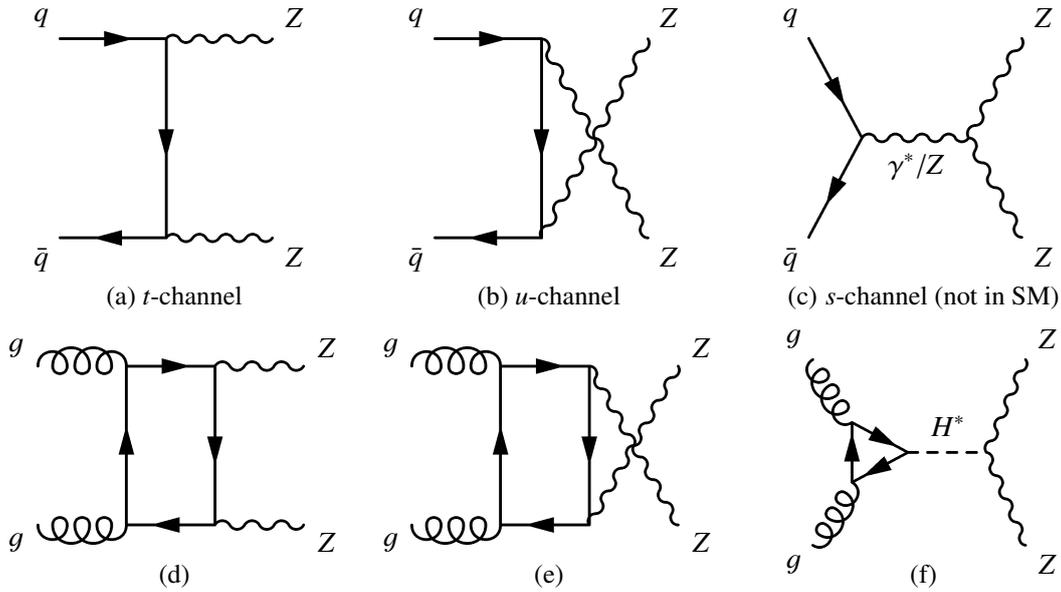
\begin{figure}[!h]
\centering
  \begin{subfigure}[b]{0.3\textwidth}
    \centering
  %\subfloat[t-channel]{\label{FigChap2ZZtChannel}
    \begin{fmffile}{figures/ZZtchannel}
      \begin{fmfgraph*}(100,75)
       \fmfleft{i1,i2}
        \fmfright{o1,o2}
%    	\fmfleftn{i}{2}
%    	\fmfrightn{o}{2}
    	\fmf{fermion}{v1,i1}
    	\fmf{fermion}{i2,v2}
    	\fmf{fermion,tension=0}{v2,v1}
    	\fmf{photon}{v1,o1}
    	\fmf{photon}{v2,o2}
    	\fmflabel{$\bar{q}$}{i1}
    	\fmflabel{$q$}{i2}
    	\fmflabel{$Z$}{o1}
    	\fmflabel{$Z$}{o2}
    \end{fmfgraph*}
    \end{fmffile}
    \vspace{1em}
    \caption{$t$-channel}
    \label{fig:ZZtchannel}
  \end{subfigure}
  %%%%%%%%%%%%%%%%%%%%%%%%%%%%%%%%%%%%%%%%%
  \begin{subfigure}[b]{0.3\textwidth}
    \centering
    \begin{fmffile}{figures/ZZuchannel}
      \begin{fmfgraph*}(100,75)
        \fmfleft{i1,i2}
        \fmfright{o1,o2}
        \fmf{fermion}{i2,v2}
        \fmf{phantom}{v1,o1} % Invisible rubber band
        \fmf{fermion}{v1,i1}
        \fmf{phantom}{v2,o2} % also invisible rubber band
        \fmf{fermion, tension=0}{v2,v1}
        % These are visible, but have no tension.
        \fmf{photon,tension=0}{v1,o2}
        \fmf{photon,tension=0}{v2,o1}
        %\fmfdot{v1,v2}
        \fmflabel{$\bar{q}$}{i1}
        \fmflabel{$q$}{i2}
        \fmflabel{$Z$}{o1}
        \fmflabel{$Z$}{o2}
      \end{fmfgraph*}
    \end{fmffile}
    \vspace{1em}
    \caption{$u$-channel}
    \label{fig:ZZuchannel}
   \end{subfigure}
   \begin{subfigure}[b]{0.3\textwidth}
     \centering
     \begin{fmffile}{figures/ZZschannel}
       \begin{fmfgraph*}(100,75)
         \fmfleft{i1,i2}
         \fmfright{o1,o2}
         \fmflabel{$\bar{q}$}{i1}
         \fmflabel{$q$}{i2}
         \fmflabel{$Z$}{o1}
         \fmflabel{$Z$}{o2}
         \fmf{fermion}{v1,i1}
         \fmf{fermion}{i2,v1}
         \fmf{photon,label=$\gamma^{*}/Z^{*}$}{v1,v2}
         \fmf{photon}{v2,o1}
         \fmf{photon}{v2,o2}
       \end{fmfgraph*}
     \end{fmffile}
     \vspace{1em}
     \caption{$s$-channel (not in SM)}
     \label{fig:ZZschannel}
   \end{subfigure}\\ \vspace{1em}
%%%%%%%%%%%%%%%%%%%%%%%%%%%%%%%%%%%%%%%%%%%%
%%%%
%%%%%%%%%%%%%%%%%%%%%%%%%%%%%%%%%%%%%%%%%%%%
   \begin{subfigure}[b]{0.3\textwidth}
     \centering
     \begin{fmffile}{figures/ZZgg1}
       \begin{fmfgraph*}(100,75)
         \fmfbottom{i1,d1,o1}
          \fmftop{i2,d2,o2}
          \fmf{gluon}{i1,v1}
          \fmf{gluon}{v3,i2}
          \fmf{photon}{v2,o1}
          \fmf{photon}{v4,o2}
          \fmf{fermion}{v3,v4}
          \fmf{fermion}{v2,v1}
          \fmf{fermion,tension=0}{v1,v3}
          \fmf{fermion,tension=0}{v4,v2}
          \fmflabel{$g$}{i2}
          \fmflabel{$g$}{i1}
          \fmflabel{$Z$}{o1}
          \fmflabel{$Z$}{o2}
       \end{fmfgraph*}
     \end{fmffile}
     \caption{}
     \label{fig:ZZgg1}
   \end{subfigure}
%%%%%%%%%%%%%%%%%%%%%%%%%%%%%%%%%%%%%%%%
   \begin{subfigure}[b]{0.3\textwidth}
     \centering
     \begin{fmffile}{figures/ZZgg2}
       \begin{fmfgraph*}(100,75)
         \fmfbottom{i1,d1,o1}
         \fmftop{i2,d2,o2}
         \fmf{gluon}{i1,v1}
         \fmf{gluon}{v3,i2}
         \fmf{phantom}{v2,o1}
         \fmf{phantom}{v4,o2}
         \fmf{photon,tension=0}{v2,o2}
         \fmf{photon,tension=0}{v4,o1}
         \fmf{fermion}{v3,v4}
         \fmf{fermion}{v2,v1}
         \fmf{fermion,tension=0}{v1,v3}
         \fmf{fermion,tension=0}{v4,v2}
         \fmflabel{$g$}{i2}
         \fmflabel{$g$}{i1}
         \fmflabel{$Z$}{o1}
         \fmflabel{$Z$}{o2}
       \end{fmfgraph*}
     \end{fmffile}
     \caption{}
     \label{fig:ZZgg2}
  \end{subfigure}
   \begin{subfigure}[b]{0.3\textwidth}
     \centering
     \begin{fmffile}{figures/ZZggH}
       \begin{fmfgraph*}(100,70)
         \fmfleft{i1,i2}
         \fmfright{o1,o2,o3}
         \fmf{gluon}{i1,v1}
         \fmf{gluon}{i2,v2}
         \fmf{fermion,tension=0.7}{v1,v2}        
         \fmf{fermion,tension=0.7}{v2,v3}         
         \fmf{fermion,tension=0.7}{v3,v1}         
         \fmf{dashes,label=$H^{*}$}{v3,v4}
         \fmf{photon}{v4,o1}
         \fmf{photon}{v4,o3}
         \fmf{phantom}{i1,o1}
         \fmf{phantom}{i2,o3}
         \fmflabel{$g$}{i1}
         \fmflabel{$g$}{i2}
         \fmflabel{$Z$}{o1}
         \fmflabel{$Z$}{o3}
       \end{fmfgraph*}
     \end{fmffile}
     \caption{}
     \label{fig:ZZggH}
  \end{subfigure}
   \caption{Lowest-order Feynman diagrams for $ZZ$ production. The \subref{fig:ZZtchannel} $t$-channel 
            and \subref{fig:ZZuchannel} $u$-channel diagrams contribute to \ZZ\ production cross section, while
            the \subref{fig:ZZschannel} $s$-channel diagram is not present in the SM, 
            as it contains a neutral $ZZZ$ or $ZZ\gamma$ vertex. 
            Examples of one-loop contributions to \ZZ\ production via gluon pairs are shown in 
            \subref{fig:ZZgg1}, \subref{fig:ZZgg2} and \subref{fig:ZZggH}.}
   \label{fig:ZZProductionQQandGGDiagrams}
\end{figure}

The production of electroweak gauge boson pairs provides an
opportunity to perform precision studies of the electroweak sector by looking
for deviations from the predicted total and differential production
cross sections, which could be an indication of new resonances or couplings not
included in the Standard Model (SM). Pairs of \Z\ bosons may be produced at lowest order via
quark--antiquark ($q\bar{q}$) annihilation, as well as through gluon--gluon fusion via a quark loop.
In $\sqrt{s} = 8$ \TeV\ proton--proton ($pp$) collisions, approximately 6\%
 of the predicted total cross section is due to
gluon--gluon fusion \cite{Campbell:2011bn}. A pair of \Z\ bosons may
also be produced by the decay of a Higgs boson.  Lowest-order Feynman diagrams
for SM production of \ZZ\ dibosons are given in
\Cref{fig:ZZtchannel,fig:ZZuchannel,fig:ZZgg1,fig:ZZgg2,fig:ZZggH}. These represent the dominant mechanisms
for \ZZ\ diboson production at the Large Hadron Collider (LHC).
The self-couplings of the electroweak gauge bosons are fixed by the form
of the SM Lagrangian. Consequently, neutral triple gauge couplings such as
$ZZZ$ and $ZZ\gamma$ are not present in the SM, making the
contribution from
the $s$-channel diagram zero (\Cref{fig:ZZschannel}).

In addition to precision tests of the electroweak sector of the SM, \ZZ\ diboson measurements
motivate higher-order calculations in perturbative quantum chromodynamics (pQCD) and allow for in-depth tests of pQCD.
Production of \ZZ\ dibosons is a background to the SM Higgs boson process and
to many searches for physics beyond the SM,
and precise knowledge of the cross section is necessary to observe deviations relative to SM predictions. 

\vspace{2em}
\vspace{2em}
Many extensions to the SM predict new scalar, vector, or tensor
particles, which can decay to pairs of electroweak bosons. For example,  diboson
resonances are predicted 
in technicolour models
\cite{Lane:2002sm,Eichten:2007sx,Sannino:2005zf,Andersen:2011yj},
models with warped extra dimensions \cite{Randall:1999ee,Randall:1999vf,Davoudiasl:2000wi}, 
extended gauge models \cite{Altarelli:1989ff,Eichten:1984eu}, and grand unified theories 
\cite{Georgi:1974sy}. Furthermore, extensions to the SM 
such as supersymmetry or extra dimensions predict new particles, which can either produce boson pairs
directly, in cascade decays, or indirectly via loops. At higher orders,
loop contributions involving new particles can lead to effective anomalous neutral triple gauge couplings (aTGCS) as large as $10^{-3}$ \cite{Gounaris:2000tb}.
Any significant deviation in the observed production cross section relative to the SM predictions can indicate a potential source of new physics.
Thus, \ZZ\
production is important not only for precision tests of the electroweak sector
and pQCD, but also for searches for new physics processes.

This paper presents measurements of the fiducial, total and differential cross sections
for \ZZ\ production in $pp$ collisions at a centre-of-mass energy of $\sqrt{s} = 8$
\TeV\ using \ZZLuminosityFbOneDp~\invfb ~of data.
These have been measured by both the ATLAS \cite{Aad:2012awa} and CMS \cite{Chatrchyan:2012sga} Collaborations at 7 \TeV.  
Recently, the ATLAS Collaboration has measured the fiducial
and total cross section for \ZZ\ production at a centre-of-mass
energy of $\sqrt{s} = 13 \TeV$ \cite{Aad:2015zqe} and the cross section as a function of the
invariant mass of the four-lepton system at a centre-of-mass
energy of $\sqrt{s} = 8 \TeV$ \cite{STDM-2014-15-001}. 
The CMS Collaboration has recently measured the \ZZ\ production cross section
at 8 \TeV\ \cite{CMS:2014xja}.  

This paper also presents limits on $ZZZ$ and $ZZ\gamma$ aTGCs within the
context of an effective Lagrangian framework \cite{Hagiwara:1986vm}. 
The limits obtained by both
ATLAS \cite{Aad:2012awa} and CMS \cite{Chatrchyan:2012sga} using the full
7 \TeV\ data sets are approximately 10 to 20 times stricter than limits
set at LEP2 \cite{Alcaraz:2006mx} and the Tevatron \cite{Tevatron:d0}. 
More recently, limits on aTGCs have been set by the CMS
Collaboration using the full 8 \TeV\ data set of
\SI{\CMSLuminosityFbOneDp}{\invfb} in the \zzlmlplmlpprimed\
channel ($\ell = e, \mu, \tau$)
\cite{CMS:2014xja}. CMS has also measured
the \ZZ\ production cross section using the \zzllvv\ decay mode and set limits on aTGCs using the combination of ~\SI{5}{\invfb} of data at 7 \TeV\ and
\SI{\CMSLuminosityFbOneDp}{\invfb} of data at 8 \TeV\ \cite{Khachatryan:2015pba}.

The paper is organized as follows. An overview of the ATLAS detector is given
in \Cref{sec:detector}. \Cref{sec:phasespace} defines the phase space
in which the cross sections are measured, while
\Cref{sec:smpredictions} gives the SM predictions. The simulated signal 
and background samples used for this analysis
are given in \Cref{sec:simulation}. Data samples, reconstruction of
leptons, jets and \met\ , and event selection for each final state are presented in
\Cref{sec:selection}. The estimation of background contributions to
the \zzlmlplmlpprimed\ 
and \zzllvv\ channels, using a combination of simulation-based and data-driven
techniques, is discussed in \Cref{sec:background}. The observed and expected
event yields are presented in \Cref{sec:yields}, while \Cref{sec:acceptance}
describes the correction factors and detector acceptance for this measurement.
\Cref{sec:systematics} describes the experimental and theoretical systematic
uncertainties considered. \Cref{sec:results} presents the results of the
total and differential cross-section measurements. Limits on aTGCs are discussed in \Cref{sec:atgc} 
in the context of an effective Lagrangian framework. Finally, \Cref{sec:conclusion} 
presents the conclusions.

%% file: ATLASDetector.tex
\section{The ATLAS detector}
\label{sec:detector}
The ATLAS detector~\cite{bib:ATLASDetectorPaper} is a multi-purpose
particle physics detector with a forward-backward symmetric cylindrical geometry.
It consists of
inner tracking devices	surrounded by a superconducting solenoid, which
provides  a 2\,T axial magnetic field, 
electromagnetic and hadronic sampling calorimeters and a muon spectrometer (MS)
with a toroidal magnetic field. 

The inner detector (ID) provides  tracking of
charged particles in the pseudorapidity\footnote{ATLAS uses a right-handed coordinate system
  with its origin at the nominal interaction point (IP) in the centre
  of the detector and the $z$-axis along the beam pipe. The $x$-axis
  points from the IP to the centre of the LHC ring and the $y$-axis points
  upward. Cylindrical coordinates ($r$,$\phi$) are used in the transverse
  plane, $\phi$ being the azimuthal angle around the beam pipe. The
  pseudorapidity is defined in terms of the polar 
  angle $\theta$ as 
  $\eta$ = $- \mathrm{ln}~ [\tan\left(\theta/2\right)]$.}
range
$|\eta|<2.5$. It consists
of three layers of silicon pixel detectors and eight layers of silicon
microstrip detectors surrounded by a straw-tube transition radiation tracker 
in the region $|\eta|<2.0$, which contributes to electron identification.
 
The high-granularity electromagnetic (EM) calorimeter utilizes
liquid argon (LAr) as the sampling medium and lead as an absorber, covering the pseudorapidity range
$|\eta|<3.2$. A steel/scintillator-tile calorimeter provides hadronic
coverage for $|\eta|<1.7$. The endcap and forward regions of the
calorimeter system, extending to  $|\eta| = 4.9$, are instrumented with
copper/LAr and tungsten/LAr modules for both the EM and hadronic
measurements.

The MS consists of three large
superconducting toroids, each comprising eight coils, and a system of
trigger chambers and tracking chambers that provide
triggering and tracking capabilities in the ranges $|\eta|<2.4$ and
$|\eta|<2.7$, respectively. 

The ATLAS trigger system \cite{ATLAS_Trigger} consists of a
hardware-based Level-1 trigger followed by a software-based High-Level
Trigger (HLT). It selects events to be recorded for offline
analysis, reducing their rate to about \SI{400}{Hz}.

%% file: PhaseSpace.tex
\section{Phase-space definitions}
\label{sec:phasespace}
This analysis measures the cross section of \ZZ\ diboson production
in a region of kinematic phase space very close to the 
geometric acceptance of the full detector. 
Fiducial cross sections are measured for the \eeee, ~\eemm\ and ~\mmmm\ 
final states in the \zzlmlplmlpprimed\ channel and for the ~\eevv\ and ~\mmvv\ final states in the 
\zzllvv\ channel.   
Final states with leptonic $\tau$ decays are not included as signal in any of the final states considered. 

The information from each final state in both channels is combined to measure the total \ZZ\ production cross section
in a kinematic phase space, referred to as the total phase space, defined by
$66 < m_{\ell^{-}\ell^{+}} < 116 \GeV$, where $m_{\ell^{-}\ell^{+}}$ is the
invariant mass of each  charged lepton pair.  Where there is ambiguity in the choice 
of lepton pairs, the pairing procedure described in Section~\ref{subsubsec:zz4lSelection} is used.  

The kinematic properties of final-state electrons and muons include the contributions from final-state radiated photons
within a distance in the ($\eta, \phi$) plane of $\Delta R = 0.1$ 
around the direction of the charged lepton.\footnote{Angular
separations between particles or reconstructed objects are measured
in the ($\eta, \phi$) plane using $\Delta R$ = $\sqrt{\left(\Delta\phi\right)^2 +
\left(\Delta\eta\right)^2}$.}

\subsection{\texorpdfstring{\zzlmlplmlpprimed}{ZZ -> l-l+l'-l'+} channel}
\label{subsec:zz4lFiducialPhaseSpace}
Three different fiducial phase-space regions are used for the \zzlmlplmlpprimed\
channel of the analysis, one for each decay mode, and selected to increase 
the geometric acceptance by using the forward regions of the detector while controlling 
backgrounds. 
The \Z\ boson pairs are required to decay to \eeee, \eemm, or \mmmm, where the
invariant mass of each opposite-sign, same-flavour lepton pair is required to be within $66 <
m_{\ell^{-}\ell^{+}} < 116 \GeV$. 
The transverse momentum, \pT, of each lepton must be at least
7 \GeV. In the \mmmm\ decay mode, the muons must fall within a pseudorapidity
range $|\eta| < 2.7$. In the \eeee\ decay mode, three electrons
are required to have $|\eta| < 2.5$ and the fourth electron is required to
lie in the pseudorapidity range $|\eta| < 4.9$. In the \eemm\ decay mode,
both muons are required to be within $|\eta| < 2.7$, while for the
electrons, one electron must be central ($|\eta| < 2.5$), while the
second must fall within $|\eta| < 4.9$. The minimum angular separation between any 
two of the four charged leptons must be $\Delta R > 0.2$.

\subsection{\texorpdfstring{\zzllvv}{ZZ -> l-l+vvbar} channel}
\label{subsec:zzllvvFiducialPhaseSpace}
The fiducial phase space for the \zzllvv\ channel is defined by
requiring one \Z\ boson to decay to neutrinos (invisible) and one
\Z\ boson to decay to an $e^{-}e^{+}$ or $\mu^{-}\mu^{+}$ pair.
The invariant mass of the charged lepton pair must lie within $76 <
m_{\ell^{-}\ell^{+}} < 106 \GeV$. Each charged lepton used to form
\Z\ candidates must have transverse momentum $\pT\ > 25 \GeV$ and  
$|\eta| < 2.5$. The charged leptons must be separated by more than 
$\Delta R = 0.3$.
 The axial missing transverse momentum in the event (axial-\met),
which expresses the projection of the transverse momentum  of the 
neutrino pair of the invisibly decaying \Z\ boson ($\vec{p}_{\mathrm{T}}^{~\nu\bar{\nu}}$) onto the direction of the
transverse momentum of the \Z\ boson decaying to charged leptons ($\vec{p}_{\mathrm{T}}^{~Z}$), is defined as
$-\pT^{\nu\bar{\nu}}\cdot\cos(\Delta\phi(\vec{p}_{\mathrm{T}}^{~\nu\bar{\nu}},\vec{p}_{\mathrm{T}}^{~Z}))$.
The axial-\met ~is required to be greater than 90 \GeV.
The \pT-balance between the two \Z\ bosons, defined as 
$|p_{\mathrm{T}}^{\nu\bar{\nu}} - p_{\mathrm{T}}^Z|/ p_{\mathrm{T}}^Z$,
must be less than 0.4. There must be no particle-level jets with $\pT
> 25 \GeV$, $|\eta | < 4.5$ and each jet must have a minimum distance of
$\Delta R =0.3$ from any prompt electron. Particle-level jets are constructed
from stable particles with a lifetime of $\tau >
30$ ps, excluding muons and
neutrinos, using the anti-$k_{t}$  algorithm~\cite{Cacciari:2008gp}
with a radius parameter of $R = 0.4$. 

The definitions of the fiducial phase space for each of the five $ZZ$ final
states under study are summarized in Table \ref{tab:PS}.

\begin{table}[htbp]
\centering
\begin{tabular}{|l|c|c|c|c|c|}
\hline
 \multicolumn{6}{c}{Fiducial Phase Space} \\\hline
Selection &~~~~ \eeee\ ~~~~&~~~ \mumumumu\ ~~~ &~~~ \eemm\ ~~~ &~~~ \eevv\ ~~~&~~~ \mumuvv\ ~~~\\ \hline
Lepton \pt & \multicolumn{3}{|c|}{$> 7$ \GeV} & \multicolumn{2}{|c|}{$>25$ \GeV} \\ \hline

Lepton $|\eta|$ & $ |\eta|_{e_{1}, e_{2}, e_{3}}< 2.5$ & $ |\eta|_{\mu}< 2.7$ & $|\eta|_{e_{1}}<2.5$, $|\eta|_{e_{2}} < 4.9$ & $|\eta|_{e} < 2.5$ & $|\eta|_{\mu}<2.5$\\
%Lepton |\eta| & $ |\eta|_{e_{1}, e_{2}, e_{3}}< 2.5$ & $ |\eta|_{\mu}< 2.7$ & $|\eta|_{e_{1}}<2.5$, $|\eta|_{e_{2}} < 4.9$ & $|\eta|_{e} < 2.5$ & $|\eta|_{\mu}<2.5$\\ 
       &  $|\eta|_{e_{4}}< 4.9$ &                         &$|\eta|_{\mu}<2.7$ &   & \\ \hline
$\Delta R(\ell,\ell^{\prime})$ &  \multicolumn{3}{|c|}{$> 0.2$ } & \multicolumn{2}{|c|}{$> 0.3$} \\ \hline
$m_{\ell^{-}\ell^{+}}$  & \multicolumn{3}{|c|}{$66 < m_{\ell^{-}\ell^{+}} < 116$ \GeV} & \multicolumn{2}{|c|}{$76 < m_{\ell^{-}\ell^{+}} < 106$ \GeV} \\ \hline
Axial-\met  & \multicolumn{3}{|c|}{-} &  \multicolumn{2}{|c|}{$> 90$ \GeV} \\ \hline
\pT-balance &  \multicolumn{3}{|c|}{-} & \multicolumn{2}{|c|}{ $<0.4$} \\ \hline
Jet veto & \multicolumn{3}{|c|}{-}   &
\multicolumn{2}{|c|}{\pT$_{\mathrm{jet}}>25$ \GeV, $|\eta|_{\mathrm{jet}} < 4.5$,}\\
  & \multicolumn{3}{|c|}{}   & \multicolumn{2}{|c|} {and $\Delta R(e, {\mathrm{jet}}) > 0.3$}  \\ \hline

\hline
\end{tabular}
\caption{\label{tab:PS} Fiducial phase-space definitions for each of the five \ZZ\ final states under study.}
\end{table}

%% file: SMPredictions.tex
\section{Standard Model predictions}
\label{sec:smpredictions}

The fiducial and total cross-section predictions for SM \ZZ\
production reported in this paper are evaluated with
\powhegbox\ \cite{Alioli:2010xd,Melia:2011tj} at
next-to-leading order (NLO) in QCD and are supplemented with
predictions from \ggtwovv~\cite{Kauer:2012hd,Kauer:2013qba} to account for \ZZ\ production via
gluon--gluon fusion at leading order (LO) in the gluon-induced process. Interference effects with
SM Higgs boson production via gluon--gluon fusion as well as off-shell Higgs boson
production effects are              
considered, based on recent calculations \cite{Kauer:2013qba}.
The contribution of the gluon--gluon
initial state to the fiducial cross sections is about 6\% for the
\zzlmlplmlpprimed\ channel and about 3\% for the
\zzllvv\ channel. 
All computations are performed using dynamic
renormalization and factorization scales ($\mu_{\mathrm{R}}$ and $\mu_{\mathrm{F}}$)
equal to the invariant mass of the $ZZ$ system ($m_{\ZZ}$) as the baseline,
and the CT10 parton distribution function (PDF) set \cite{Lai:2010vv}. 

The results from \powhegbox\ are corrected for virtual NLO electroweak (EW)
effects~\cite{Bierweiler:2013dja}, 
applied as reweighting factors on an event-by-event basis, following the method described in
Ref.~\cite{Gieseke:2014gka}. 
As a result, the fiducial cross-section predictions for the
\zzlmlplmlpprimed\ and \zzllvv\ channels are reduced by $~4$\% and
$~9$\% respectively. 
\begin{table}[!htbp]
    \begin{centering}
    \setlength{\tabcolsep}{.16667em}
    \renewcommand{\arraystretch}{1.5}
    \begin{tabular}{lcrll} \toprule\toprule
        $\sigma^{\mathrm{fid}}_{ZZ \to \eeee}$      &~~=~~& 6.2  & $^{+0.6}_{-0.5}$  &\fb \\
        $\sigma^{\mathrm{fid}}_{ZZ \to \eemm}$      &=& 10.8 & $^{+1.1}_{-1.0}$  &\fb \\
        $\sigma^{\mathrm{fid}}_{ZZ \to \mmmm}$      &=& 4.9  & $^{+0.5}_{-0.4}$  &\fb \\
        $\sigma^{\mathrm{fid}}_{ZZ \to \eevv}$      &=& 3.7  & $\pm$ 0.3  &\fb \\
        $\sigma^{\mathrm{fid}}_{ZZ \to \mumuvv}$      &=& 3.5  & $\pm$ 0.3  &\fb \\
        \midrule
        $\sigma_{pp \to ZZ}^{\mathrm{total}}$    &=& 6.6 & $^{+0.7}_{-0.6}$  &\pb\\
        \bottomrule
        \bottomrule
    \end{tabular}
    \caption{Predicted fiducial and total \ZZ\ production cross sections.
             The considered systematic uncertainties and the accuracy in pertubation 
             theory are detailed in the text.\label{tab:xsecPredicts}}
\end{centering}
\end{table}

The SM predictions for the fiducial and total \ZZ\ production cross
sections in the regions defined in 
\Cref{sec:phasespace} and including the EW corrections are summarized in \Cref{tab:xsecPredicts}. 
The systematic uncertainties shown in the table include a PDF uncertainty of
$^{+4.2\%}_{-3.3\%}$~\cite{Baglio:2013toa} applied to the results from
both the \powhegbox\ and \ggVV\ generators. For the \powhegbox\ contribution, a scale uncertainty of 
$^{+3.1\%}_{-2.3\%}$~\cite{Baglio:2013toa} is included.
For the gluon--gluon fusion contribution, recent
publications~\cite{Li:2015jva,Caola:2015psa,Passarino:2013bha} suggest 
an increase of the \ZZ\ production cross section by up to a
factor of about two, when the calculation is performed at higher orders in QCD.
This calculation is sensitive to the choice of PDF set and even more to the
$\mu_{\mathrm{R}}$ and $\mu_{\mathrm{F}}$ scales. 
As this correction is not available differentially for all distributions and all final states
 analysed in this paper, no reweighting  is applied to the prediction
of \ggVV. In order to account for these higher-order QCD effects, the scale uncertainty for \ggVV is set
to $\pm 60\%$. 
PDF and scale uncertainties are added linearly following the
recommendation of Ref.~\cite{Dittmaier:2011ti}. 
The jet veto uncertainty obtained using the Stewart and Tackmann method \cite{Stewart:2011cf} is shown in Table~\ref{tab:sys} and 
is added in quadrature to the systematic uncertainty of the fiducial cross sections for each
\zzllvv\ final state.
This method uses samples for \ZZ\ and \Z\ production when varying the QCD scale to estimate the uncertainty associated with
selecting events with zero jets by examining the uncertainty for selecting events with one or more jets. 
This approach is conservative and it covers further
uncertainties from  higher-order QCD effects.

The contribution to the cross section predicted with \powhegbox\ is known to increase by approximately 5\% when considering NNLO QCD effects ~\cite{Grazzini:2015hta,Cascioli:2014yka}.
This enhancement is not considered in the theoretical prediction used in this paper.

%% file: SimulationSamples.tex
\section{Simulated event samples}
\label{sec:simulation}
Simulated samples \cite{Aad:2010ah} are used to correct the measured
distributions for detector effects and acceptance and to determine or validate
some background contributions. Production and subsequent decays of \ZZ\ pairs are
simulated using \powhegbox\ at NLO in the $q\bar{q}$ process, and
\ggtwovv\ at LO in the gluon-induced
process, both interfaced to \pythia
8~\cite{Sjostrand:2007gs} for parton showering and underlying-event modelling,
with the CT10 PDF set. 
In each case, the simulation includes the interference
terms between the \Z\ and $\gamma^{*}$ diagrams.
The NLO EW corrections are applied to the
\powhegbox predictions as explained in the previous section.

Moreover, the \powhegbox generator interfaced to \herwig
\cite{Corcella:2000bw} and \Jimmy \cite{Butterworth:1996zw} is
used to estimate systematic uncertainties due to the choice of parton shower
and underlying-event modelling. The LO multi-leg generator \sherpa~\cite{Gleisberg:2008ta}
with the CT10 PDF set is used to assign
systematic uncertainties due to the choice of event generator as well as to 
generate signal samples with  $ZZZ$ and $ZZ\gamma$ aTGCs.  

The LO generator \alpgen~\cite{Mangano:2002ea} using the CTEQ6L1
PDFs~\cite{Pumplin:2002vw} and interfaced to \pythia~\cite{Sjostrand:2006za} is used to simulate $Z$+jets and $W$+jets background
samples. The same generator interfaced to \herwig is used to model the
$W\gamma$ process. The diboson production processes $WW$ and $WZ$ are generated
with \powhegbox interfaced to \pythia 8 using the CT10 PDFs. Top quark pair
production ($t\bar{t}$) is simulated with \mcatnlo~\cite{Frixione:2002ik} using
the CT10 PDFs. Single-top production, including $Wt$ production, is modelled
with \mcatnlo \cite{Frixione:2005vw}, interfaced to \herwig, and AcerMC \cite{Kersevan:2004yg} using 
the CTEQ6L1 PDFs. The LO generator \madgraph~\cite{Stelzer:1994ta} using the 
CTEQ6L1 PDFs is used to model the $ZZZ^{*}$, $ZWW^{*}$ and $t\bar{t}Z$
processes.  Events with two hard interactions in a $pp$ collision (double proton
interactions, DPI) that each produce a \Z\ boson decaying to leptons are
simulated using \pythia 8 with the CTEQ6L1 PDF set.   

The signal and background generated Monte Carlo (MC) samples are
passed through the ATLAS detector simulation \cite{Aad:2010ah} based on \textsc{GEANT4}
\cite{Agostinelli:2002hh}. Additional inelastic $pp$
interactions (pile-up) 
are included in the simulation. The MC events are reweighted to reproduce
the distribution of the mean number of interactions per bunch crossing observed
in data.

%% file: ObjectSelectionAndEvtReco.tex
\subsection{Data samples}
\label{subsec:dataSamples}

The measurement presented in this paper uses the full data set of $pp$
collisions at a centre-of-mass energy of $\sqrt{s} = 8 \TeV$ collected with
the ATLAS detector at the LHC in 2012.  
The data corresponds to a total integrated luminosity of
\SI{20.3}{\ifb}, with an uncertainty of 1.9\% \cite{Lumi8TeVPaper}. 
The absolute luminosity scale and its uncertainty are derived from
beam-separation scans performed in November 2012. 
All events were required to satisfy basic
quality criteria indicating stable beams and good operating characteristics of the
detector during data taking. 
The data analysed were selected using
single-lepton triggers \cite{Aad:2014sca,ATLAS-CONF-2012-048} with
isolation requirements and thresholds of 24 \GeV\ for the
transverse momentum (energy) of muons (electrons). 

During each bunch crossing, several $pp$ collisions take place, 
which results in multiple vertices being reconstructed. To
ensure that the objects analysed originate from the products of the hard-scattered $pp$
collision, and to reduce contamination from cosmic rays, the
primary vertex is chosen to be the vertex with the highest sum of the squared
transverse momenta of the associated ID tracks. 

\subsection{Reconstruction of leptons, jets, and \met\ }
\label{subsec:physicsObjectReco}
Muon candidates are identified by tracks, or track segments,
reconstructed in the MS and matched to tracks
reconstructed in the ID \cite{Aad:2014rra}.  
Muons within $|\eta| < 2.5$ are referred to as ``central muons''.
Muons within $2.5 < |\eta| < 2.7$, where there is no ID
coverage and they are reconstructed only in the MS, are referred to as ``forward muons''. 
In order to recover efficiency at $| \eta | < 0.1$ where $\phi$ coverage in the MS is reduced 
due to mechanical supports and services,
``calorimeter-tagged'' muons are reconstructed using
calorimeter energy deposits to tag ID tracks.
In the \zzlmlplmlpprimed\ channel all three types of muons, 
``central'' with $\pT > 7 \GeV$, ``forward'' with $\pT > 10 \GeV$ and ``calorimeter-tagged'' with $\pT > 20 \GeV$ are
used, while in the \zzllvv\ channel, only ``central'' muons 
with $\pT\ > 25 \GeV$ are used.
For muons with a track in the ID (``central'' and
``calorimeter-tagged'' muons), the ratio of the transverse impact
parameter, $d_0$, with respect to the primary vertex, to its
uncertainty ($d_0$ significance), must be smaller than $3.0$ and the
longitudinal impact parameter, $|z_0| \times \sin\theta$, must be less
than $0.5$ mm. Isolated muons are then selected based on track or
calorimeter requirements. 
Track isolation is imposed on ``central'' and ``calorimeter-tagged''
muons, by requiring the scalar sum of the \pT\ of the tracks originating from the primary vertex inside a
cone of size $\Delta R = 0.2$ around the muon to be less than
15\% of the muon \pT. Similarly, calorimeter isolation requires the
sum of the calorimeter transverse energy in a cone of size $\Delta R = 0.2$ 
around the muon candidate to be less than 15\% of the muon \pT. 
For the \zzllvv\ channel, both track and calorimeter isolation are
imposed on muons, while for the  \zzlmlplmlpprimed\ channel, for
``central'' muons, calorimeter isolation is not required, as it 
does not offer any extra background rejection, and for ``forward'' muons, where track isolation is
not possible, only calorimeter isolation is required.    

Electron candidates in the central region are reconstructed from energy clusters in the
calorimeter matched to an ID track \cite{ATLAS-CONF-2014-032}. 
The lateral and transverse shapes of the cluster must be consistent
with those of an electromagnetic shower. 
The transverse energy of the electron, \ET, must 
be greater than $7 \GeV$ for the \zzlmlplmlpprimed\ channel and
greater than $25 \GeV$ for the \zzllvv\ channel, while the pseudorapidity  
of the electromagnetic cluster for both channels must be $|\eta| < 2.47 $.
To ensure that electron candidates originate from the primary
vertex, the $d_0$ significance of the electron must be smaller than $6.0$ and the
longitudinal impact parameter, $|z_0|\times \sin\theta$, must be less than $0.5$ mm.
The electron candidates must be isolated; therefore, the scalar sum of
the transverse momentum of all the tracks inside a cone of size 
$\Delta R = 0.2$ around the electron must be less than $15\%$ of the
\pT\ of the electron. 
Calorimeter isolation requires the total transverse energy, \ET,
corrected for pile-up effects in an isolation cone of size $\Delta R =
0.2$ to be less than $15\%$ of the electron \pT\ and is required only for
the \zzllvv\ channel.  

To further increase the detector acceptance in the
\zzlmlplmlpprimed\ channel, ``forward'' electrons are used,
extending 
the pseudorapidity coverage to $2.50 < |\eta| < 3.16$ and $3.35 < |\eta|
< 4.90$ \cite{Aad:2014fxa}. These ``forward'' electrons have $\ET\ > 20 \GeV$, without
any track or calorimeter isolation requirements.
Beyond $|\eta| = 2.5$ there is no ID coverage for tracking, so these electrons are
reconstructed from calorimeter information alone. No calorimeter isolation is used for electrons in this region as the calorimeter 
segmentation is too coarse.  

The missing transverse momentum, with magnitude \met, is defined as the negative vector sum of
the transverse momenta of reconstructed muons, electrons, and jets as well as
calorimeter cells not associated to objects.
Calorimeter cells are calibrated to the jet energy scale (JES) if they are associated
with a jet and to the electromagnetic energy scale otherwise ~\cite{ATLAS-CONF-2013-082}. 

Jets are reconstructed using
the anti-$k_{t}$ algorithm \cite{Cacciari:2008gp} with a 
radius parameter $R = 0.4$, using topological clusters of energy
deposition in the calorimeter.  
Jets arising from detector noise or non-collision events are 
rejected. The jet energy is corrected to account for detector and
pile-up effects and is calibrated to account for the different
response of the calorimeters to electrons and hadrons, using a
combination of simulations and in situ techniques
~\cite{ATLAS_Jets1,ATLAS_Jets3,ATLAS-CONF-2013-083}.    
In order to reject jets from pile-up, 
the summed scalar \pT\
of tracks associated with both the jet and the primary vertex
is required to be greater than 
50\% of the  summed scalar \pT ~of all tracks associated with the 
jet. This criterion is only applied to
jets with $\pT < 50 \GeV$ and $| \eta | < 2.4$.  
Jets used in this analysis are required to have $| \eta | <
4.5$ and $\pT > 25 \GeV$. 
Jets that are within  $\Delta R = 0.3$ to an electron or muon that passes the
selection requirements are not considered in the analysis.

%% file: EventSelection.tex
\subsection{Event selection}
\label{subsec:zzEventSelection}
\subsubsection{\texorpdfstring{\zzlmlplmlpprimed}{ZZ -> l-l+l'-l'+} selection}
\label{subsubsec:zz4lSelection}
The \zzlmlplmlpprimed\ events are characterized by two pairs of oppositely charged,
same-flavour leptons. Events fall into three categories: \eeee, \eemm\ and
\mmmm. Selected events are required to have exactly four isolated
leptons above the \pT\ threshold. At least one lepton with $\pT > 25 \GeV$ must
be matched to a trigger object. In the ~\eeee\ and ~\mmmm\ decay modes, there is an
ambiguity when pairing leptons to form \Z\ candidates. A pairing procedure to
form the candidates is used, which minimizes the quantity $|m_{\ell^{-}\ell^{+}}
- m_{Z}| + |m_{\ell^{\prime\, -}\ell^{\prime\, +}} -  m_{Z}|$, where
$m_{\ell^{-}\ell^{+}}$, and $m_{\ell^{\prime\, -}\ell^{\prime\, +}}$ are the 
invariant masses of the two lepton pairs of a given pairing from the quadruplet, and $m_{Z}$ is the \Z\  mass
\cite{Agashe:2014kda}. The two \Z\ candidates must have masses in the range $66
< m_{\ell^{-}\ell^{+}} < 116 \GeV$. All leptons are required to be separated by
$\Delta R > 0.2$. 
Each event is allowed to have a maximum of one extension lepton per
category (forward electron, forward muon, or calorimeter-tagged muon)
and each lepton pair may only have one extension lepton. 
In this way, an event must contain at least two central leptons and may contain
 two extension leptons of different types, as long as they are each paired with a central
lepton. Events with a forward electron have the additional requirement that the
central electron that is paired with the forward electron must have a transverse
momentum of at least 20 \GeV\ instead of 7 \GeV.
\subsubsection{\texorpdfstring{\zzllvv}{ZZ -> l-l+vv} selection}
\label{subsubsec:zzllvvSelection}
In the \zzllvv\ channel, final states with
electron or muon pairs and large \met\ are considered. Candidate events must
have exactly two opposite-sign, same-flavour isolated leptons of $\pT
> 25 \GeV$. At least one of the two leptons must be matched to a trigger
object. The invariant mass of the leptons must
be in the range $76 < m_{\ell^{-}\ell^{+}} < 106 \GeV$. 
The mass-window requirement is stricter than in the \zzlmlplmlpprimed\ channel in
order to suppress backgrounds, which could produce real or fake lepton
pairs close to the \Z\ mass. 
Leptons are also required to have an angular separation of $\Delta R > 0.3$.
The selection of \zzllvv\ candidate events requires that the \metvec\
be highly anti-collinear with the $\vec{p}_{\mathrm{T}}$ of the \Z\ candidate decaying to charged leptons. The quantity
used is referred to as axial-\met\ and is given by
$-\met\cdot\cos(\Delta\phi(\metvec,\vec{p}_{\mathrm{T}}^{~Z}))$, where $\vec{p}_{\mathrm{T}}^{~Z}$ is the
transverse momentum of the \Z\ candidate.
The axial-\met\ is required to be above
90 \GeV. This requirement is particularly effective in removing
\Z\ +jets background, as mismeasured \met\ would in general not have the \metvec\ 
anti-parallel to the  $\vec{p}_{\mathrm{T}}$  of the
\Z\ candidate. The \pT-balance, defined by $| \met -
p_{\mathrm{T}}^{Z}|/ p_{\mathrm{T}}^{Z}$, is required to be less than
0.4 in order to distinguish the signal \zzllvv\ from
the background, such as \Z\ + jets. In order to suppress the \ttbar\ and single-top-quark  backgrounds, events are required not
to have any reconstructed jet with $\pT> 25 \GeV$ and $|\eta | <
4.5$.  This requirement is referred to as the ``jet veto''. Finally, to suppress
$WZ$ background, a veto on a third electron (muon) with $\pT> 7 \GeV$
(6 \GeV) is applied.

%% file: BackgroundEstimate.tex
\section{Background estimation}
\label{sec:background}

\subsection{\texorpdfstring{\zzlmlplmlpprimed}{ZZ -> l-l+l'-l'+} backgrounds}
\label{subsec:zz4lBkgd}
Backgrounds to the \zzlmlplmlpprimed\ channel are events in which four objects identified as isolated, prompt
leptons have paired-lepton invariant masses in the signal region $66 < m_{\ell^{-}\ell^{+}} < 116$.
The leptons of background events in the \zzlmlplmlpprimed\ channel can either
be ``true'' leptons from the decays of \Z\ bosons, $W^{\pm}$ bosons, or top
quarks or they can be ``fake'' leptons that are defined as jets which
are misidentified as leptons or leptons that come from hadronic decays.
Background events in which all four leptons are true leptons are called the ``irreducible background''
 as these events have the same signature as the signal events in this channel. In the SM, there are few final states with significant cross sections
that can produce four true leptons.  The largest sources of irreducible backgrounds are $t\bar{t}Z$ and $ZZZ^{*}/ZWW^{*}$ production and events with
 DPI that separately produce \Z\ bosons that each decay to two leptons.  The contributions from each of these background sources are
estimated from MC simulations that have been scaled to
\ZZLuminosityFbOneDp~\invfb ~and can be found in Table \ref{table:bck_irred4l}.
The systematic uncertainty for the irreducible background is neglected.
The cross sections for these processes are much smaller than for the signal, and their overall contribution to the total background is small.

\begin{table}[htbp]
  \renewcommand{\arraystretch}{1.07}
  \centering
  \begin{tabular}{lcccc}
    \toprule\toprule  
    Source              &   \multicolumn{1}{c}{\eeee}   & \multicolumn{1}{c}{\mmmm}  &   \multicolumn{1}{c}{\eemm} & \llll \\
    \midrule
    $ZZZ^{*}/ZWW^{*}$   &       0.12 $\pm$ 0.01  &  0.19 $\pm$ 0.01 &   0.28 $\pm$ 0.02  & 0.58 $\pm$ 0.02 \\
    DPI                 &       0.13 $\pm$ 0.01  &  0.15 $\pm$ 0.01 &   0.29 $\pm$ 0.01  & 0.57 $\pm$ 0.02 \\
    $t\tbar~Z$          &       0.15 $\pm$ 0.03  &  0.16 $\pm$ 0.03 &   0.35 $\pm$ 0.05  & 0.66 $\pm$ 0.07 \\
    \hline
    Total irreducible background  &     0.40 $\pm$ 0.04  &  0.50 $\pm$ 0.04 &   0.93 $\pm$ 0.05  & 1.82 $\pm$ 0.08 \\
    \hline\hline
  \end{tabular}
  \caption{Number of events from the irreducible background SM sources that can produce four true leptons
      scaled to \ZZLuminosityFbOneDp~\invfb. The full event selection is applied along with all
           corrections and scale factors. The errors shown are statistical only. }

  \label{table:bck_irred4l}
\end{table}

Background events containing one or more fake leptons, constitute the
``reducible background''. The dominant reducible background
contributions to \zzlmlplmlpprimed\ production are $Z$ + jets, $WW$ + jets, 
and top quark (\ttbar\ and single-top quark)
events in which two prompt leptons are paired with two jets or leptons from a
heavy-flavour decay which are misidentified as isolated leptons.
Additional background arises from $WZ$+jets events
containing three true leptons and one fake lepton.
To estimate backgrounds containing fake leptons, the data-driven method employed in the ATLAS measurement at 7 \TeV\ \cite{Aad:2012awa}
is used and only a summary of the relevant parameters is given here.

The data-driven background estimate requires identifying events with two or three selected leptons, with the remaining leptons satisfying
a relaxed set of criteria.  The relaxed set of criteria is defined for each lepton type. 
For muons,  the relaxed criteria give fully selected muons except that they either fail the isolation requirement or fail
the impact parameter requirement but not both. For electrons with $|\eta| <
2.47$, the relaxed criteria give clusters in the electromagnetic
calorimeter matched to ID tracks that fail either the strict identification requirement  or the isolation requirement but not both.
For electrons with  $|\eta| > 2.5$, the relaxed criteria give electromagnetic
clusters that are reconstructed as electrons but fail the identification
requirement. All events are otherwise required to satisfy the full event selection.

The expected number of reducible background \llll\ events, $N(\mathrm{BG})$, is calculated as:
\begin{linenomath}
\begin{equation}
    N(\textrm{ BG}) = [ N_{\mathrm{data}}(\ell\ell\ell j) - N_{\ZZ}(\ell\ell\ell j) ] \times f - [ N_{\mathrm{data}}(\ell\ell j j) - N_{\ZZ}(\ell\ell j j) ] \times f^2
\label{eq:zz4l_bkg_formula}
\end{equation}
\end{linenomath}
where double counting from $\ell \ell \ell j$ and $\ell \ell j j$ events is
accounted for, and the terms $N_{ZZ}(\ell\ell\ell j)$ and $N_{\ZZ}(\ell\ell j
j)$ are  MC estimates correcting for contributions from signal
\zzlmlplmlpprimed\ events having one or two real leptons that instead satisfy the relaxed lepton selection criteria ($j$).

The factor $f$ is calculated as a function of the \pt\ and \eta\ of the fake lepton and is the ratio of the probability for a fake lepton to
satisfy the full lepton selection criteria to the probability of the fake lepton only satisfying the relaxed lepton criteria.  It is measured in a 
control sample of data events that contains a \Z\ boson candidate consisting of a pair of isolated same-flavour opposite-sign electrons or muons.  In these
events, $f$ is measured using the leptons and relaxed leptons not assigned to the \Z\ boson and is found to vary
from $0.082\pm0.001$ ($0.33\pm0.01$) for $\pT<10 \GeV$ to  $0.027\pm0.001$ ($0.72\pm0.11$) for $\pT>40 \GeV$ for electrons (muons).
The quoted uncertainties are statistical.  The weighted number of data
events for each of the ingredients in \Cref{eq:zz4l_bkg_formula} can be
found in Table~\ref{table:zzfakes}.

The systematic uncertainty in the reducible background 
is estimated using two additional and independent methods.
The maximum difference between each additional estimate and the nominal estimate is taken as the systematic uncertainty.
The first additional method is to count the number of events in data with one pair of
opposite-sign, same-flavour leptons and another pair of same-sign, same-flavour leptons ($\ell^{+}\ell^{-}\ell^{\prime \pm}\ell^{\prime \pm}$)
that satisfy the complete selection criteria while subtracting the number of $ZZ$ events that have one lepton with misidentified charge from MC simulation.
The second additional method removes the parameterization of the factor $f$ in \pt\ and \eta\ and uses \Cref{eq:zz4l_bkg_formula} 
to recalculate the background estimate. 
The systematic uncertainty is estimated to be $\pm$2.8 events (63\%) in the \eeee final state,
$\pm$0.9 events (48\%) in the \mumumumu final state, $\pm$3.9 events (43\%) in the \eemm final state and $\pm$7.1 events (46\%) in the combined \llll channel.

\begin{table}
\begin{center}
\begin{tabular}{lS[table-format=-1.2,table-figures-uncertainty=1]S[table-format=-1.2,table-figures-uncertainty=1]S[table-format=-1.2,table-figures-uncertainty=1]S[table-format=-1.2,table-figures-uncertainty=1]}
\hline \hline
Ingredients in Eq. \ref{eq:zz4l_bkg_formula}             &  \multicolumn{1}{c}{\eeee}   & \multicolumn{1}{c}{$\mumumumu$}  &   \multicolumn{1}{c}{$\eemm$} &  \multicolumn{1}{c}{Combined ($\llll$)} \\ \hline
$(+) N_{\mathrm{data}}(\ell\ell\ell j) \times f $  & 8.6  \pm 0.7  & 4.8  \pm 2.4  &  16.0 \pm  3.5  &  29.3 \pm 4.3 \\
$(-) N_{ZZ}(\ell\ell\ell j) \times f $    & 0.58 \pm 0.01 & 1.96 \pm 0.02 &  2.82 \pm  0.02 &  5.36 \pm 0.03 \\
$(-) N_{\mathrm{data}}(\ell\ell j j) \times f^{2}$ & 3.6  \pm 0.1  & 1.0  \pm 0.4  &  4.1  \pm  0.6  &  8.8  \pm 0.8 \\
$(+) N_{ZZ}(\ell\ell j j) \times f^{2}$   & 0.00 \pm 0.01 & 0.02 \pm 0.08 &  0.02 \pm  0.02 &  0.04 \pm 0.02 \\
\hline
Background estimate, & \multicolumn{1}{r}{4.4 $\pm$ 0.7 (stat)\ } & \multicolumn{1}{r}{1.8  $\pm$ 2.4 (stat)\ }  &  \multicolumn{1}{r}{9.0   $\pm$  3.6 (stat)\ } &  \multicolumn{1}{c}{15.2   $\pm$ 4.4 (stat)\ }\\
$N(\textrm{ BG})$            & \multicolumn{1}{r}{$~~~~\pm$ 2.8 (syst)} & \multicolumn{1}{r}{$~~~~\pm$ 0.9 (syst)} & \multicolumn{1}{r}{$~~~~\pm$ 3.9 (syst)} &\multicolumn{1}{c}{$~~~~~\pm$ 7.1 (syst)} \\

\hline \hline
\end{tabular}
\end{center}
\caption{The number of \ZZ \ background events from sources with fake leptons estimated using the data-driven fake-factor method in \ZZLuminosityFbOneDp~\invfb ~of data. The uncertainties quoted are statistical only, unless otherwise indicated, and combine the statistical uncertainty in the number of observed events of each type and the statistical uncertainty in the associated fake factor. The systematic uncertainty is shown for the background estimate in each final state. }
\label{table:zzfakes}
\end{table}

\subsection{\texorpdfstring{\zzllvv}{ZZ -> l-l+vv} backgrounds}
\label{subsec:zz2lvBkgd}   
The main background sources for the \zzllvv\ channel are processes
with two true isolated leptons and \MET\ in the event. Such processes can be diboson $WZ$ events, as
well as \zzlmlplmlpprimed\ , $\ttbar$, $W^{-}W^{+}$, $Wt$, $ZZ\to\tau\tau\nu\nu$ and $Z \to
\tau^{-}\tau^{+}$. Additionally, processes such as the
production of a $Z$ or a $W$ boson in association with jets ($Z$ + jets,
$W$+ jets), as well as multijets, may satisfy the \zzllvv\ ~event selection criteria and contribute to the background. The
backgrounds from diboson $WZ$ and \zzlmlplmlpprimed\ production are
estimated from MC simulations, while, for all other background sources
mentioned above, a combination of data-driven techniques and MC
simulation is used for their estimation. 
\subsubsection{Backgrounds from leptonic $WZ$ decays and \zzlmlplmlpprimed\ decays}
\label{subsubsec:Z4lWZBkgd}
Background events with multiple true isolated leptons may be $WZ$ events in which both bosons decay
leptonically and one of the three leptons is not 
reconstructed in the detector, and \zzlmlplmlpprimed\ events in which two
of the four leptons are not reconstructed. After all selections, the $WZ$ events constitute
the dominant background for the \zzllvv\ channel. 
Although this background is estimated only from MC simulation, the
simulation is validated using events in dedicated 
control regions, $eee$, $\mu\mu\mu$, $\mu\mu e$ and $ee\mu$, in which a third lepton is required in addition to the full selection criteria.  
No significant difference between data and MC simulation is observed in the three-lepton control regions and
therefore no scaling is applied to the MC prediction in the signal region.  The background due to $WZ$ events is 
estimated to be $16.7\pm1.1 \mathrm{(stat)} \pm 1.7\mathrm{(syst)}$ events in
the \eevv\ final state and $18.5\pm1.0\mathrm{(stat)}\pm1.5\mathrm{(syst)}$
events in the \mumuvv\ final state, and constitutes more than 50\% of the total background. The background due to \zzlmlplmlpprimed\ is small, contributing
less than 2\% to the total background as shown in \Cref{tab:backgrounds-llvv}.
The dominant uncertainties of this background source are
theoretical, followed by uncertainties in the reconstruction
correction factors applied to the simulated events. The dominant
theoretical uncertainty is in the choice of QCD scale (about $7\%$), while the PDF uncertainties are less than 1\%.

\subsubsection{Backgrounds from \texorpdfstring{\ttbar}{tt}, \texorpdfstring{$W^{-}W^{+}$}{W-W+}, \texorpdfstring{$Wt$}{Wt}, 
\texorpdfstring{$ZZ\to\tau\tau\nu\nu$}{ZZ -> tau tau nu nu} and \texorpdfstring{$Z \to \tau^{-}\tau^{+}$}{Z -> tau-tau+} }
\label{subsubsec:ttbarWtZtauBkgd}
The background contribution from these processes is measured by
extrapolating from a control region formed by events with one electron
and one muon (instead of two electrons or two muons), which otherwise
satisfy the full \zzllvv\ selection. This $e\mu$ region is free from
signal events. The extrapolation from the $e\mu$ control region to the
$ee$ or $\mu\mu$ signal regions takes into account the relative
branching fractions (2 : 1 : 1 for $e\mu : ee : \mu\mu $ ), as well as the
ratio of the efficiencies $\epsilon_{ee}$ or $\epsilon_{\mu\mu}$, for
the $ee$ or $\mu\mu$ selections to the efficiency $\epsilon_{e\mu}$
for the $e\mu$ selection. These efficiency ratios are not equal to unity
because of the difference in electron and muon reconstruction and
trigger efficiencies
\cite{Aad:2012awa}.  
This background is estimated to be $13.3\pm 3.2 \mathrm{(stat)}\pm 0.2
\mathrm{(syst)}$ events in the \eevv\ final state 
and $15.4\pm 3.6\mathrm{(stat)} \pm 0.3\mathrm{(syst)}$ events in the \mumuvv\ final
state, and accounts for the 41\% and 46\% of the total background in the
\eevv\ and \mumuvv\ final states, respectively.
The dominant uncertainty for these background contributions is
statistical because of the limited number of events in the control region,
while additional uncertainties are due to systematic uncertainties in
the normalization of the simulated samples used to correct the $e\mu$
contribution in data and the systematic uncertainty in the efficiency
correction factors.

\subsubsection{$W$+jets and multijet background}
\label{subsec:multijet}
Leptons originating from semileptonic decays of heavy-flavour hadrons may also contribute in the
electron or muon final states. However, this background is highly
suppressed because of the dilepton mass requirement in the signal
selection. 
The $W$+jets and multijet background is estimated using the ``matrix method'' technique
\cite{Aad:2012mza}. The fraction of events in the signal region that contain at least one fake lepton is estimated by extrapolating from a background-dominated control region to the signal region using factors measured in data.
The contribution of this background to the total background is 8\% in the \eevv\ final state and negligible in the \mumuvv\ final state.   
The dominant systematic uncertainty for this background
is  due to the uncertainty in the extrapolation factors and the limited number of 
events in the control regions.

\subsubsection{$Z$+jets background}
\label{subsec:zplusjets}
Occasionally, events with one \Z\ boson produced in association with jets
or with a photon ($Z$+jets, or $Z + \gamma$) may mimic signal events
if they have large \MET\ due to the
mismeasurement of the jets or the photon. 
This background of events with a \Z\ boson and jets is estimated by selecting events in data with a high-\pt\ photon and jets, and reweighting these events to account 
for differences in the \Z\ boson and photon \pt\ spectra and reconstruction efficiencies.  
These weights are determined in a low-\MET\ control region. 
To remove contamination to single-photon events, subtraction of
non-($\gamma$ + jet) events (e.g. $Z(\to\nu\bar{\nu}) + \gamma$) is performed.
The full signal selection is applied to the single-photon plus jets
events, and the background is estimated by reweighting these events using weights determined from the low-\MET\ control region. 
The procedure is repeated in bins of $\pT^{Z}$ in order to obtain the
\pt\ distribution of the $Z$+jets and $Z + \gamma$ backgrounds. 
As shown in \Cref{tab:backgrounds-llvv}, this background is negligible in both the \eevv\ and \mumuvv\ final states. 
 The dominant uncertainty for this background is due to the statistical
uncertainty of non-($\gamma$+jet)
events, which are subtracted from the $\gamma + $\ jets sample. 

\subsubsection{Background summary for \zzllvv}

A summary of both the simulation-based and data-driven backgrounds in the \zzllvv\ channel is given 
in \Cref{tab:backgrounds-llvv}. The largest background contributions come from $WZ$ and
\ttbar\ , $W^{-}W^{+}$,  $Wt$, $ZZ\to\tau\tau\nu\nu$, and $Z \to \tau^{-}\tau^{+}$. 
Several of the techniques used to determine the data-driven backgrounds require subtraction of  non-background processes so that negative background estimates may result when extrapolating to the signal region.
Background estimates are required to have a minimum value of zero but are allowed to fluctuate positively within their uncertainty 
bounds during the cross-section extraction.

\begin{table}[htbp]
    \renewcommand{\arraystretch}{1.12}%
  \centering
  \begin{tabular}{lrr}
    \toprule
    \toprule
    Source              &     \multicolumn{1}{c}{\eevv}       &  \multicolumn{1}{c}{\mumuvv}       \\
    \midrule
    $WZ$         		        		& $16.7\pm1.1\pm1.7 $ &  $18.5\pm1.0\pm1.5$             \\
    \zzlmlplmlpprimed  					&  $0.6\pm0.1\pm0.1$   &  $0.6\pm0.1\pm0.1 $ \\
    $t\bar{t}$, $W^{-}W^{+}$, $Wt$, $ZZ\rightarrow\tau\tau\nu\nu$, $Z \to \tau^{-}\tau^{+}$& $13.3\pm3.2\pm0.2$  &  $15.4\pm3.6\pm0.3$     \\
    $W+\mathrm{jets}$              				& $2.6\pm1.1\pm0.5$   &  $-0.9\pm0.7\pm1.0$          \\
    $Z+\mathrm{jets}$               		& $-0.7\pm3.5\pm2.7$  &  $-0.5\pm3.8\pm2.9$             \\
    \midrule
    Total background 					& $32.4\pm5.5\pm3.3$  & $33.2\pm6.0\pm3.4$               \\
    \bottomrule
    \bottomrule
 \end{tabular}
 \caption{Number of background events for simulation-based and data-driven estimates in the \zzllvv\ channel 
(\eevv\  and \mumuvv). The first uncertainty is statistical and the
   second systematic. 
   The exact treatment of
   background estimates for the cross-section extraction is
   discussed in the text. 
  }
 \label{tab:backgrounds-llvv}
\end{table}

%% file: EventYields.tex
\begin{figure}[htbp]
  \centering
  \begin{subfigure}{0.48\textwidth}
    \includegraphics[width=\textwidth]{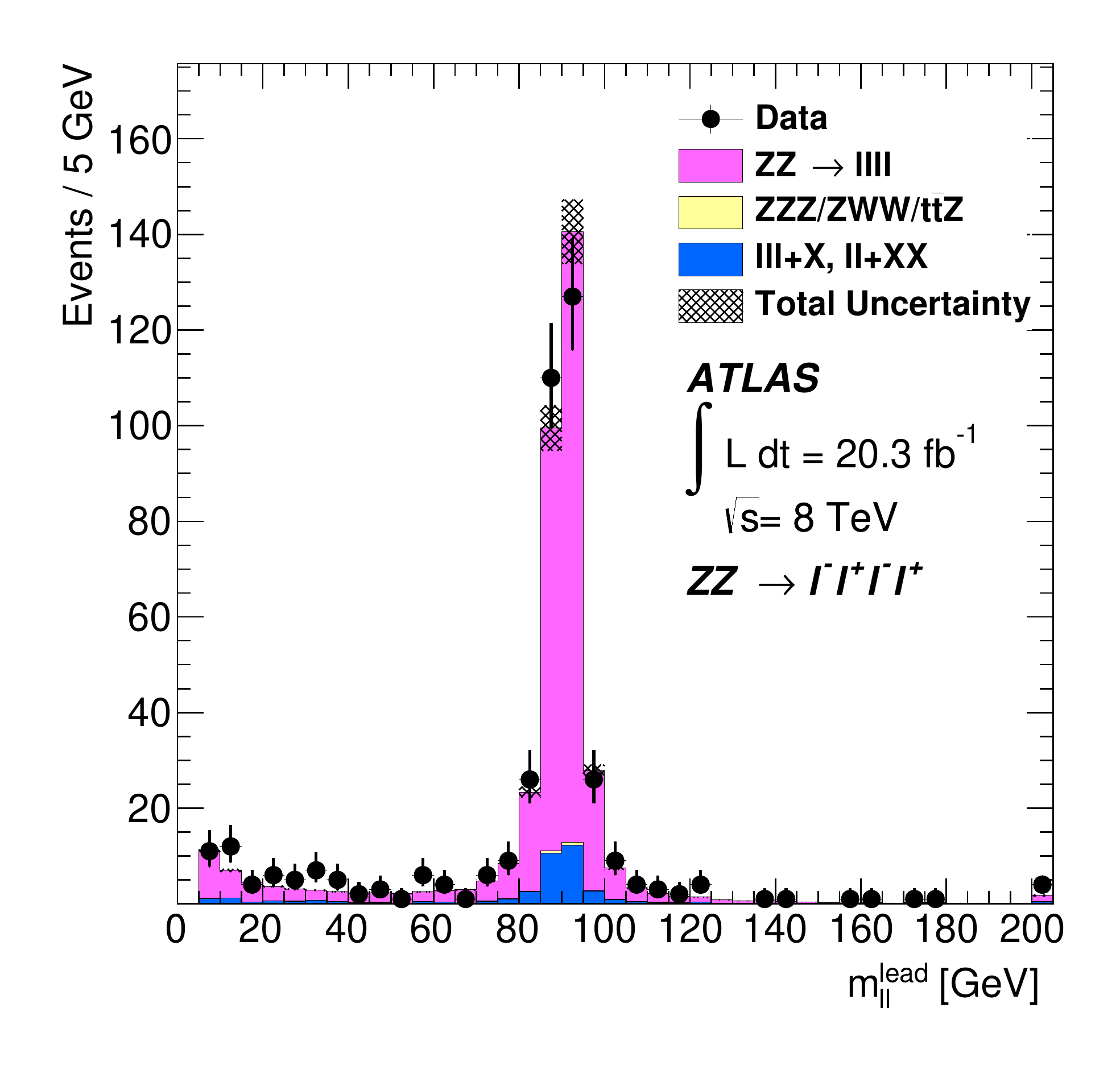}
    \caption{}
    \label{fig:evt_yield_signal_zz4l_leading_zmass_4l}
  \end{subfigure}
  \begin{subfigure}{0.48\textwidth}
    \includegraphics[width=\textwidth]{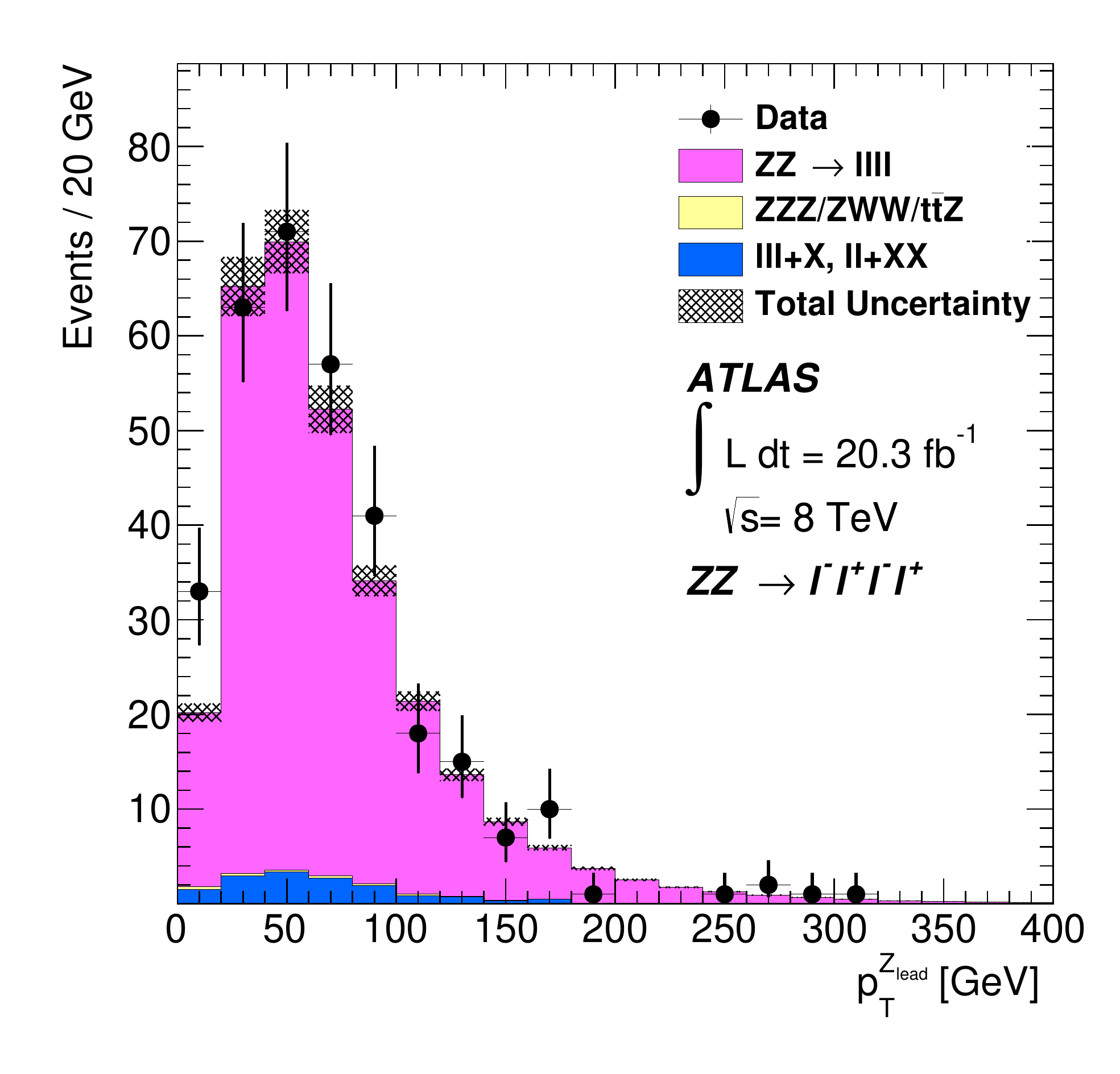}
    \caption{}
    \label{fig:evt_yield_leading_zpt_4l}
  \end{subfigure}
  \begin{subfigure}{0.48\textwidth}
    \includegraphics[width=\textwidth]{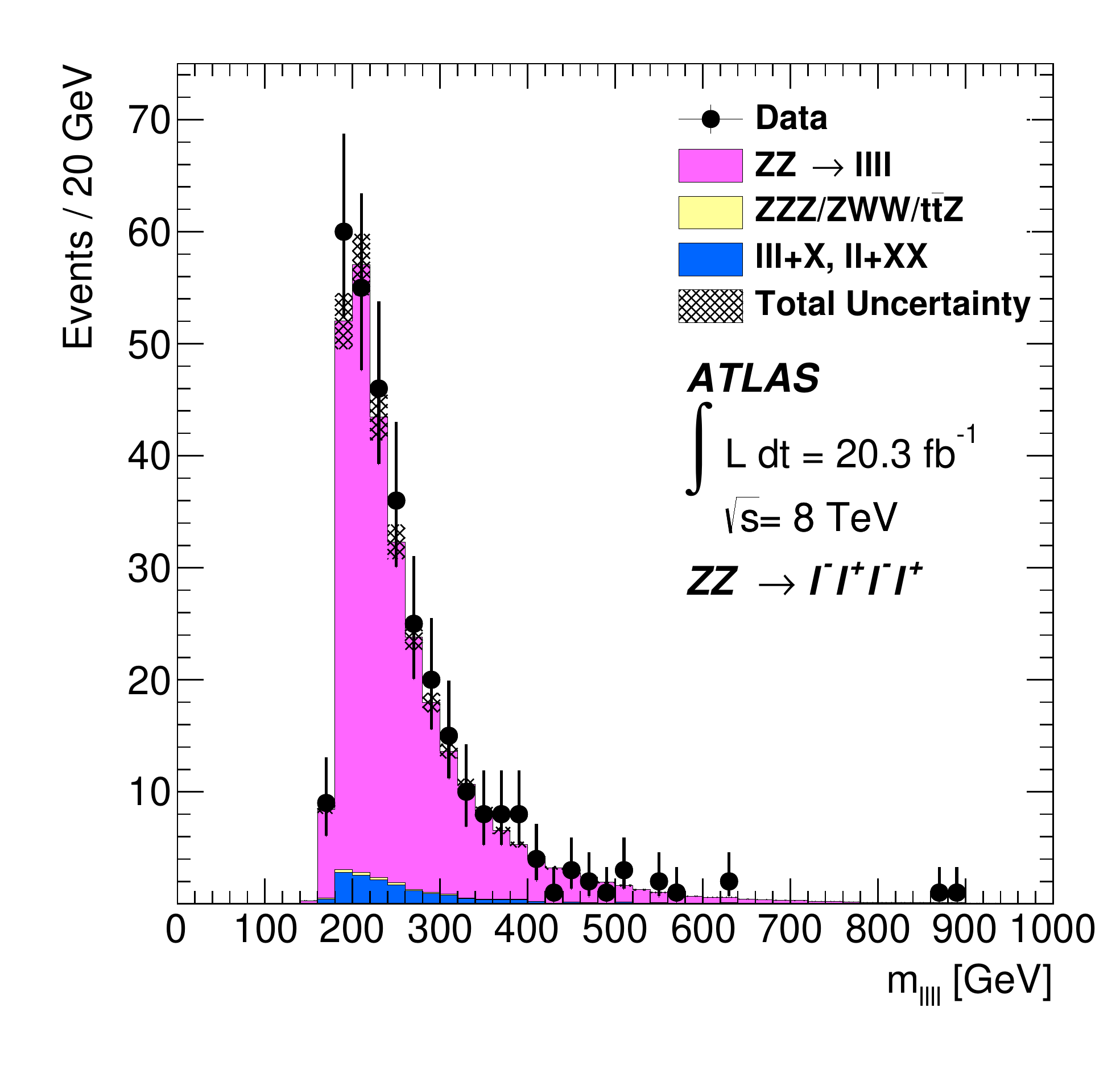}
    \caption{}
    \label{fig:evt_yield_signal_zz4l_zzmass_4l}
  \end{subfigure} 
  \begin{subfigure}{0.48\textwidth}
    \includegraphics[width=\textwidth]{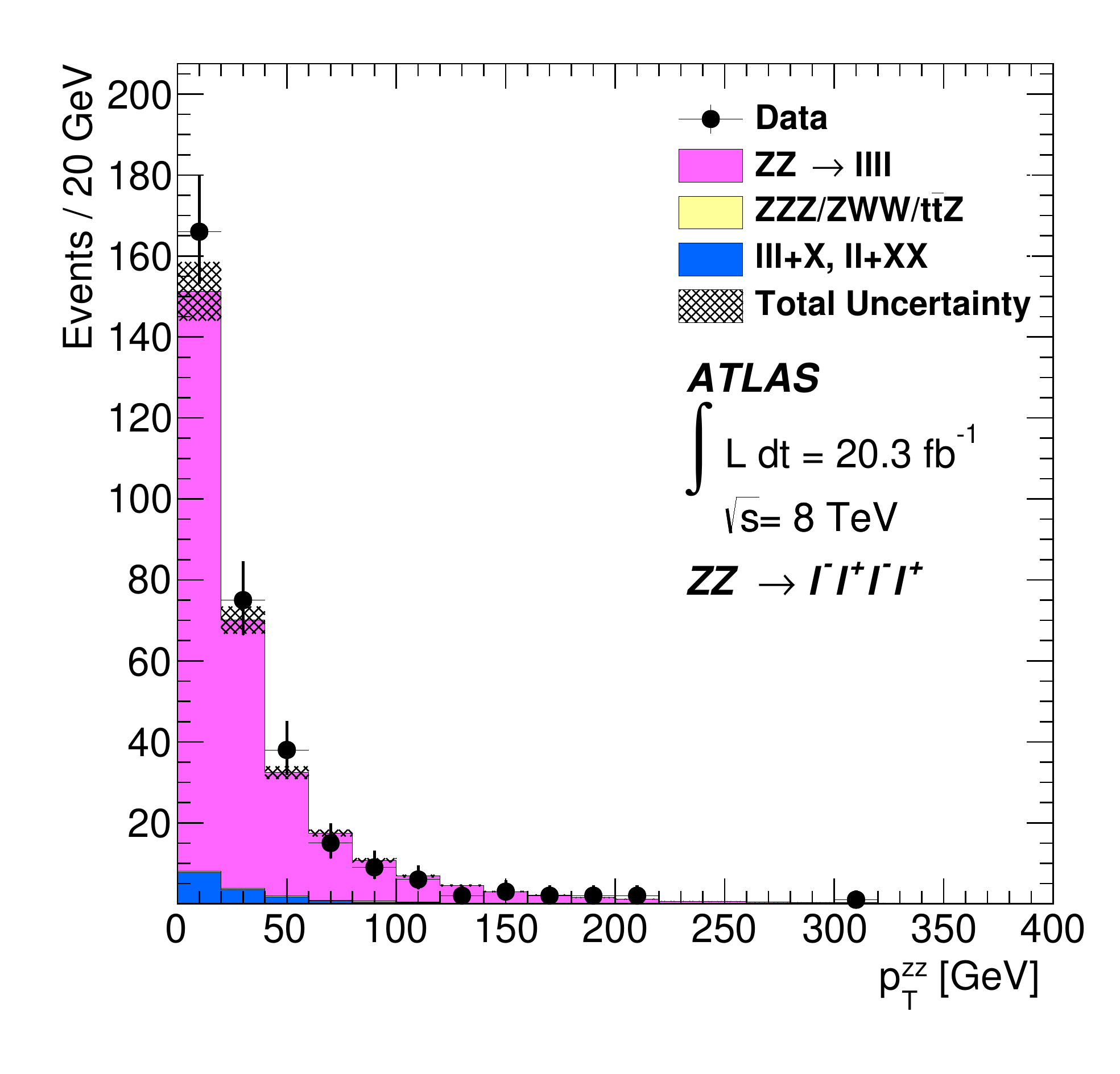}
    \caption{}
    \label{fig:evt_yield_signal_zz4l_zzpt_4l}
  \end{subfigure}
  \caption{Kinematic distributions for \zzlmlplmlpprimed\ candidates 
         in all four-lepton final states: \subref{fig:evt_yield_signal_zz4l_leading_zmass_4l}
         $m_{\ell^{-}\ell^{+}}^{\mathrm{lead}}$, 
         \subref{fig:evt_yield_leading_zpt_4l}
         $\pT^{Z_{\mathrm{lead}}}$, 
         \subref{fig:evt_yield_signal_zz4l_zzmass_4l}
         $m_{\ell^{-}\ell^{+}\ell^{\prime\, -}\ell^{\prime\, -}}$ 
         and \subref{fig:evt_yield_signal_zz4l_zzpt_4l} 
         $\pT^{ZZ}$. 
%  The  points represent the observed data and the histograms show the  prediction from simulation, as well as the data-driven background estimate. 
  The points represent the observed data and the histograms show the expected number of ZZ signal events and the background estimate.  
  The shaded band shows the combined statistical and
  systematic uncertainties in the prediction and the background.  No
    selection on the leading lepton pair mass is required
    for \subref{fig:evt_yield_signal_zz4l_leading_zmass_4l}, while the
    full selection is applied for the other distributions. }
\label{fig:zz4lplots} 
\end{figure}
\begin{figure}[htbp]
\centering
\includegraphics[width=0.50\columnwidth]{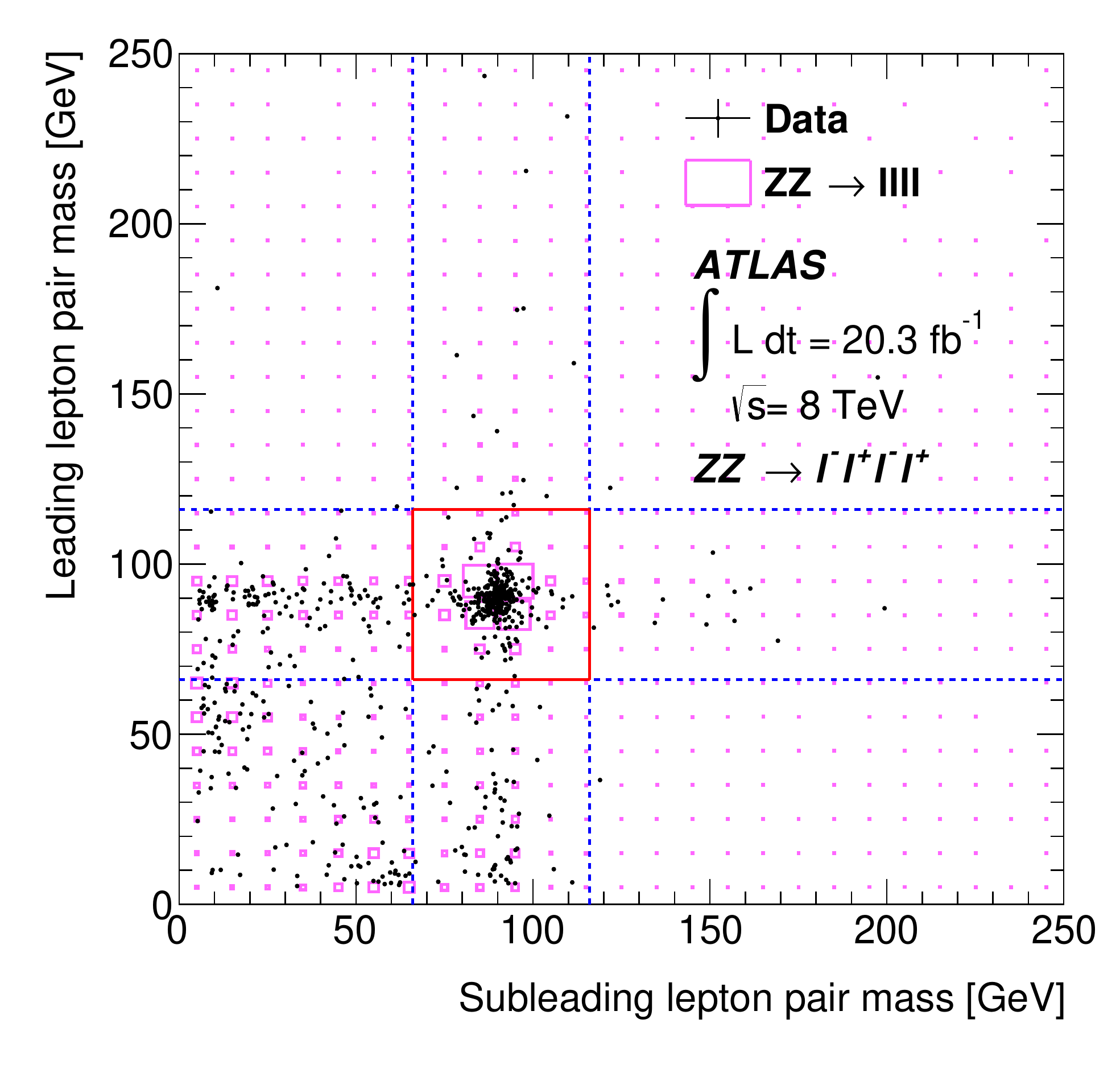}
\caption{The mass of the leading lepton pair versus the mass of the
  subleading lepton pair. The events observed in the data are shown as
  solid circles and the \zzlmlplmlpprimed\  signal prediction from
  simulation, normalized to the luminosity of the data, as pink
  boxes. The size of each box is proportional to the number of events in each
  bin. The region enclosed in the solid red box indicates the signal
  region defined by the requirements on the lepton pair masses for
  $ZZ$ events.} 
\label{fig:zzleadsublead} 
\end{figure}

The observed \zzlmlplmlpprimed\ and \zzllvv\ number of candidates in the data,
the total background estimates and the expected signal for the individual
decay modes, as well as their combinations, are shown in \Cref{tab:selected_data_MC}.
\begin{table}
\centering
  \begin{tabular}{lrrrrr}
    \toprule
    \toprule
     \zzlmlplmlpprimed   & \multicolumn{1}{c}{\eeee} & \multicolumn{1}{c}{\mumumumu} & \multicolumn{1}{c}{\eemm} &  \multicolumn{1}{c}{\llll} \\
     \midrule
     Observed data & 	\multicolumn{1}{c}{64}	&	\multicolumn{1}{c}{86}	& 	\multicolumn{1}{c}{171}	& 	\multicolumn{1}{c}{321}	 \\
     Expected  signal & $62.2\pm0.3\pm2.6$ & $83.7\pm0.4\pm3.2$ & $141.6\pm0.6\pm4.0$ & $287.0\pm0.8\pm8.1$ \\ 
     Expected background  &  $4.8\pm0.7\pm2.8$ & $2.3\pm2.4\pm1.0$ & $10.0\pm3.6\pm3.9$ & $17.1\pm4.4\pm 7.1$\\
    \midrule  \\ \midrule
    \zzllvv              & \eevv\ & \mumuvv\ & \multicolumn{2}{c}{\llvv} \\
    \midrule
    Observed data       & \multicolumn{1}{c}{102} & \multicolumn{1}{c}{106} & \multicolumn{2}{c}{208} \\
    Expected signal      & $ 51.1\pm0.9\pm2.6$ & $55.1\pm1.0\pm2.9$ & \multicolumn{2}{c}{$106.2\pm1.3\pm3.9$} \\
    Expected background  & $32.4\pm5.5\pm3.3$ & $33.2\pm6.0\pm3.4$ & \multicolumn{2}{c}{$~~65.6\pm8.1\pm4.7$} \\
    \bottomrule
    \bottomrule
  \end{tabular}

  \caption{
           Summary of observed \zzlmlplmlpprimed\ and \zzllvv\ candidates in the data, total background estimates and expected signal
           for the individual decay modes and for their combination
           (last column). The first uncertainty quoted is statistical,
           while the second is systematic. The uncertainty in the
           integrated luminosity (1.9\%) 
           is not included. 
          }
  \label{tab:selected_data_MC}
\end{table}
The kinematic distributions of the leading lepton pair mass
(the pair with the larger transverse momentum of the two pairs of leptons), 
$m_{\ell^{-}\ell^{+}}^{\mathrm{lead}}$,  the transverse momentum of
the leading $Z$ boson (the \Z\ boson that decays to the leading lepton
pair), $p_\mathrm{T}^{Z_{\mathrm{lead}}}$,  the mass of the four
leptons, 
$m_{\ell^{-}\ell^{+}\ell^{\prime\, -}\ell^{\prime\, -}}$, as well as
the transverse momentum of the $ZZ$ system, $p_\mathrm{T}^{ZZ}$, for
the \zzlmlplmlpprimed\ candidates 
in all four-lepton final states, are shown in 
\Cref{fig:zz4lplots}. \Cref{fig:zzleadsublead} shows the mass of the 
leading lepton pair versus the mass of the subleading lepton pair for
the data and predicted signal events in the \zzlmlplmlpprimed\  channel. 

\begin{figure}[htbp]
  \centering
  \begin{subfigure}{0.48\textwidth}
    \includegraphics[width=\textwidth]{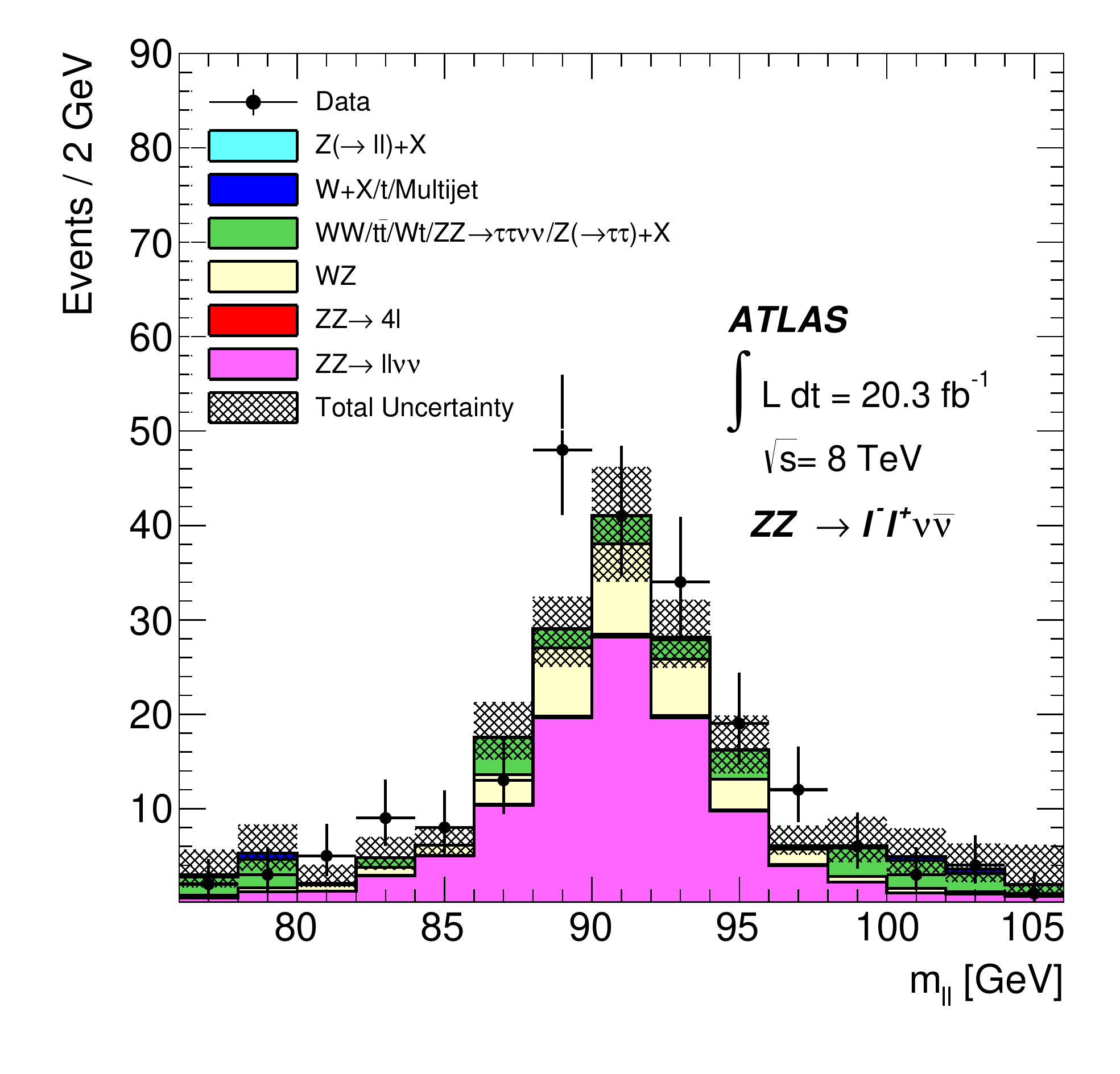}
    \caption{}
    \label{fig:evt_yield_papernoratio_2l2nu_DDperfPlot_ll_mz}
  \end{subfigure}
  %%%%%%%%%%%%%%%%%%%%%%%%%%%%%%%%%%%%%%%%%%
  \begin{subfigure}{0.48\textwidth} 
    \includegraphics[width=\textwidth]{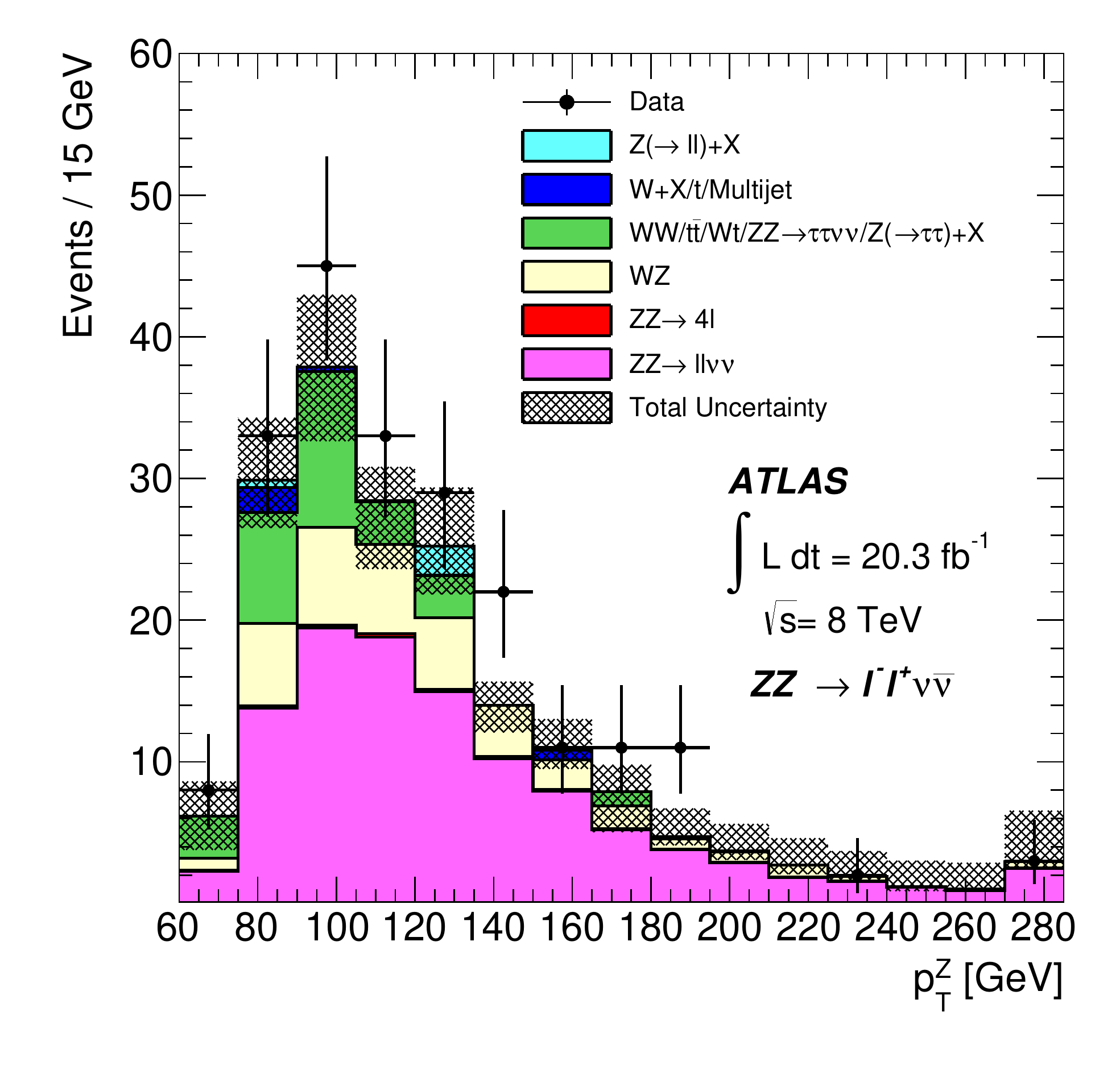}
    \caption{}
    \label{fig:evt_yield_papernoratio_2l2nu_DDperfPlot_ll_ptz}
  \end{subfigure}
  %%%%%%%%%%%%%%%%%%%%%%%%%%%%%%%%%%%%%%%%%%
  \begin{subfigure}{0.48\textwidth} 
    \includegraphics[width=\textwidth]{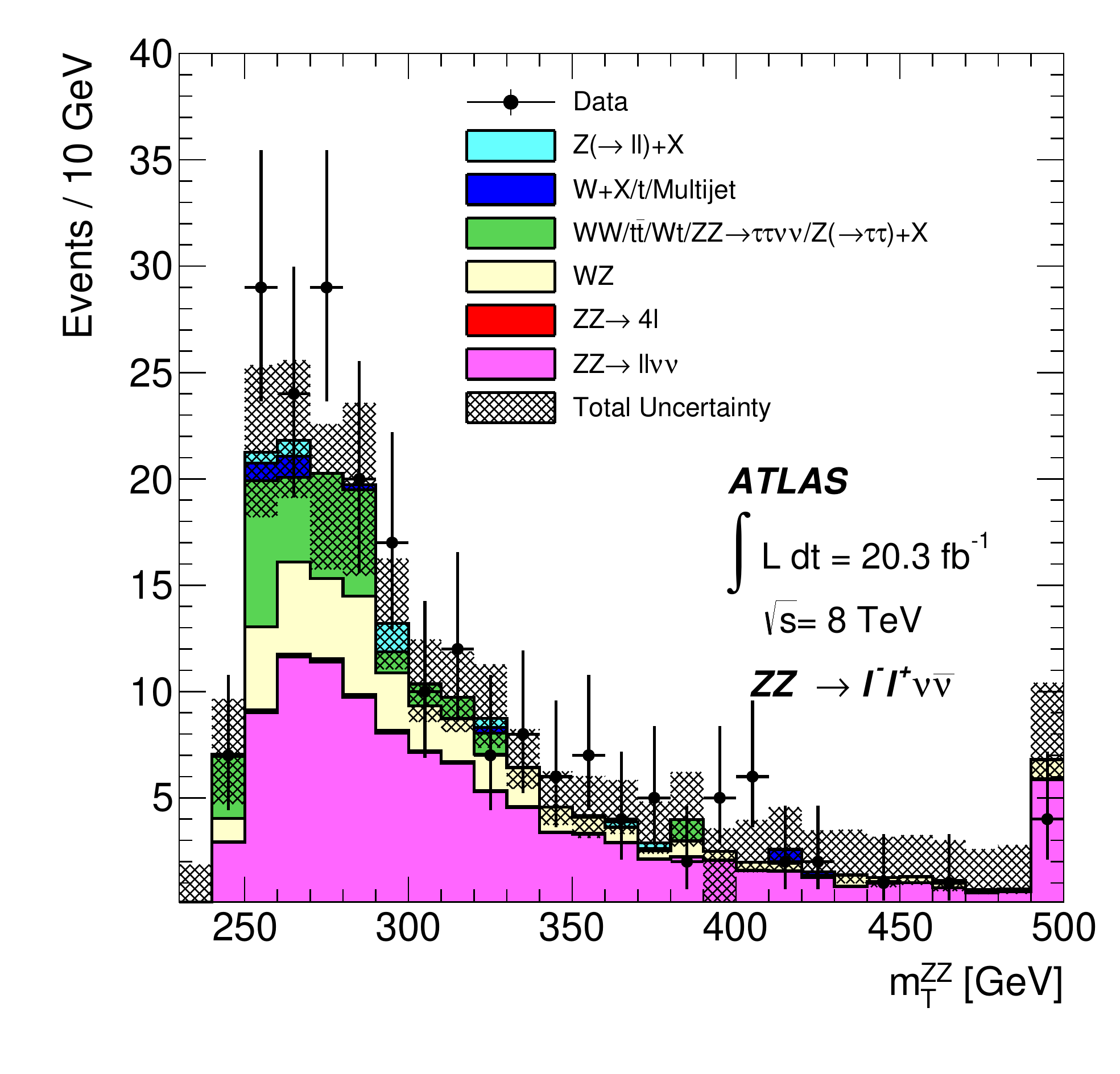}
    \caption{}
    \label{fig:evt_yield_papernoratio_2l2nu_DDperfPlot_ll_mt}
  \end{subfigure}
  %%%%%%%%%%%%%%%%%%%%%%%%%%%%%%%%%%%%%%%%%%
  \begin{subfigure}{0.48\textwidth}
    \includegraphics[width=\textwidth]{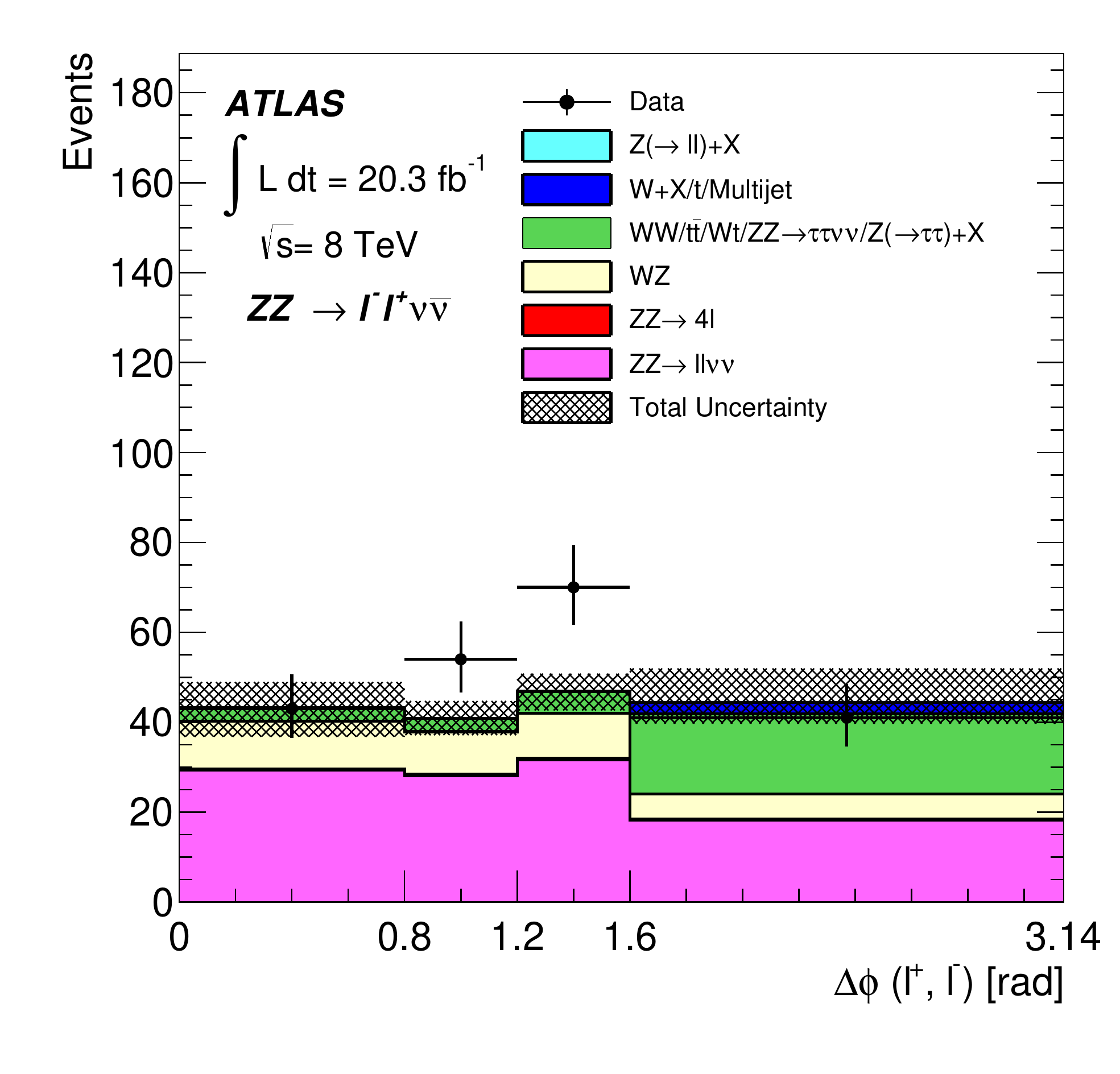}
    \caption{}
    \label{fig:papernoratio_2l2nu_DDperfPlot_ll_dphi_unfold}
  \end{subfigure}
  \caption{Kinematic distributions for \zzllvv\ candidates in both lepton final states: 
           \subref{fig:evt_yield_papernoratio_2l2nu_DDperfPlot_ll_mz} $m_{\ell^{-}\ell^{+}}$, 
           \subref{fig:evt_yield_papernoratio_2l2nu_DDperfPlot_ll_ptz}
    $p_{\mathrm{T}}^{Z}$, \subref{fig:evt_yield_papernoratio_2l2nu_DDperfPlot_ll_mt}
    $m_{\mathrm{T}}^{ZZ}$
    and \subref{fig:papernoratio_2l2nu_DDperfPlot_ll_dphi_unfold} $\Delta\phi(\ell^{+},\ell^{-})$. 
  The points represent the observed data and the histograms show the
  expected number of ZZ signal events and the background estimate. The shaded band shows the combined statistical and
  systematic uncertainties in the prediction and the background. The
  last bin
  in \subref{fig:evt_yield_papernoratio_2l2nu_DDperfPlot_ll_ptz}
  and \subref{fig:evt_yield_papernoratio_2l2nu_DDperfPlot_ll_mt}
  distributions, contains the overflow events.}
  \label{fig:zz2l2vplots}
\end{figure}

The kinematic distributions of the lepton pair mass,
$m_{\ell^{-}\ell^{+}}$, the $\pT^{Z}$, the transverse mass\footnote{The transverse mass, $m_{\mathrm{T}}^{ZZ}$, is defined as:
$m_{\mathrm{T}}^{ZZ} = \sqrt {\left(\sqrt{p_{\mathrm{T}}^{2}+m_{Z}^{2}} + \sqrt{E_{\mathrm{T}}^{\text{miss}~2}+m_{Z}^{2}}\right)^{2} - ({p}_{\mathrm{T}}+{E}_{\mathrm{T}}^{\text{miss}})^{2} }$,
where $\pT$ is the transverse momentum of the dilepton pair and
$m_{Z}=91.1876$ \GeV,  the mass of the $Z$ boson~\cite{Agashe:2014kda}. 
 } of the $ZZ$ system, $m_{\mathrm{T}}^{ZZ}$, and the
azimuthal angle between the two leptons (electrons or    
muons) originating from the $Z$ boson, $\Delta\phi(\ell^{+},\ell^{-})$, for
the \zzllvv ~candidates in both lepton final states, are shown
in \Cref{fig:zz2l2vplots}. 

%% file: SignalAcceptance.tex
\section{Correction factors and detector acceptance}
\label{sec:acceptance}
The fiducial cross section as measured in a given phase space for a
given final state, \zzlmlplmlpprimed\ or \zzllvv\ , where $\ell$ and $\ell^{'}$
are either an electron or a muon, may be expressed as:
\begin{equation}
 \sigma^{\mathrm{fid}} = \frac{N_{\mathrm{data}} - N_{\mathrm{bkg}}}{\mathcal{L}\cdot C_{ZZ}},
 \label{eq:fidCrossSectionExpr}
\end{equation}
where $N_{\mathrm{data}}$ is the number of observed candidate events in data
passing the full selection, $N_{\mathrm{bkg}}$ is the estimated number
of background events, $\mathcal{L}$ is the integrated luminosity, and
$C_{ZZ}$ is the correction factor applied to the measured cross
section to account for detector effects. 
This factor corrects for detector
inefficiencies and resolution
and is defined as:    
\begin{equation}
  C_{ZZ} = \frac{N_{ZZ}^{\text{reco}}}{N_{ZZ}^{\text{fid}}},
  \label{eq:CZZDef}
\end{equation}
where the numerator, $N_{ZZ}^{\text{reco}}$, is the expected yield of
reconstructed \ZZ\ events in the signal region after the full selection is applied,
and the denominator, $N_{ZZ}^{\text{fid}}$, is the generated
yield of \ZZ\ events in the fiducial phase space defined for a given final
state. 
It is determined using simulated \ZZ\ production samples.
The numbers of events $N_{ZZ}^{\text{reco}}$
and $N_{ZZ}^{\text{fid}}$ found in each sample (\powhegbox and \ggVV) are weighted by the relative
cross sections of the two samples in order to combine them in the
ratio.  In the calculation of $C_{ZZ}$ for \zzlmlplmlpprimed\ final
states, pairs of oppositely charged leptons produced from decays of 
$Z\rightarrow
\tau^{+}\tau^{-}\rightarrow\ell^{+}\ell^{-}\nu\bar{\nu}\nu\bar{\nu}$
are included in $N_{ZZ}^{\text{reco}}$, as those decays have the same final state as the signal and are not subtracted as background but are excluded from $N_{ZZ}^{\text{fid}}$  because the fiducial regions are defined only with \ZZ\ decays directly to electrons, muons or neutrinos, depending on the channel.

The total cross section as measured in a particular final state may be expressed as:
\begin{equation}
 \sigma^{\mathrm{tot}} = \frac{N_{\mathrm{data}} - N_{\mathrm{bkg}}}{\mathcal{L}\cdot C_{ZZ}\cdot A_{ZZ} \cdot \mathrm{BF}} = \frac{\sigma^{\mathrm{fid}}}{A_{ZZ}\cdot\mathrm{BF}},
 \label{eq:totalCrossSectionExpr}
\end{equation}
where $\mathrm{BF}$ is the branching fraction of \ZZ\ to a particular
final state (0.113\% for \eeee\ and \mmmm\ final states, 0.226\% for the \eemm\
final state and 2.69\% for the \llvv\ channel) and $A_{ZZ}$ is the detector
acceptance as measured in a particular decay mode and is determined at particle level. The acceptance factor is
defined as:
\begin{equation}
  A_{ZZ} = \frac{ N_{ZZ}^{\text{fid}} }{ N_{ZZ}^{\text{tot}} },
  \label{eq:AZZDef}
\end{equation}
where the numerator, $N_{ZZ}^{\text{fid}}$, is again the number of $ZZ$
events predicted in the fiducial phase space, and the denominator, $N_{ZZ}^{\text{tot}}$, is
the number of $ZZ$ events predicted in the total phase space. 

According to \Cref{eq:totalCrossSectionExpr}, the acceptance for the
total phase-space events in the signal region 
is given by the quantity $C_{ZZ} \cdot A_{ZZ} \cdot \mathrm{BF}$.
The purpose of this factorization is to separate the term that is
sensitive to theoretical uncertainties ($A_{ZZ}$) from the term representing
primarily detector efficiency ($C_{ZZ}$). 

The $C_{ZZ}$ and $A_{ZZ}$ factors are shown in 
Table \ref{tab:czzazztable} for all decay modes considered here. The acceptance in the  \zzllvv\ channel
is much smaller than the one in the \zzlmlplmlpprimed\ channel mainly
due to the axial-\met\ and jet veto requirements, which reduce the number of
selected events by about 86$\%$ and 40$\%$ respectively. 

\begin{table}
\begin{centering}
  \renewcommand{\arraystretch}{1.4}
  \begin{tabular}{lccc} 
    \toprule
    \toprule
    Channel & $C_{ZZ}$ & $A_{ZZ}$\\
    \midrule
    \eeee      & \ZZCzzEEEE & \ZZAzzEEEE\\
    \eemm      & \ZZCzzEEMM & \ZZAzzEEMM\\
    \mumumumu      & \ZZCzzMMMM & \ZZAzzMMMM\\
    \eevv      & \ZZCzzEENN & \ZZAzzEENN\\
    \mumuvv      & \ZZCzzMMNN & \ZZAzzMMNN\\
   \bottomrule
    \bottomrule
  \end{tabular}
  \caption{The $C_{ZZ}$ and $A_{ZZ}$ factors for each of the \zzlmlplmlpprimed\ and \zzllvv\ decay modes. 
           The total uncertainties (statistical and systematic) are shown and a description of the systematic uncertainties can be found in Section~\ref{sec:systematics}.}
  \label{tab:czzazztable}
\end{centering}
\end{table}

%% file: SystematicUncertainties.tex
\section{Systematic uncertainties}
\label{sec:systematics}
Systematic uncertainties arise from theoretical and experimental sources, which affect the correction factor, $C_{ZZ}$, the detector acceptance,
 $A_{ZZ}$, the number of expected background events, and the extracted aTGCs limits.  
These uncertainties are also propagated through the unfolding procedure (\Cref{sec:diffxsec}) to obtain the 
differential distributions. A summary of these uncertainties is shown in \Cref{tab:sys}.

The dominant experimental uncertainties depend on both the channel and final state under study.  
In the \zzlmlplmlpprimed\ channel, the lepton reconstruction uncertainty along with the isolation and impact parameter uncertainties
have the largest effect, while in the 
\zzllvv\ channel, the modelling of the jets and the measurement of the \MET\ are the dominant uncertainties. 
The systematic uncertainties due to lepton reconstruction are
estimated using the $Z\rightarrow \ell^{+} \ell^{-}$ and $W\rightarrow \ell \nu$ 
processes as described in Refs. \cite{Aad:2014nim,ATLAS-CONF-2014-032,Aad:2014rra}.   
For final states with electrons, the electron reconstruction uncertainty is about 4.0\%, 2.0\% and 1.7\% 
in the \zzeeee, \zzeemm\ and \zzeevv\ final states, respectively. Modelling of the isolation of muons along with their reconstructed 
impact parameter relative to the reconstructed collision vertex are the dominant effects on $C_{ZZ}$ for final states with muons, having contributions 
of 3.4\% and 3.2\% in the \zzmmmm\ and \zzmmvv\ final states, respectively.  

Uncertainties in the modelling of the jets and \MET\ are significant in the \zzllvv\ channel due to the jet veto requirement and the
 axial-$\met > 90$ \GeV\ selection.  
The JES uncertainty\footnote{The JES uncertainty is fully parameterized  by 56 nuisance parameters
resulting from various estimation techniques including $Z$+jets, $\gamma$+jets
and multijet balance.} corresponding to the local cluster weighting calibration scheme
is obtained using data from test-beams, LHC
collision data and simulations \cite{ATLAS-CONF-2015-037,ATLAS-CONF-2015-017}
 and is provided in bins of jet \pT\ and $|\eta|$. The jet energy
 resolution (JER) and its uncertainty are 
determined using in situ techniques based on the transverse momentum
balance in dijet events. The impact due to the uncertainty on the resolution is evaluated by
smearing the \pT\ of the jets within its uncertainty. %its uncertainty window.
The reconstruction of the \MET\ is affected by uncertainties
associated with the leptons, JES and 
JER that are propagated to the \MET\ determination.   
As there are no requirements on either jet reconstruction or \MET\
for the \zzlmlplmlpprimed\ channel, the impact of these uncertainties is negligible for these final states.

The uncertainty in the integrated luminosity is 1.9\% \cite{Lumi8TeVPaper}.  This affects the overall normalization of \ZZ\ production for the total 
cross-section measurement and the unfolded differential distributions.  

In addition to experimental uncertainties, the measurements
are subject to sources of theoretical
uncertainty. The correction factor and detector acceptance for \zzlmlplmlpprimed\ and \zzllvv\ final
states are calculated using \powhegbox\ interfaced to
\pythia\ for the \qqbar\ component, and using \ggVV\ for the $gg \to \ZZ$ component.
These calculations are sensitive to the choice of $\mu_{\mathrm{R}}$ and $\mu_{\mathrm{F}}$ scales, 
as they are missing higher terms from the perturbative expansion.
The uncertainty associated with this choice is estimated by comparing the 
detector acceptance, $A_{ZZ}$, when the $\mu_{\mathrm{R}}$ and $\mu_{\mathrm{F}}$ scales are increased and
decreased by a factor of two, with the nominal.  
The uncertainty associated with the jet veto in the \zzllvv\ final state is determined via the Stewart and Tackmann method \cite{Stewart:2011cf} 
using the jet veto efficiency for each sample generated 
with different $\mu_{\mathrm{R}}$ and $\mu_{\mathrm{F}}$ scales. 

The choice of the underlying-event modelling and  parton shower, which includes initial and final state radiation effects, is
one of the smaller sources of theoretical uncertainty and its effect
is estimated in two ways.
First, $A_{ZZ}$ is recalculated from MC samples generated with \powhegbox\ but interfaced with  \herwig\ for the 
parton showering instead of \pythia as is done for the nominal samples.   
The uncertainty is estimated from the difference in $A_{ZZ}$ for the \herwig\ and \pythia\ showered samples.
The second method uses \ZZ\ samples generated using \sherpa\ to calculate both 
$C_{ZZ}$ and $A_{ZZ}$. \sherpa\ is formally a LO generator with
respect to the $\qqbar$ process, and does
not include the gluon diagrams. However, \sherpa\ uses its own matrix-element
generation and parton shower algorithms, and can be used to provide an estimate of
the effects of the uncertainty due to the choice of parton shower. As in the first method, the uncertainty is estimated
using the difference in $C_{ZZ}$ and $A_{ZZ}$ calculated using the nominal and \sherpa\ samples.

As described in \Cref{sec:smpredictions}, the predicted cross sections for the \ZZ\ final states  are corrected for
virtual NLO EW effects by applying a  reweighting factor to each event.  
The uncertainty in this reweighting procedure is estimated by combining the uncertainty
in the theoretical predictions  used to estimate the NLO EW effects
and the statistical uncertainty from its prediction. These uncertainties are added in quadrature.

The choice of PDF represents an
additional source of uncertainty. To estimate this
theoretical uncertainty, the eigenvectors
of the CT10 PDF set are varied within their $\pm 1 \sigma$
uncertainties. The same procedure is followed for the backgrounds estimated from
simulation where the CT10 PDF set is used.

\begin{table}[htbp]
\centering
\begin{tabular}{lS[table-format=2.2]S[table-format=2.2]S[table-format=2.2]S[table-format=2.2]S[table-format=2.2]}
\hline
Source & \eeee\ & \mumumumu\ & \eemm\  &\eevv\ & \mumuvv\  \\\hline
 & \multicolumn{4}{c}{$C_{ZZ}$} \\\hline
Electron rec. and ID efficiency 	& 4.0\% &   \multicolumn{1}{c}{--} & 2.0\%  & 1.7\% & \multicolumn{1}{c}{--}   \\
Electron energy/momentum 		& 0.4\% & 0.01\%  & 0.2\%  & 2.0\% & 0.1\%	\\
Electron isolation/impact parameter 	& 1.4\% & \multicolumn{1}{c}{--} & 0.7\%  & 0.3\% & \multicolumn{1}{c}{--}	\\
Muon rec. and ID efficiency 			& \multicolumn{1}{c}{--} & 1.8\%   & 0.9\%  & \multicolumn{1}{c}{--} & 0.7\%	\\
Muon energy/momentum 			& \multicolumn{1}{c}{--} & 0.03\%  & 0.04\% & \multicolumn{1}{c}{--} & 0.3\%	\\
Muon isolation/impact parameter 	& \multicolumn{1}{c}{--} & 3.4\%   & 1.7\%  & \multicolumn{1}{c}{--} & 3.2\%	\\
Jet+\MET\ modelling 	  		& \multicolumn{1}{c}{NA}   & \multicolumn{1}{c}{NA}  & \multicolumn{1}{c}{NA}     & 4.7\% & 5.3\%	\\
Trigger efficiency    			& 0.1\% & 0.2\%   & 0.1\%  & 0.1\% & 0.5\%	\\
PDF and parton shower         		& 0.2\% & 0.1\%   & 0.1\%  & 0.9\% & 2.2\%	\\
\hline
 & \multicolumn{4}{c}{$A_{ZZ}$} \\
\hline
Jet veto                 		& \multicolumn{1}{c}{NA} &\multicolumn{1}{c}{NA}  & \multicolumn{1}{c}{NA} 	   & 1.8\%& 1.6\%  \\ 
Electroweak Corrections                 & 0.03\% & 0.03\% & 0.02\% & 0.9\% & 1.0\%  \\
PDF and scale            & 0.7\% &  0.9\%  & 0.8\% & 3.1\% & 2.1\% \\
Generator modelling and parton shower & 2.0\% & 3.0\% & 2.3\% & 4.3\% & 4.1\% \\
\hline
\end{tabular}
\caption{\label{tab:sys}A summary of the systematic uncertainties, as relative percentages
  of the correction factor $C_{ZZ}$ and the detector acceptance $A_{ZZ}$ is shown.  For rows with multiple sources, the uncertainties are added in quadrature.  
 Dashes indicate uncertainties which are smaller than 0.01\% and uncertainites with NA are not applicable for that specific final state.}
\end{table}

%% file: CrossSectionExtraction.tex
\label{sec:expXsec} 

\begin{figure}[h!]
  \begin{center}
    \includegraphics[width=.9\textwidth]{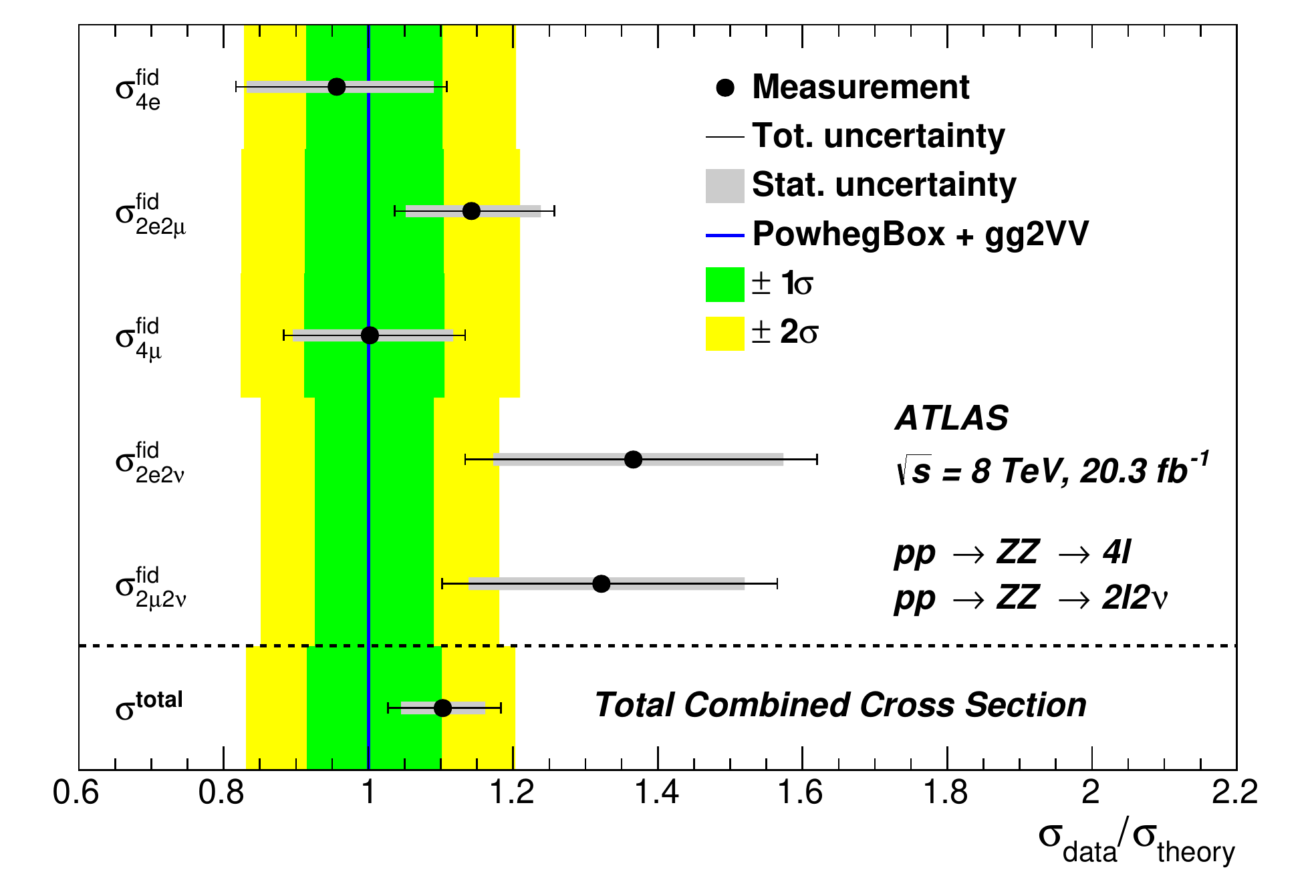}
    \caption{\label{fig:theComp}
    The ratio of the measured \ZZ\ cross sections in the fiducial
    phase space to the SM prediction from \powhegbox\ and \ggtwovv\ in
    each of the five decay modes considered. The ratio between the total
    combined cross section and the SM prediction is also shown. The
    inner grey error bars on the data points represent the statistical
    uncertainties, while the outer black error bars represent the
    total uncertainties. The green and yellow bands represent the
    $1\sigma$ and $2\sigma$ uncertainties, respectively, associated with the SM prediction.
} 
  \end{center}
\end{figure}

Two types of cross sections, fiducial and total, are extracted using \Cref{eq:fidCrossSectionExpr,eq:totalCrossSectionExpr}. A fiducial
cross section is extracted for every final state in both 
the \zzlmlplmlpprimed\ and \zzllvv\ channels. The information from these
final states is combined to measure a single $pp \to ZZ$ total cross section in
the total phase space ($66 < m_{\ell^{-}\ell^{+}} < 116 \GeV$) using
the detector acceptance and branching fraction of \ZZ\ to a given four-lepton
or dilepton + $\nu\overline{\nu}$ final state. 
For each measurement, a likelihood method is used to extract the expected \ZZ\ event rate according to a Poisson probability distribution, as described
in Ref. \cite{Aad:2015rka}.  
The likelihood is maximized with respect to
the cross section. For fiducial (total) cross-section measurements, sources of
systematic uncertainties affecting backgrounds, object reconstruction and
identification efficiencies, detector acceptance and luminosity are
included as nuisance parameters and the affected terms are allowed to
fluctuate according to Gaussian probability
distributions with widths equal to the uncertainties. The measured cross sections for
the \zzlmlplmlpprimed\ and \zzllvv\ channels are
given in \Cref{tab:xsecResults} and the ratios of these measurements
with respect to the SM predictions are shown in \Cref{fig:theComp}.

\begin{table}
    \begin{centering}
        \setlength{\tabcolsep}{.16667em}
    \renewcommand{\arraystretch}{1.5}
    \begin{tabular}{lcrrrrrrr}
    \toprule \toprule
                                            & &\multicolumn{5}{c}{Measurement} & \multicolumn{2}{c}{Prediction} \\
\hline
        $\sigma^{\mathrm{fid}}_{ZZ \to \eeee}$  &=& \ZZfidEEEEXsecTOneDp  & $\pm$ \ZZfidEEEEXsecStatOnlyErrUpOneDp\ (stat) &$\pm$ \ZZfidEEEEXsecAllSystNoLumiErrUpOneDp\ (syst) &$\pm$ \ZZfidEEEEXsecLUMIErrUpOneDp\ (lumi)  &fb~~~~~~~~ & $6.2$~& $^{+0.6}_{-0.5}$ fb\\
        $\sigma^{\mathrm{fid}}_{ZZ \to \eemm}$  &=& \ZZfidEEMMXsecTOneDp  & $\pm$ \ZZfidEEMMXsecStatOnlyErrUpOneDp\ (stat) &$^{+\ZZfidEEMMXsecAllSystNoLumiErrUpOneDp}_{\ZZfidEEMMXsecAllSystNoLumiErrDwOneDp}$ (syst) &$^{+\ZZfidEEMMXsecLUMIErrUpOneDp  }_{\ZZfidEEMMXsecLUMIErrDwOneDp  }$ (lumi)  &fb~~~~~~~~ & $10.8$~& $^{+1.1}_{-1.0}$ fb\\
        $\sigma^{\mathrm{fid}}_{ZZ \to \mmmm}$  &=& \ZZfidMMMMXsecTOneDp  &$^{+\ZZfidMMMMXsecStatOnlyErrUpOneDp}_{\ZZfidMMMMXsecStatOnlyErrDwOneDp}$ (stat) &$^{+\ZZfidMMMMXsecAllSystNoLumiErrUpOneDp}_{\ZZfidMMMMXsecAllSystNoLumiErrDwOneDp}$ (syst) &$ \pm \ZZfidMMMMXsecLUMIErrUpOneDp$ (lumi)  &fb~~~~~~~~ & $4.9$~& $^{+0.5}_{-0.4}$ fb\\
        $\sigma^{\mathrm{fid}}_{ZZ \to \eevv}$  &=& \ZZfidXsecEENuNuOneDp &$^{+\ZZfidXsecStatErrUpEENuNuOneDp  }_{\ZZfidXsecStatErrDwEENuNuOneDp  }$ (stat) &$^{+\ZZfidXsecSysErrUpEENuNuOneDp        }_{\ZZfidXsecSysErrDwEENuNuOneDp        }$ (syst) &$\pm$ \ZZfidXsecLumiErrUpEENuNuOneDp\ (lumi)  &fb~~~~~~~~ & $3.7$~& $\pm0.3$ fb\\
        $\sigma^{\mathrm{fid}}_{ZZ \to \mumuvv}$  &=& \ZZfidXsecMMNuNuOneDp &$\pm$ \ZZfidXsecStatErrUpMMNuNuOneDp\ (stat) &$^{+\ZZfidXsecSysErrUpMMNuNuOneDp}_{\ZZfidXsecSysErrDwMMNuNuOneDp}$ (syst) &$\pm$ \ZZfidXsecLumiErrUpMMNuNuOneDp\ (lumi)  &fb~~~~~~~~ & $3.5$~& $\pm0.3$ fb\\
        \midrule
        $\sigma_{pp \to ZZ}^{\mathrm{total}}$     &=& \ZZTotalCrossSectionTOneDp &$\pm$ \ZZTotalCrossSectionStatOnlyUpOneDp\ (stat) &$\pm$ \ZZTotalCrossSectionAllSystNoLumiUpOneDp\ (syst) & $^{+\ZZTotalCrossSectionLUMIUpOneDp}_{\ZZTotalCrossSectionLUMIDwOneDp}$ (lumi)  &pb~~~~~~~~ & $6.6$~& $^{+0.7}_{-0.6}$ pb\\
        \bottomrule \bottomrule
    \end{tabular}
    \caption{ The measured fiducial
        cross sections and the combined total cross section 
        compared to the SM predictions.  For experimental results, the statistical, systematic, and luminosity uncertainties are shown. For the theoretical predictions, the combined statistical and systematic uncertainty is shown.
}
\label{tab:xsecResults}
\end{centering}
\end{table}

%% file: DifferentialCrossSections.tex
The differential cross sections presented in
this section allow a more detailed comparison of the measurement to
current and future theoretical predictions.
The measured kinematic distributions
are unfolded
back to the underlying distributions, accounting for the effect of
detector resolution, efficiency and acceptance. The unfolding as a
function of different kinematic variables is
performed separately for the two channels. More specifically, it is
performed within the fiducial phase space of the \zzllvv\ measurement 
and within the total phase space of the \zzlmlplmlpprimed\ measurement, defined in 
Section~\ref{sec:phasespace}. This different approach between the
channels is chosen to benefit from the extended fiducial phase space
for leptons in the \zzlmlplmlpprimed\ channel.
 
The unfolding procedure is based on a Bayesian iterative
algorithm~\cite{bib:UnfoldingPaper}. In the unfolding of binned data,
the effects of the experimental acceptance and resolution are
expressed in terms of a two-dimensional response matrix, $A_{ij}$, where each element
corresponds to the probability of an event in the $i$-th generator-level
bin being reconstructed in the $j$-th measurement bin. The unfolding
algorithm combines the measured spectrum with the response matrix to
form a likelihood, takes as input a prior for the specific kinematic
variable and iterates using the posterior distribution as prior for
the next iteration. The SM prediction calculated using the \powhegbox and
\ggtwovv\ generators is used as the initial prior and three iterations are
performed. The number of iterations is optimized to find a balance
between too many iterations, causing high statistical uncertainties
associated with the unfolded spectra, and too few iterations, which
increases the dependency on the MC prior. 

The statistical uncertainty of the unfolded distribution is tested via
toy-MC tests. Each measured data-point is Poisson
fluctuated and the full nominal unfolding procedure is applied. This
is repeated 2000 times and the root mean square of the resulting unfolded values
is taken as the unfolded distribution's statistical uncertainty.

The systematic uncertainties are estimated as follows: for each scale,
efficiency or resolution systematic uncertainty, a new
response matrix is produced reflecting a variation by that systematic
uncertainty. The measured data distribution is then unfolded for all 
instances separately, leading to one distribution for
each systematic uncertainty. The difference between each of the distributions
that correspond to the different systematic uncertainties and the
nominal distribution, where no variation has been applied, is defined
as the systematic uncertainty in each bin. 

Uncertainties on the unfolding due to imperfect description of the
kinematic properties of the data by the MC are evaluated using a
data-driven method~\cite{Unfolding_closuretest}, where the MC differential 
distribution is corrected to match the 
data distribution and the resulting weighted MC distribution at
reconstruction level is unfolded with the response matrix used in the
actual data unfolding. The new unfolded distribution is compared to
the weighted MC distribution at generator level and the difference is
taken as the systematic uncertainty. Moreover, in
the \zzlmlplmlpprimed\ channel, as the unfolding is performed within
the total phase space, theoretical uncertainties due to this
extrapolation are also considered. These uncertainties include the choice
of $\mu_{\mathrm{R}}$ and $\mu_{\mathrm{F}}$ scales, which access the impact of higher-order contributions from QCD, the PDF set, and the parton shower
modelling.
The latter is estimated by comparisons with \sherpa\ $ZZ$ samples. 

The bin limits and bin widths of the differential kinematic distributions are chosen to
balance the need of finer bins, in order to 
provide detailed information, against the limited number of events
and bin migration effects. More specifically, the fraction of
reconstructed events generated in the same bin (i.e.\ purity) is higher
than 75\%.

\subsubsection{\texorpdfstring{\zzlmlplmlpprimed}{ZZ -> l-l+l'-l'+} channel}
\begin{figure}[htbp]
\centering
 \begin{subfigure}{0.48\textwidth}
    \includegraphics[width=\textwidth]{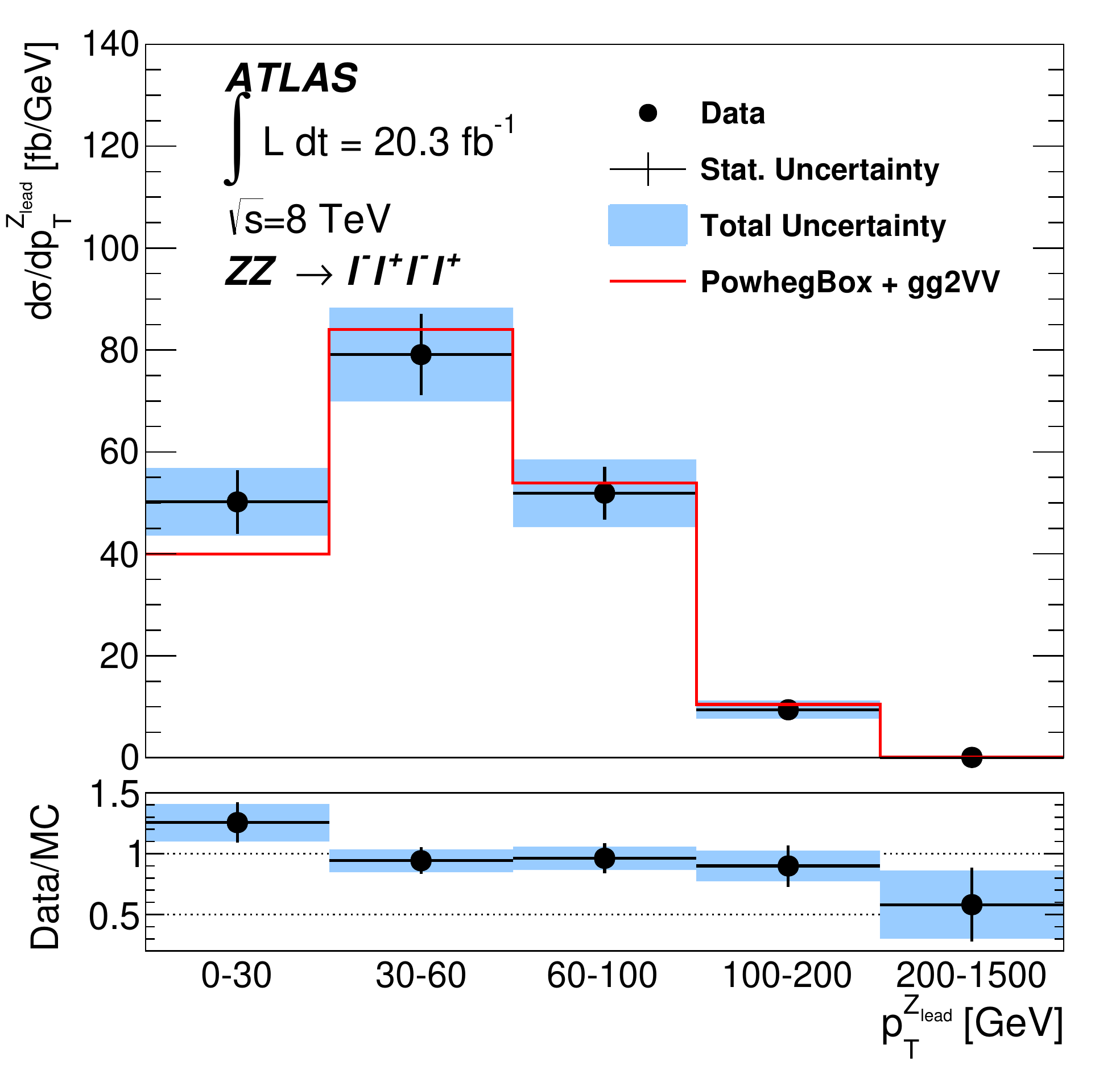}
    \caption{}
    \label{fig:unfold_4l_zpt_abs}
  \end{subfigure}
  \begin{subfigure}{0.48\textwidth}
    \includegraphics[width=\textwidth]{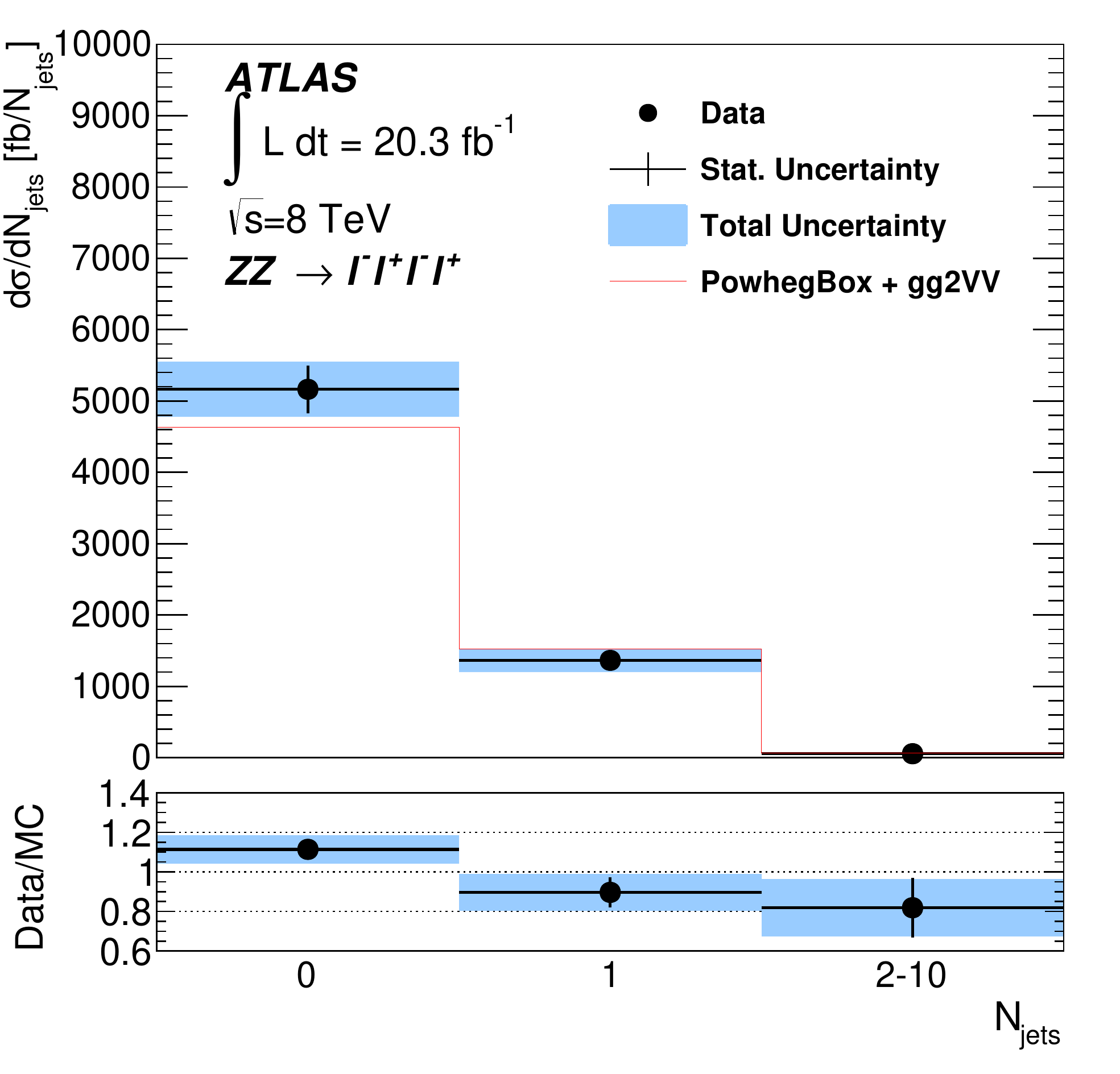}
    \caption{}
    \label{fig:unfold_4l_njets_abs}
  \end{subfigure}
  \begin{subfigure}{0.48\textwidth}
    \includegraphics[width=\textwidth]{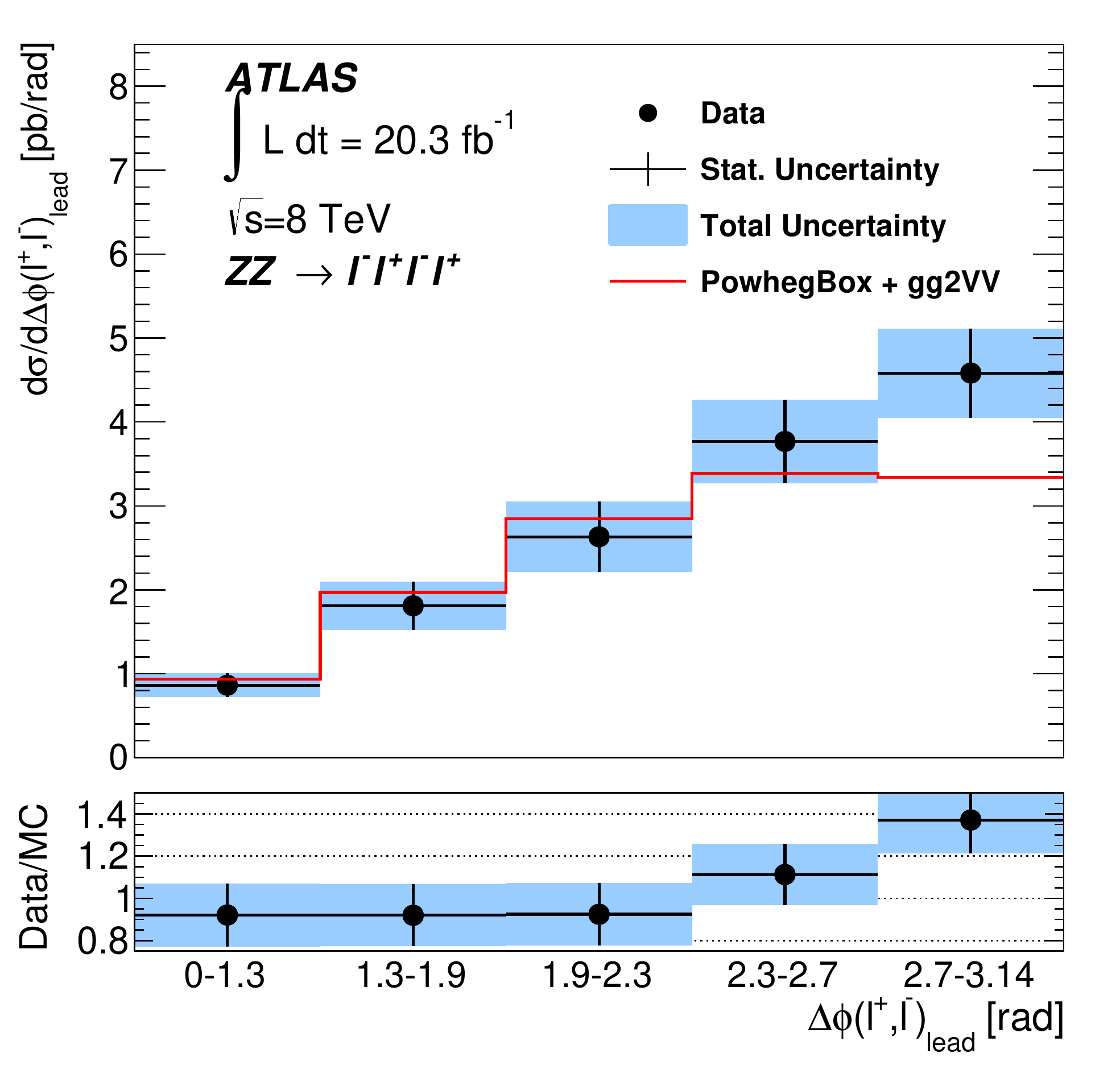}
    \caption{}
    \label{fig:unfold_4l_dphi_abs}
  \end{subfigure}
  \begin{subfigure}{0.48\textwidth}
    \includegraphics[width=\textwidth]{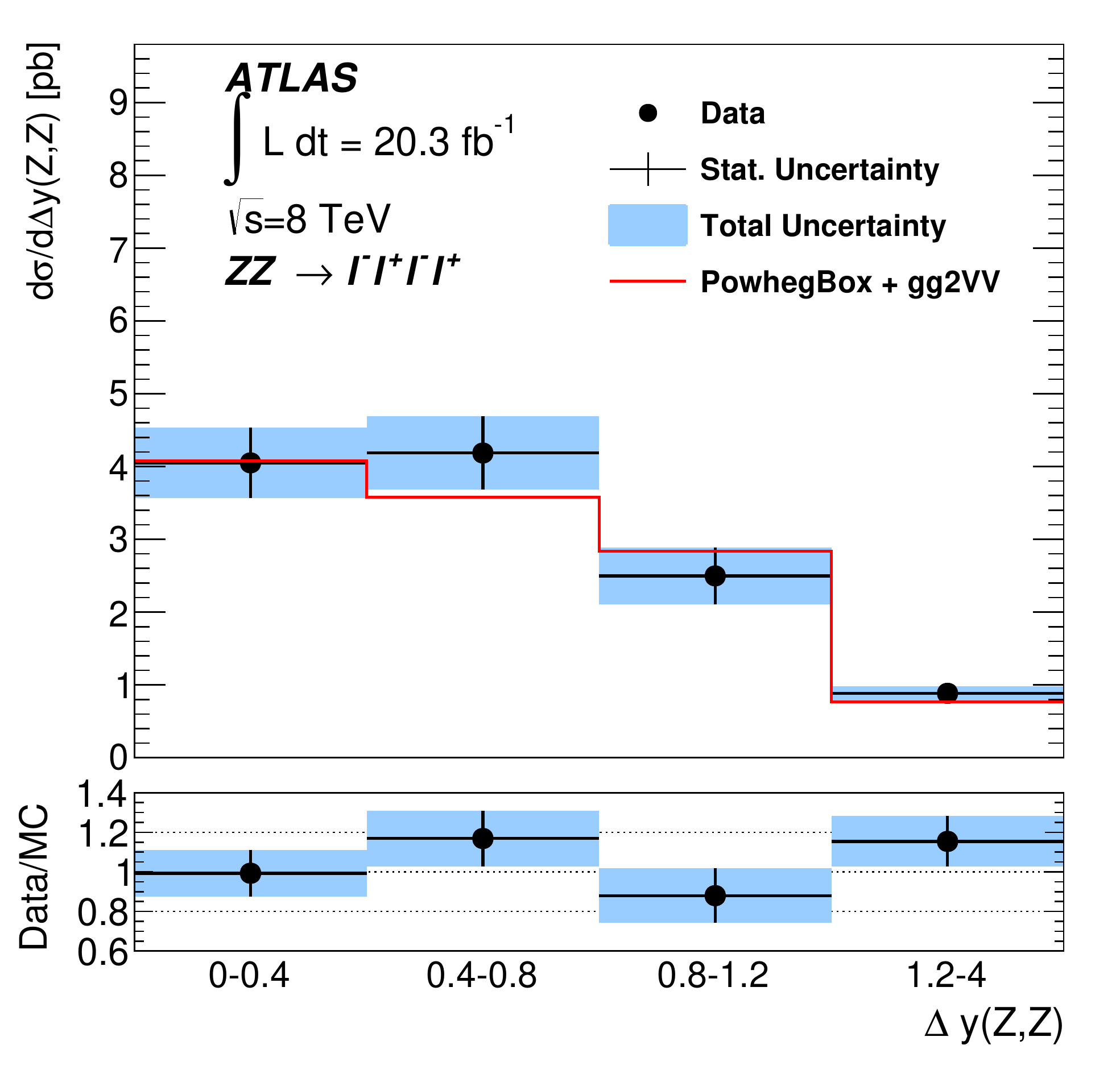}
    \caption{}
    \label{fig:unfold_4l_Yzz_abs}
  \end{subfigure}
\caption{The measured differential cross-section distributions (black
points) normalized to the bin width for \subref{fig:unfold_4l_zpt_abs}
$\pT^{Z_{\mathrm{lead}}}$, \subref{fig:unfold_4l_njets_abs}
$N_{\mathrm{jets}}$, \subref{fig:unfold_4l_dphi_abs} $\Delta\phi(\ell^{+},\ell^{-})_{\textrm{lead}}$  
and \subref{fig:unfold_4l_Yzz_abs} $\Delta y(Z,Z)$ in
the \zzlmlplmlpprimed\ 
channel, unfolded within the total   
phase space, compared to the theory
predictions of \powhegbox  and \ggtwovv\ (red line). The vertical error bars
show the respective statistical uncertainties, while the light blue error bands 
express the statistical and systematic
uncertainties of the measurements added in quadrature.}
\label{fig:unfolded4l}
\end{figure}

The kinematic distributions that are unfolded in this channel are the
$\pT^{Z_{\mathrm{lead}}}$, the number of jets in associated 
production with \zzlmlplmlpprimed\ ($N_{\mathrm{jets}}$), the azimuthal angle
between the two leptons (electrons or  
muons) originating from the leading $Z$ boson
($\Delta\phi(\ell^{+},\ell^{-})_{\textrm{lead}}$) and the difference in rapidity
between the two $Z$ bosons of the $ZZ$ system ($\Delta y(Z,Z)$).
The differential cross sections and their comparison with the SM predictions
(\powhegbox and \ggtwovv) are shown in Figure~\ref{fig:unfolded4l}. The
dominant uncertainty is the statistical uncertainty of the data, ranging from  7$\%$ to 17$\%$ in most bins. The theoretical modelling uncertainties are of
the order of 1$\%$--3$\%$. According to
Figure~\ref{fig:unfold_4l_njets_abs}, more than 70\%
of \zzlmlplmlpprimed\ events are produced without any associated high-\pT\ jets,
and this is well modelled by MC simulation. The measurement is
consistent with the SM prediction within $\SI{1}{\sigma}$ in most of
the bins.
\subsubsection{\texorpdfstring{\zzllvv}{ZZ -> l-l+vv} channel}
The kinematic distributions that are unfolded in this channel are 
the \Zpt\ of the $Z$ boson that decays to electrons or muons, the
azimuthal angle between the two leptons (electrons or    
muons) originating from the $Z$ boson ($\Delta\phi(\ell^{+},\ell^{-})$) and the
transverse mass of the $ZZ$ system ($m_{\mathrm{T}}^{ZZ}$).

\begin{figure}[htbp]
\centering
 \begin{subfigure}{0.48\textwidth}
    \includegraphics[width=\textwidth]{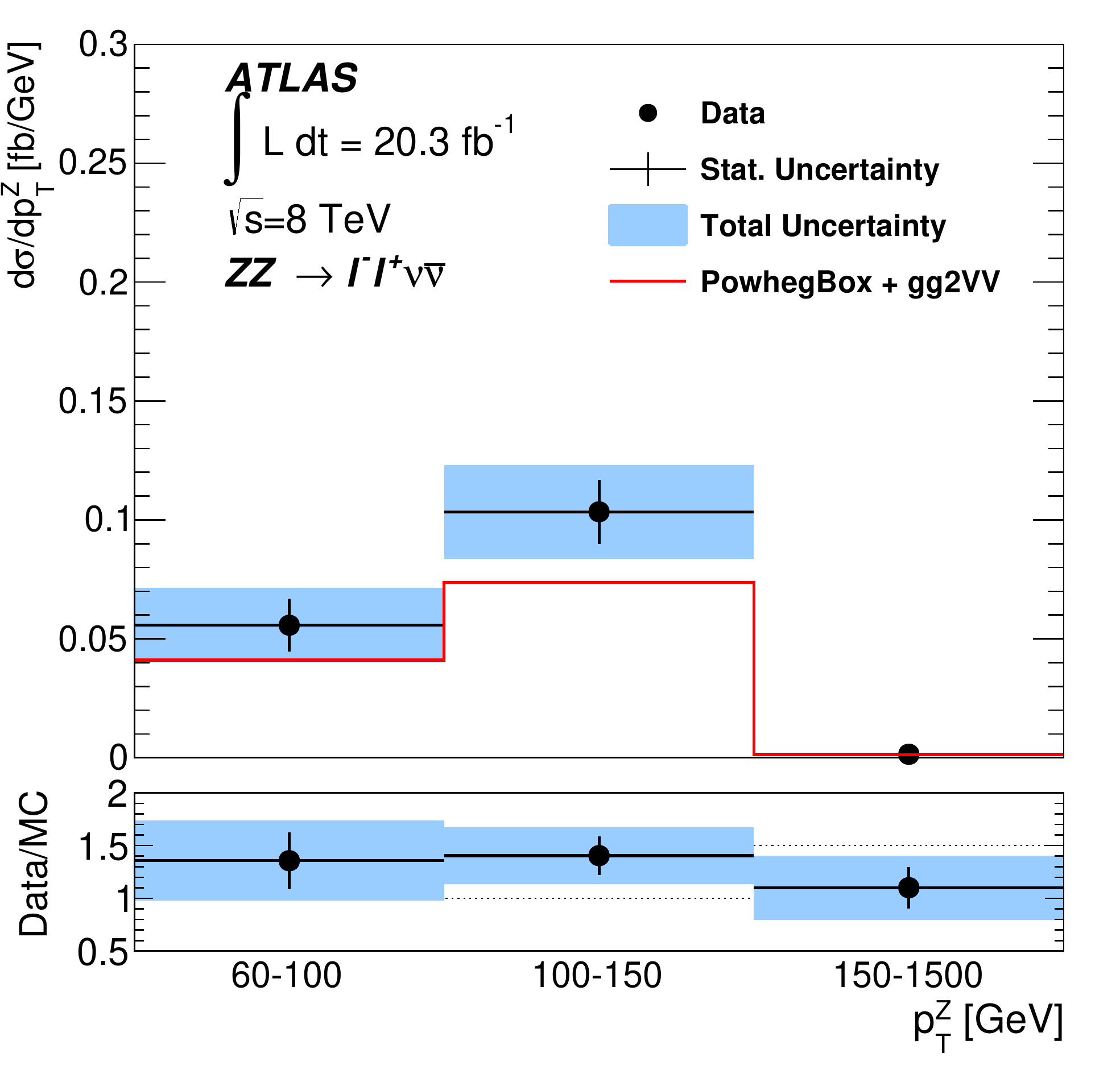}
    \caption{}
    \label{fig:unfold_2l2v_zpt_abs}
  \end{subfigure}
 \begin{subfigure}{0.48\textwidth}
    \includegraphics[width=\textwidth]{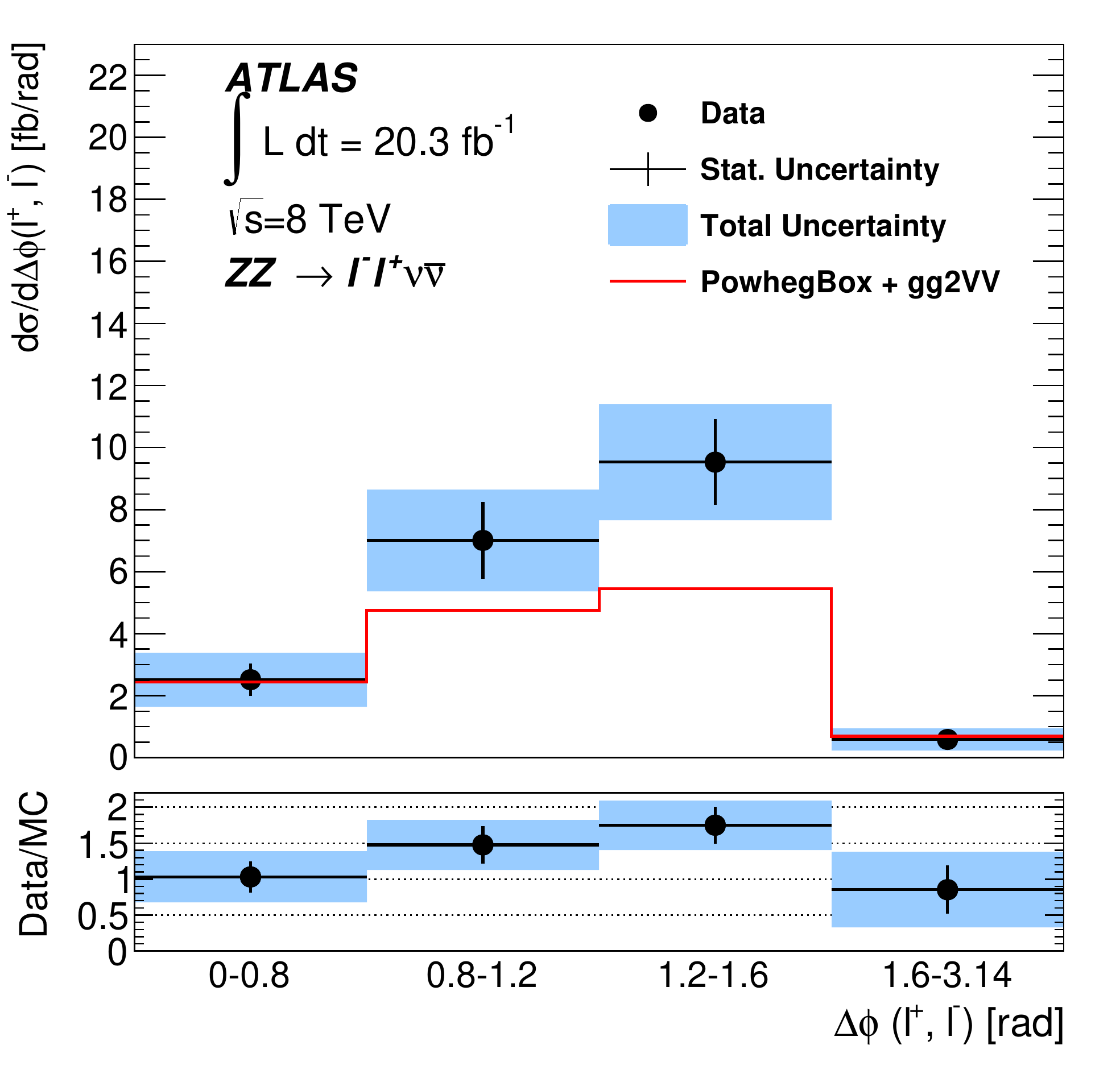}
    \caption{}
    \label{fig:unfold_2l2v_dphi_abs}
  \end{subfigure}
 \begin{subfigure}{0.48\textwidth}
    \includegraphics[width=\textwidth]{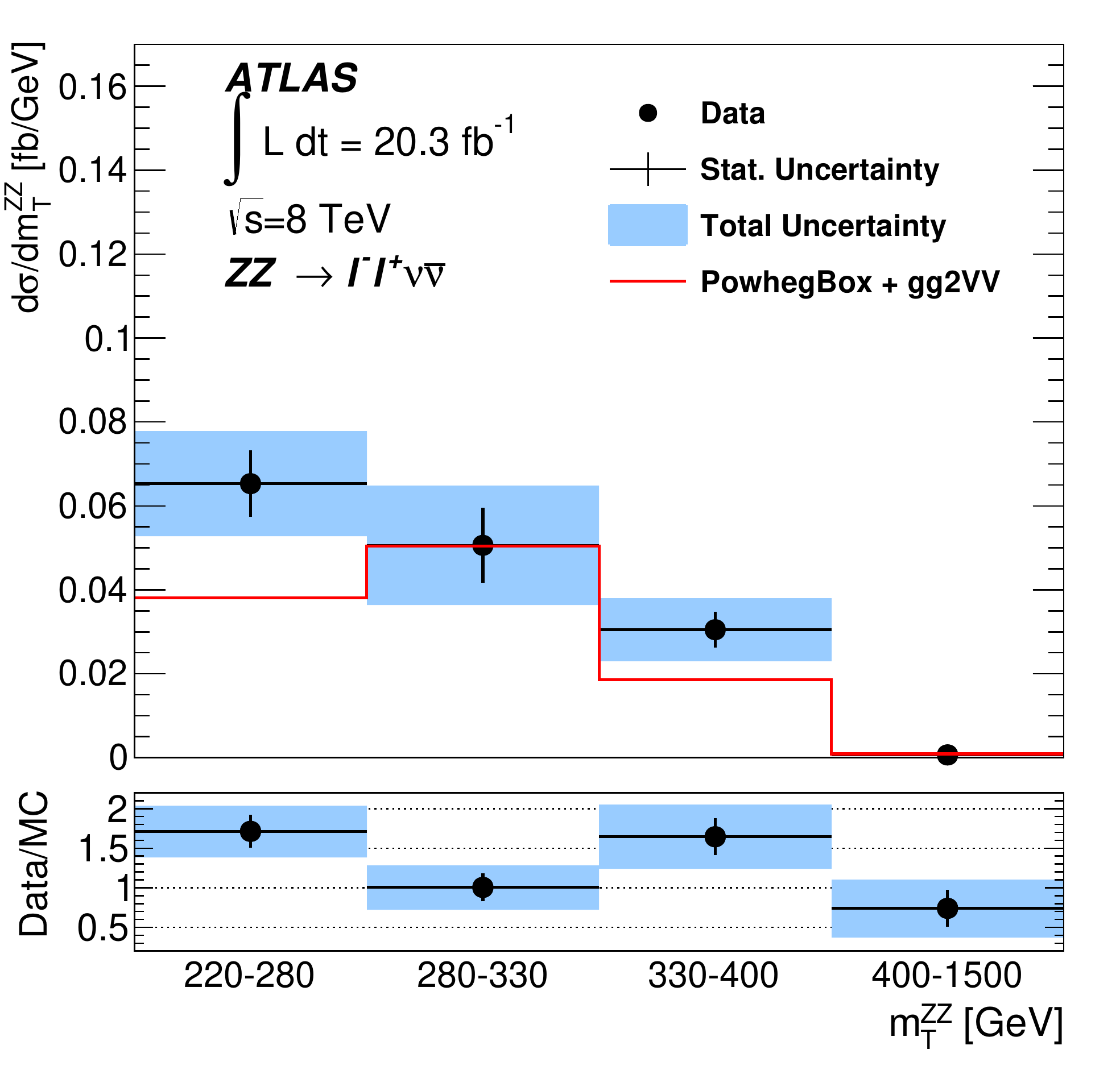}
    \caption{}
    \label{fig:unfold_2l2v_mt_abs}
  \end{subfigure}
\caption{The measured differential cross-section distributions (black
points) normalized to the bin width
for \subref{fig:unfold_2l2v_zpt_abs} \Zpt
~, \subref{fig:unfold_2l2v_dphi_abs} $\Delta\phi(\ell^{+},\ell^{-})$
and \subref{fig:unfold_2l2v_mt_abs} $m_{\mathrm{T}}^{ZZ}$ in
the \zzllvv\ channel, unfolded within the fiducial phase space, compared to the theory
predictions of \powhegbox  and \ggtwovv\ (red line). The vertical error bars
show the respective statistical uncertainties, while the light blue error bands 
express the statistical and systematic
uncertainties of the measurements added in quadrature.}
\label{fig:unfolded2l2vUNORMALIZED}
\end{figure}

The differential cross sections are shown in
Figure~\ref{fig:unfolded2l2vUNORMALIZED}. The measured values are 
compared with the SM predictions (\powhegbox and \ggtwovv). 
The theoretical modelling uncertainties, evaluated by the
data-driven method described in \Cref{sec:diffxsec}, are in the order of
a few percent (0.7$\%$--1$\%$ for \Zpt\ , 0.7$\%$--1.5$\%$ for
$\Delta\phi(\ell^{+},\ell^{-})$ and 3$\%$--9$\%$ for $m_{\mathrm{T}}^{ZZ}$). 
While the central values of the unfolded data differ from the prediction by up to 50$\%$ in some of the bins,
the measurement is consistent with the SM prediction within
$1$--$2\sigma$. 

%% file: aTGCs.tex
\section{Anomalous neutral triple gauge couplings}
\label{sec:atgc}
According to the SM $\mathrm{SU(2)}_{\mathrm{L}} \times \mathrm{U(1)}_{\mathrm{Y}}$ gauge symmetry,
vertices of the form $ZZZ$ and $ZZ\gamma$ are not present at
tree level.  Consequently, \ZZ\ production  does not receive a
contribution from the $s$-channel resonance diagram (\Cref{fig:ZZschannel}).
At one-loop level, fermionic triangle loops contribute to the generation of
effective neutral aTGCs at the level of $10^{-4}$ to $10^{-3}$
\cite{Gounaris:2000tb}. A typical signature of aTGCs is an enhanced cross 
section at high centre-of-mass energies. Thus, 
observables which are proportional to the invariant mass of the \ZZ\ diboson system and 
the gauge boson transverse momentum are particularly sensitive to contributions
from aTGCs.
Studies of aTGCs have been performed by the LEP Collaborations 
\cite{Barate:1999jj,Abdallah:2003dv,Acciarri:1999ug,Abbiendi:2003va,Alcaraz:2006mx}, 
as well as the CDF and D0 Collaborations. More recent studies performed by the ATLAS and CMS
Collaborations using  data collected during 2011 at 7 \TeV\
indicate that if there are any
contributions from new physics at the \TeV\ scale, they are 
at most of the order of $10^{-3}$. 
 
In this paper, an effective Lagrangian framework \cite{Baur:2000ae} is
used for the aTGCs studies,
where the most general $ZZV$ ($V = Z$ or $\gamma$) couplings, which respect gauge and
Lorentz invariance \cite{Hagiwara:1986vm} are considered. Such couplings can be
parameterized by two $CP$-violating ($f_{4}^{\gamma}$, $f_{4}^{Z}$) and two
$CP$-conserving ($f_{5}^{\gamma}$, $f_{5}^{Z}$) parameters.
The contribution of anomalous couplings to the \ZZ\ production cross section grows with the partonic
centre-of-mass energy squared, $\hat{s}$. To avoid violation of unitarity a form factor is introduced to the anomalous
couplings of the form:
\begin{equation}
   f_{i}^{V}(\hat{s}) = f_{i,0}^{V} \Big(1+\frac{\hat{s}}{\Lambda^2}\Big)^{-2},
  \label{eq:formfactor}
\end{equation}
where $f_{i,0}^{V}$ is the generic anomalous coupling value (i=4,5) at low energy and $\Lambda$ is a cutoff scale related to the energy at which the effective field theory breaks down and new physics would be observed.  For the results presented, no form factor is used as the current sensitivity is well within the unitarization constraints and $\Lambda$ is large enough that no energy dependence for the anomalous couplings needs to be considered. 
The $ZZV$ couplings in $ZZ$ production considered here are distinct from the $Z\gamma V$ couplings probed in $Z\gamma$
production in $e^{-}e^{+}$ and hadronic collisions. Additional
anomalous couplings can contribute when the \Z\ bosons are off-shell
\cite{Gounaris:2000dn}, although these couplings are highly suppressed near
the \Z\ boson resonance.

\subsection{Parameterization of signal yield}
\label{subsec:atgcParametrisation}
In order to look for the effects of $ZZV$ aTGCs,  the signal yield must be parameterized in terms
of the coupling strength. Simulated samples are produced using a generator which contains
matrix elements with aTGCs at various strengths with one reference sample
generated at the SM points of zero for all couplings, and at least two other samples with non-zero
couplings in various combinations. The signal yield is obtained as the simulated samples are
reweighted from one aTGC point to another 
using a framework \cite{Bella:2008wc} which allows the kinematics properties to be reweighted on an event-by-event basis. 
The matrix elements used for reweighting are extracted from the Baur, Han and Ohnemus (BHO)~\cite{bib:bho}
generator. The event yields are then expressed as a function of the aTGC parameters, which 
contains terms both linearly and quadratically proportional to the couplings. 
The expected number of events generated by \sherpa (only \qqZZ) is then normalized to the prediction of \powhegbox + \ggVV.  

\subsection{Confidence intervals for aTGCs}
\label{subsec:limitExtraction}
\input{aTGCExtraction}

%% file: aTGCExtraction.tex
\begin{figure}[hbtp]
  \centering
  \begin{subfigure}{0.47\textwidth}
    \includegraphics[width=\textwidth]{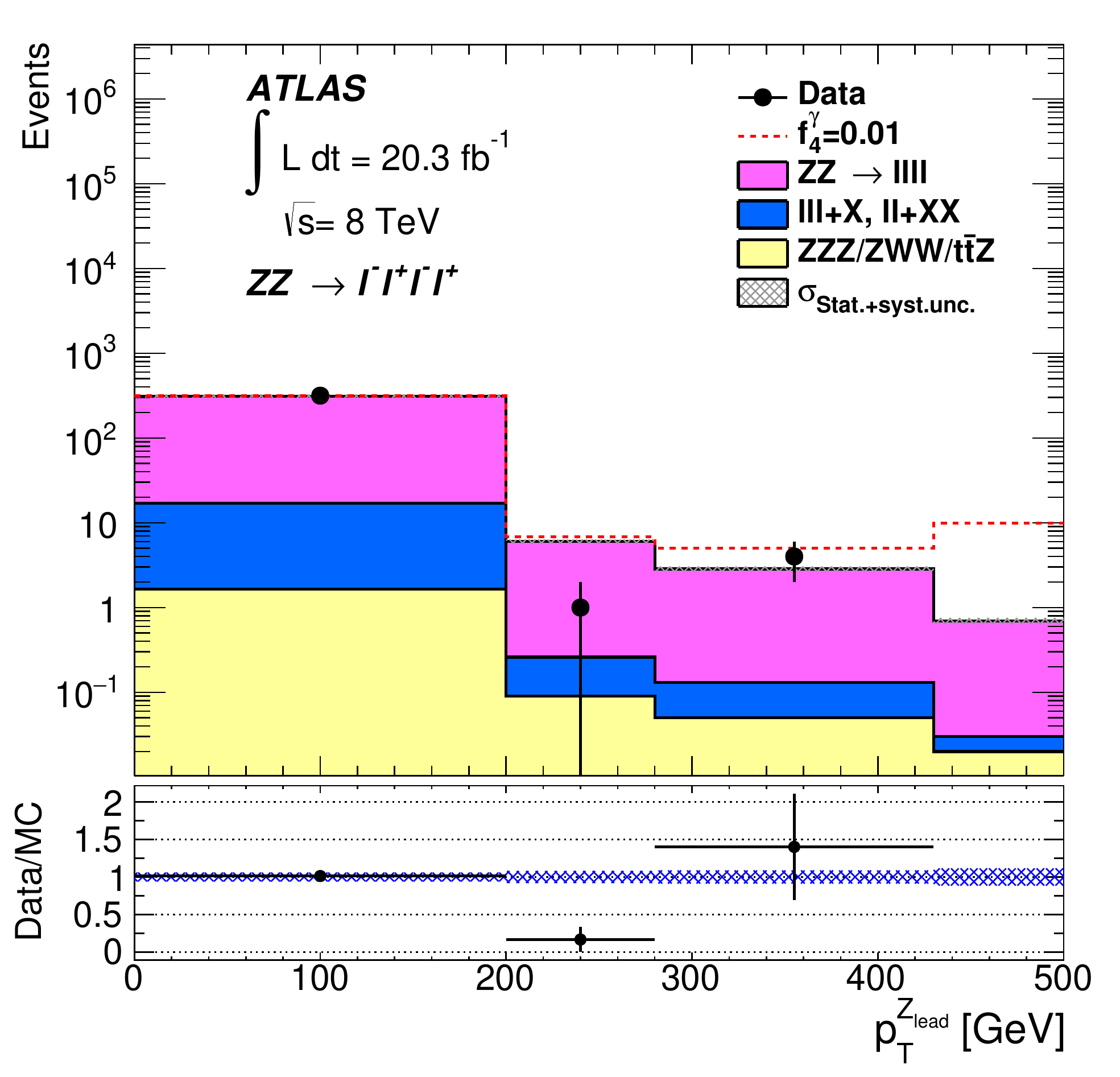}
    \caption{}
    \label{fig:aTGC_4lepton_incl_log}
  \end{subfigure}
  %%%%%%%%%%%%%%%%%%%%%%%%%%%%%%%%%%%%%%%%%%%%%%
  \begin{subfigure}{0.47\textwidth}
    \includegraphics[width=\textwidth]{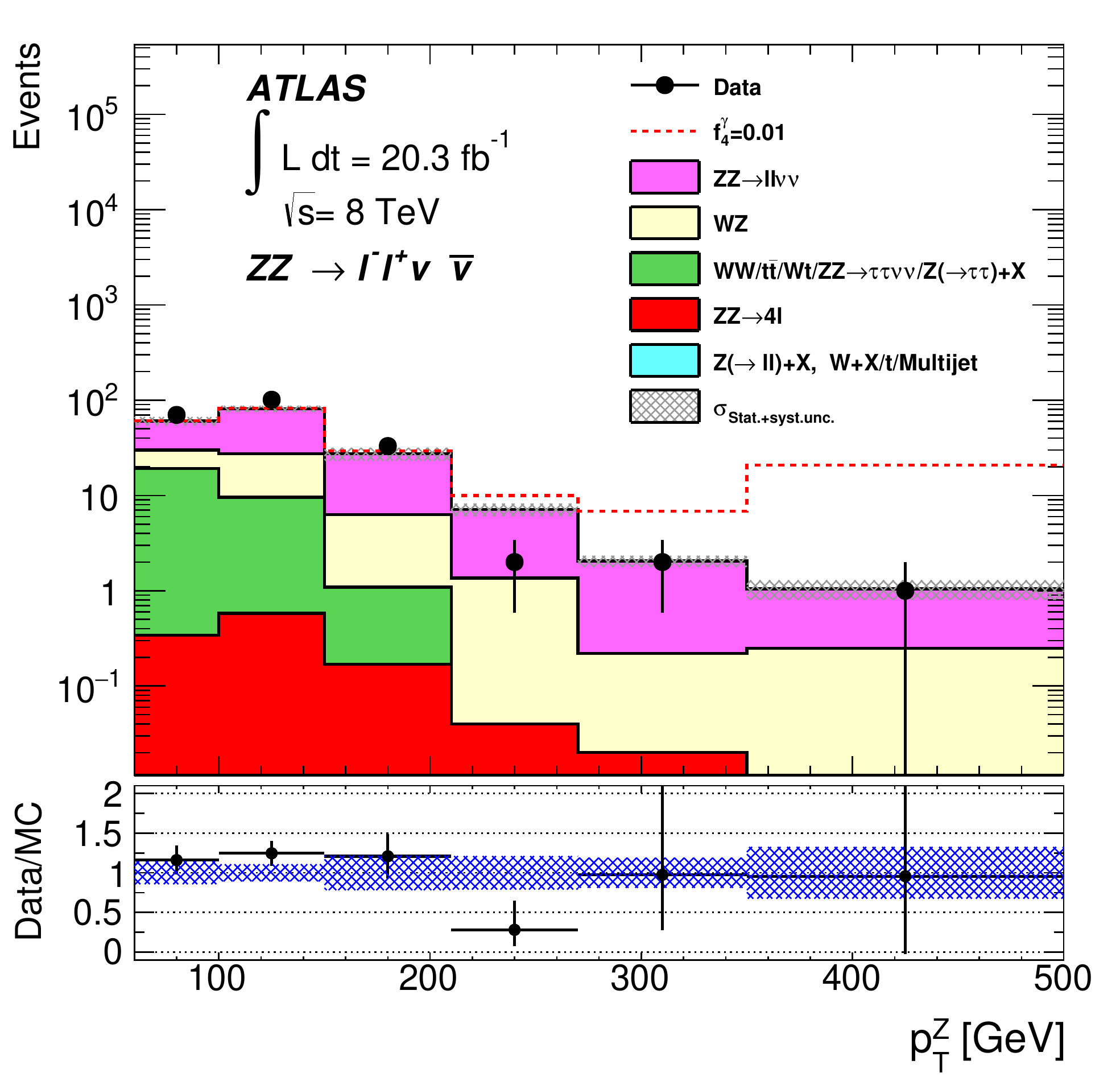}
    \caption{}
    \label{fig:aTGC_2l2nu_incl_log}
  \end{subfigure}
  \caption{Data and SM prediction of the \Zpt\ distribution for the \subref{fig:aTGC_4lepton_incl_log} \zzlmlplmlpprimed\
           and \subref{fig:aTGC_2l2nu_incl_log} \zzllvv\ channels. The expected contribution from the aTGC point with  $f_{4}^{\gamma} = 0.01$ is also shown. 
           }
    \label{fig:TGC_ZPt_dataMC}
\end{figure}

The $\pT^{Z_{\mathrm{lead}}}$ in the \zzllll\ channel and the \Zpt\ in
the \zzllvv\ channel are particularly sensitive to aTGCs and therefore
these distributions are used to probe them.
Given the limited number of events in the selected data sample, especially
in the high-\Zpt\ region, all the events of the \eeee,~\eemm\ and \mmmm\ 
final states in the \llll\ channel are combined. Likewise, all the
events of the \eevv\ and \mmvv\ final states in the  
\llvv\ channel are combined. \Cref{fig:TGC_ZPt_dataMC}
shows the data distribution comparison with the SM predictions, as well as the
prediction for a non-zero aTGC parameter point, where the
$CP$-violating parameter $f_{4}^{\gamma}$ is set to be equal to 0.01,
while all other anomalous couplings are set to zero. 
The deficit in data versus the MC prediction for bin 2
in \subref{fig:aTGC_4lepton_incl_log} is 2.2 $\sigma$ while for bin 4
in \subref{fig:aTGC_2l2nu_incl_log} it is 1.9 $\sigma$.
The data are found to be
consistent with the SM predictions, and no indication of aTGCs is observed.

 Limits on neutral aTGC parameters are determined using the expected and
 observed numbers of events in the following \Zpt\ bins:
 280--430 \GeV\ and 430--1500 \GeV\ for the \zzlmlplmlpprimed\ channel,
 and 270--350 \GeV\ and 350--1500 \GeV\ for the \zzllvv\
 channel. The binning is optimized for maximum
 sensitivity in the aTGCs. \Cref{table:TGC_yield_2bin} shows the
 expected number of events 
from non-\ZZ\ backgrounds and from SM \ZZ\ events along with the observed number of events in each bin. 

A normalization factor is applied to the expected SM $ZZ$ events,
to scale the predicted $ZZ$ fiducial cross section to the measurement. 
The uncertainty in this normalization factor is propagated
to the limit-setting procedure. Apart from the uncertainties described in
\Cref{sec:systematics}, an additional systematic uncertainty in the
modelling of the $\pt^Z$
shape for the \qqZZ\ process is taken into account by comparing the predictions from
\powhegbox and \sherpa. The difference ranges from 30\% to 80\% for the \zzllvv\ channel
and from 30\% to 40\% for the \zzlmlplmlpprimed\ channel.

The extraction of the aTGC limits is based on detector-level distributions.  A
profile-likelihood-ratio test statistic~\cite{Cowan:2010js} is used to assess whether the predictions with aTGCs are compatible with the data. Then a frequentist
method~\cite{PhysRevD.57.3873} is used to determine the 95\% confidence
level (CL) intervals for the aTGC parameters. The number of observed data events and the
predictions for the aTGC signal and background processes are used to construct the
Poissonian probability density functions, in which systematic uncertainties are
considered as nuisance parameters constrained with Gaussian functions.
The observed intervals are compared with the expected intervals by generating `Asimov' data sets,
which are representative event samples that provide both the median expectation for an experimental
result and its expected statistical variation in the asymptotic
approximation as described in Ref.~\cite{Cowan:2010js}. The
expected limits calculated with `Asimov' data sets are cross-checked
with limits obtained from 5000 pseudo-experiments generated using
the expected number of events at each point in the aTGC parameter space.

\begin{table}[htbp]
 \centering
 \begin{tabular}{cccc}
 \toprule
 \toprule
     & Expected non-\ZZ\ background & Expected SM \ZZ\ events & Observed events \\ \hline
 %% 4l
 \zzlmlplmlpprimed\ \\  \hline
 $280<\pT^{Z_{\mathrm{lead}}} < 430 \GeV $ & $0.13  \pm0.01  \pm0.04 $   & 2.7 $\pm$ 0.1 $\pm$ 0.9 &  4 \\
 $\pT^{Z_{\mathrm{lead}}} > 430 \GeV$   & $0.03  \pm0.01  \pm0.01 $   & 0.7 $\pm$ 0.1 $\pm$ 0.3 &  0 \\
 \midrule
 %% llvv
 \zzllvv\           \\
 \midrule
 $270<\pt^Z< 350 \GeV$ & $0.22 \pm 0.10 \pm 0.03 $   & $2.3 \pm 0.20 \pm 1.8 $ & 2  \\
 $\pt^Z >  350 \GeV$   & $0.25 \pm 0.12 \pm 0.03 $   & $1.0 \pm 0.13 \pm 0.4 $ & 1  \\
  \bottomrule
  \bottomrule
 \end{tabular}
 \caption{\label{table:TGC_yield_2bin} The expected background from non-\ZZ\ events and  SM \ZZ\ events,
 and the number of observed events in the two highest $\pT^{Z_{\mathrm{lead}}}$ and \Zpt\ bins for all
 final states in each \ZZ\ channel. For the expected background and
 SM \ZZ\ events, the first uncertainty is statistical and the second is systematic.}
\end{table}

\begin{figure}[htbp]
  \centering
  \begin{subfigure}{0.3\textwidth}
    \includegraphics[width=\textwidth]{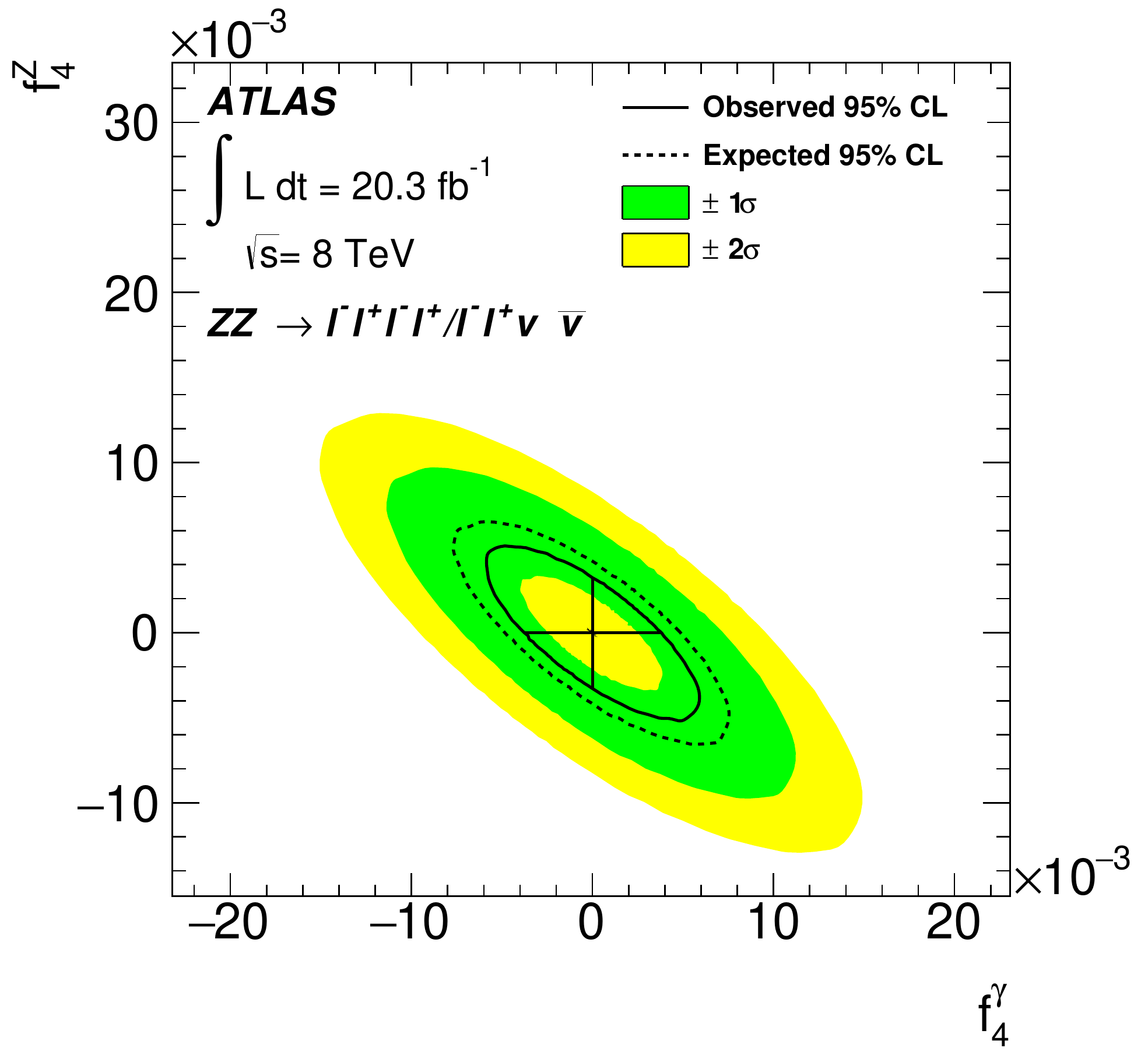}
    \caption{}
    \label{fig:aTGC_Extraction_f4g_f4z_contour}
  \end{subfigure}
  \begin{subfigure}{0.3\textwidth}
    \includegraphics[width=\textwidth]{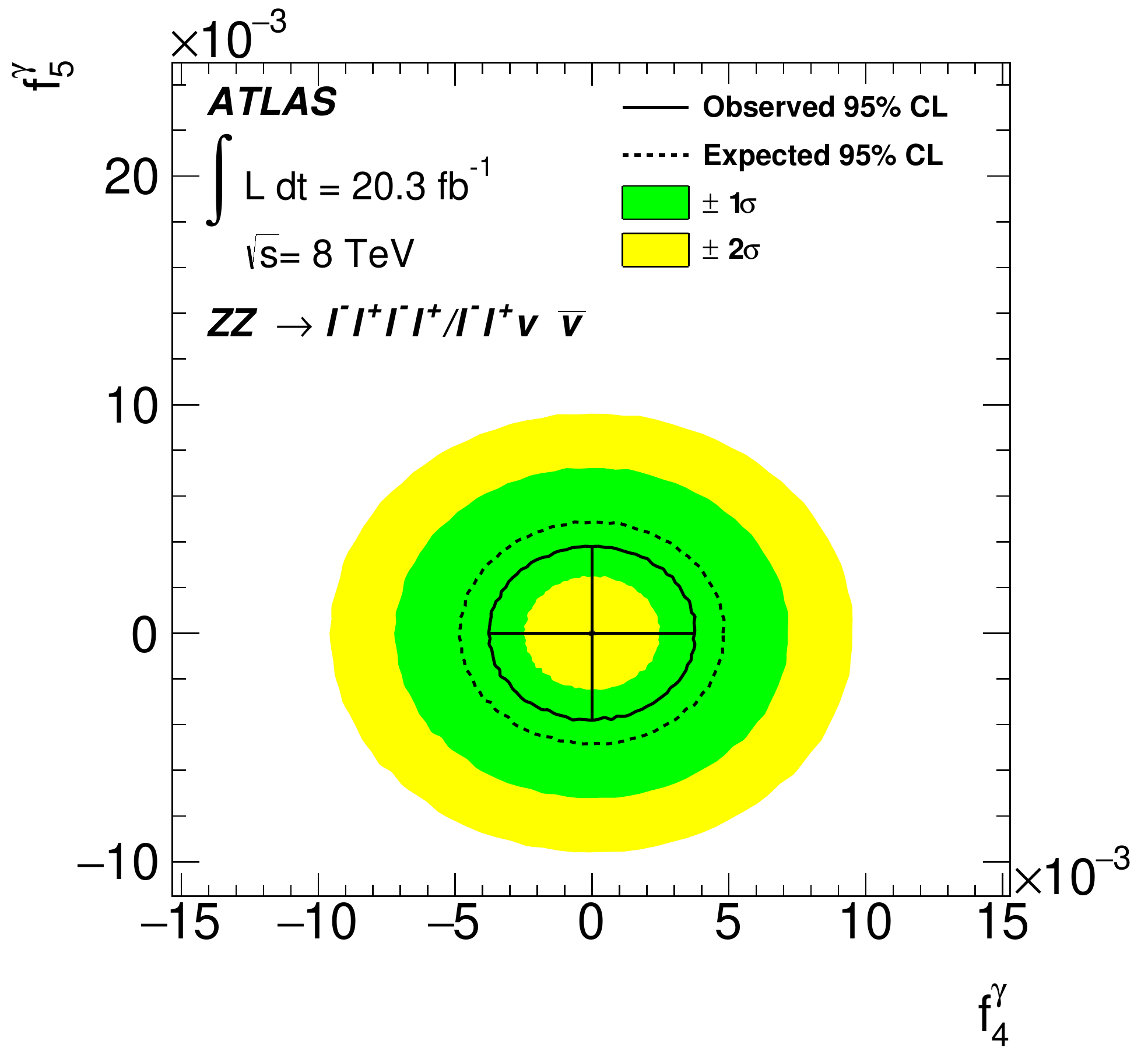}
    \caption{}
    \label{fig:aTGC_Extraction_f4g_f5g_contour}
  \end{subfigure}
  %%%%%%%%%%%%%%%%%%%%%%%%%%%%%%%%%%%%%%%%%%%%
  \begin{subfigure}{0.3\textwidth}
    \includegraphics[width=\textwidth]{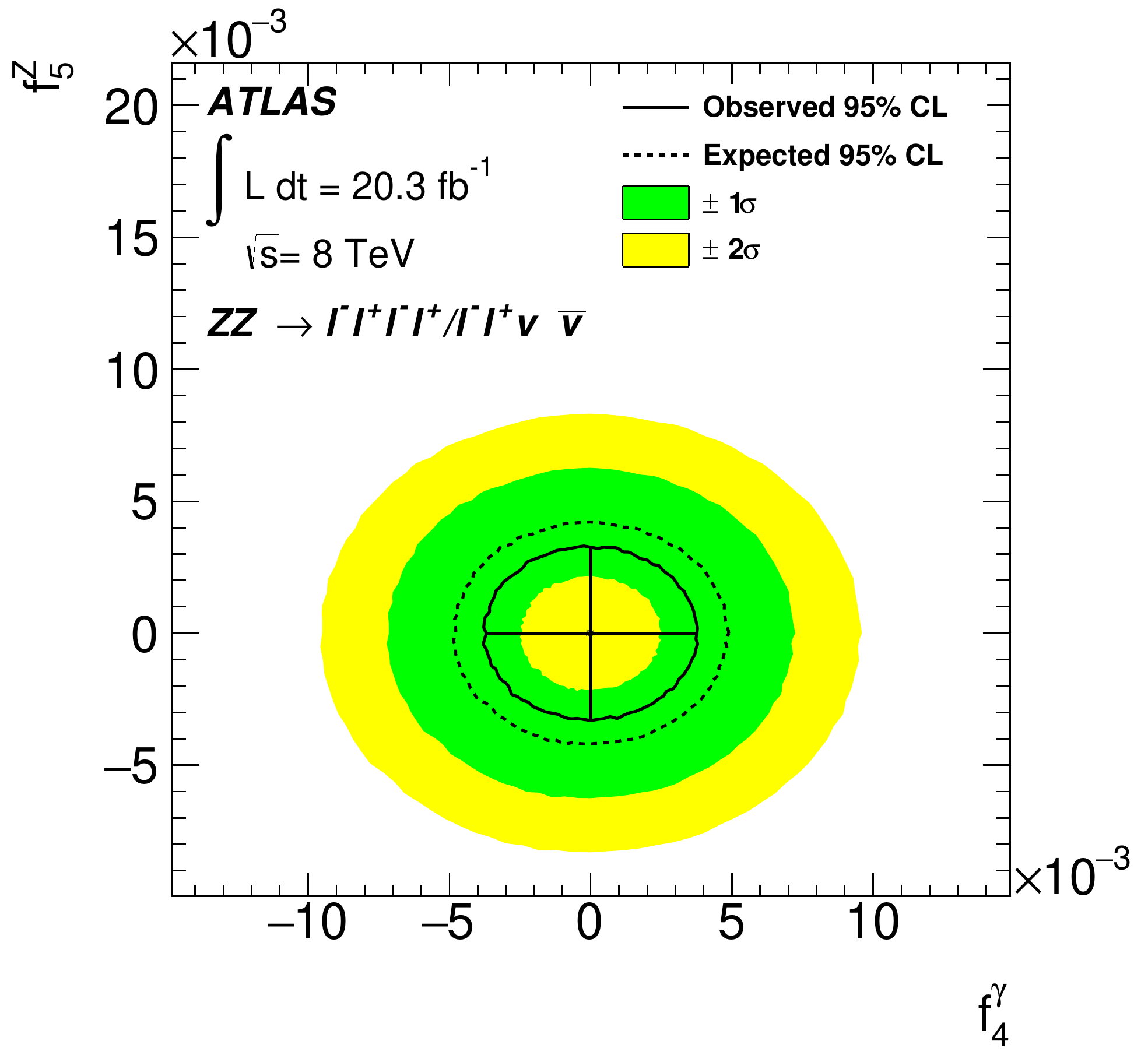}
    \caption{}
    \label{fig:aTGC_Extraction_f4g_f5z_contour}
  \end{subfigure}
  %%%%%%%%%%%%%%%%%%%%%%%%%%%%%%%%%%%%%%%%%%%%
  \begin{subfigure}{0.3\textwidth}
    \includegraphics[width=\textwidth]{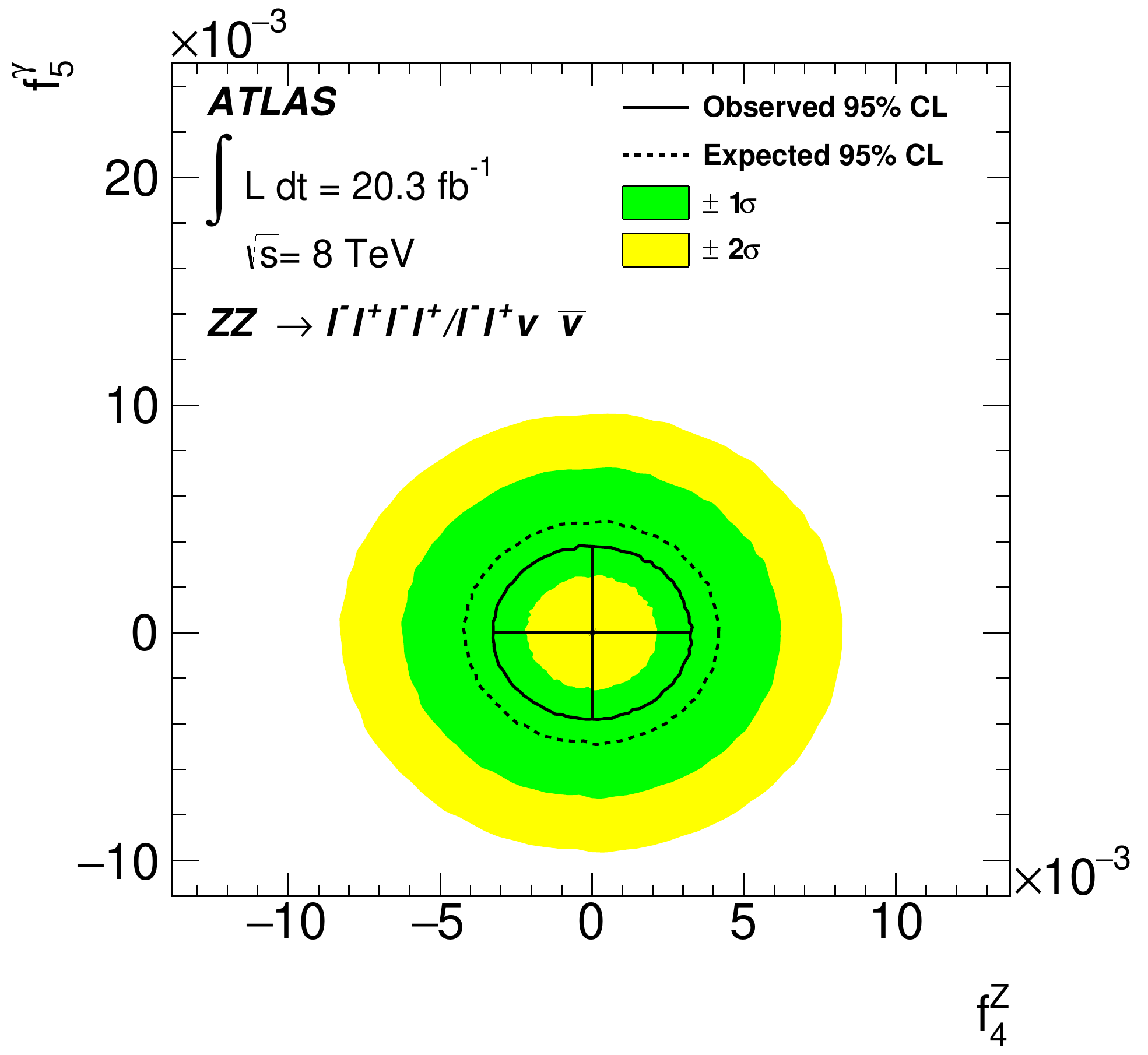}
    \caption{}
    \label{fig:aTGC_Extraction_f4z_f5g_contour}
  \end{subfigure}
  %%%%%%%%%%%%%%%%%%%%%%%%%%%%%%%%%%%%%%%%%%%%
  \begin{subfigure}{0.3\textwidth}
    \includegraphics[width=\textwidth]{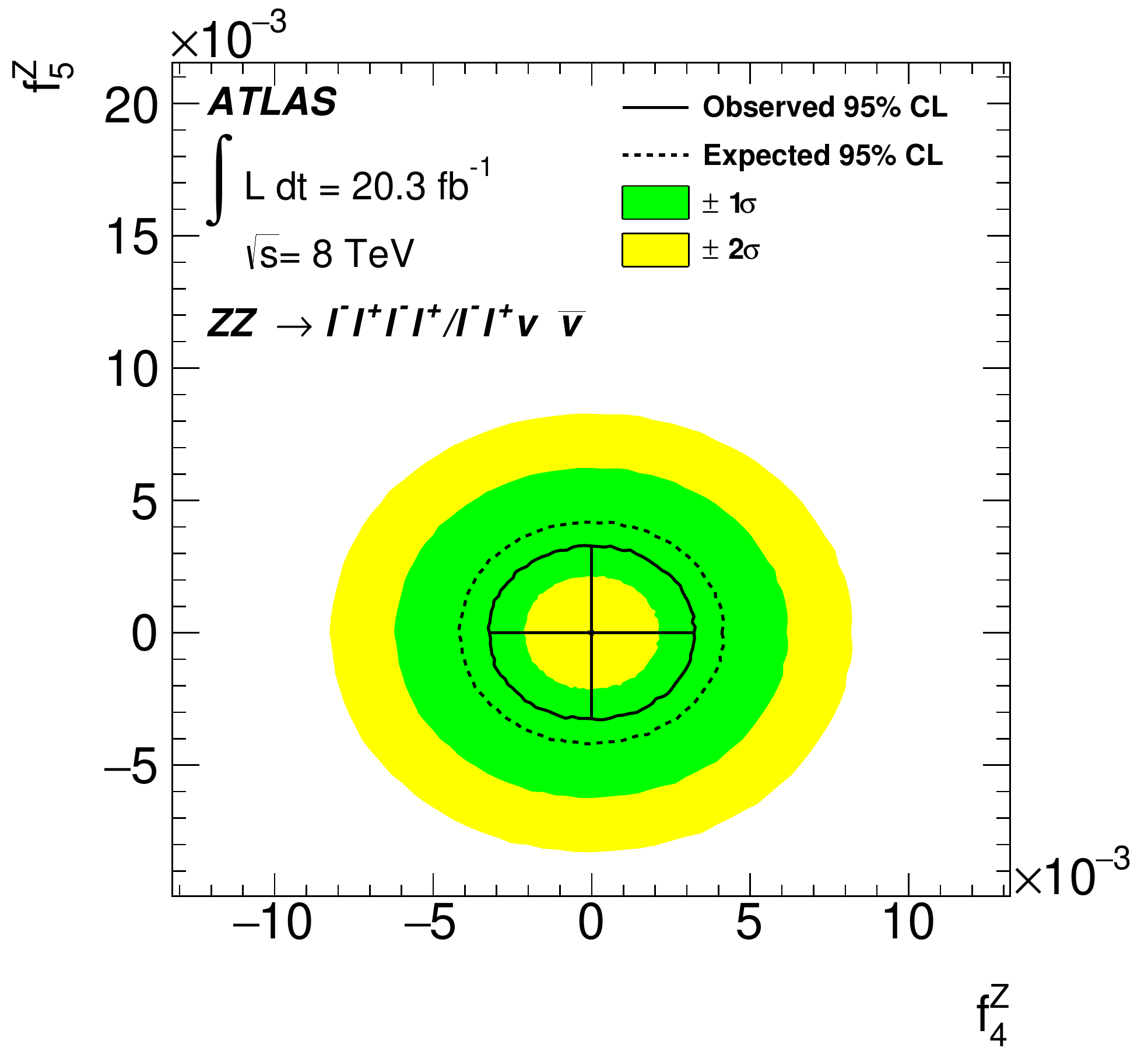}
    \caption{}
    \label{fig:aTGC_Extraction_f4z_f5z_contour}
  \end{subfigure}
  %%%%%%%%%%%%%%%%%%%%%%%%%%%%%%%%%%%%%%%%%%%%
  \begin{subfigure}{0.3\textwidth}
    \includegraphics[width=\textwidth]{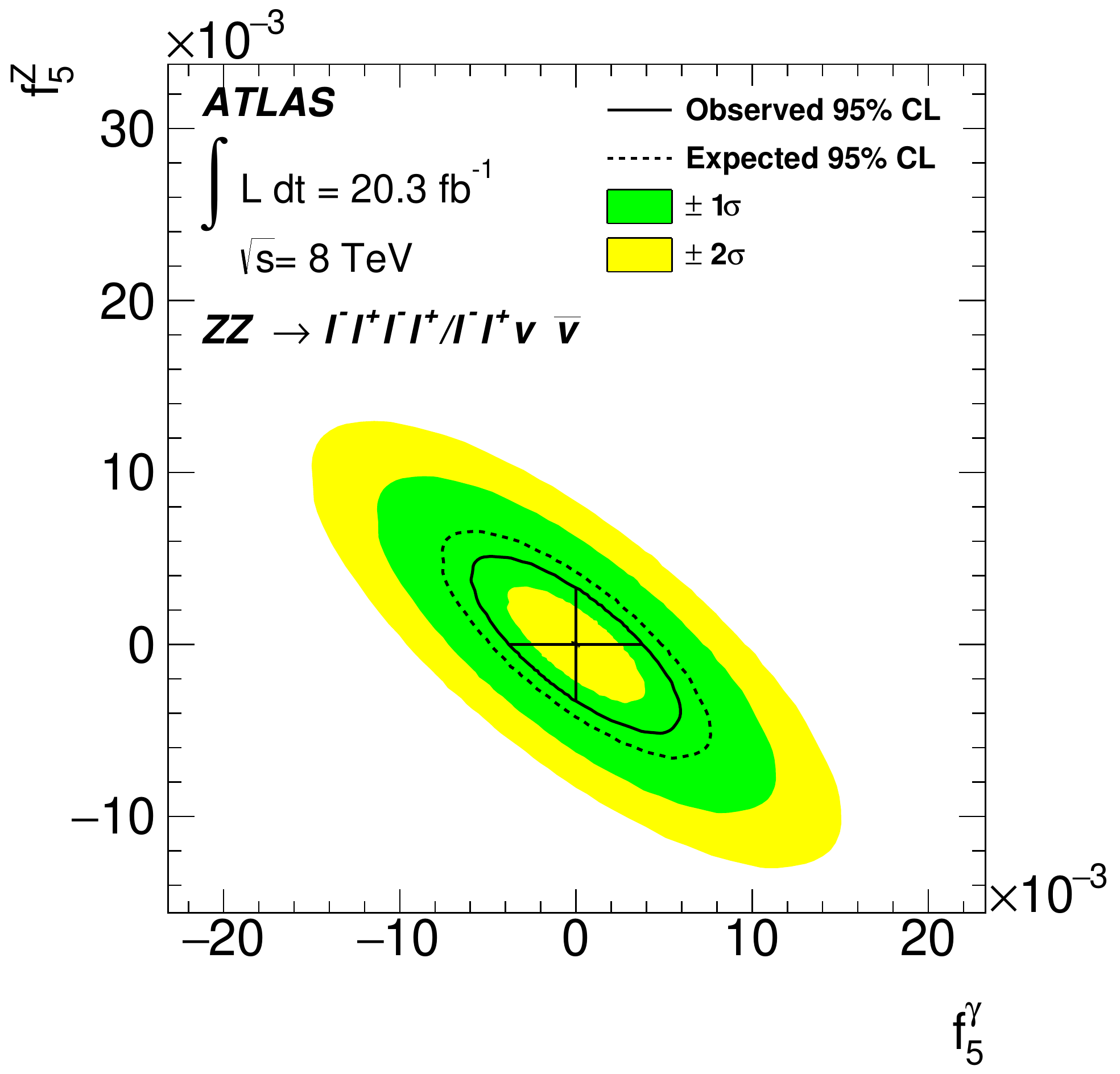}
    \caption{}
    \label{fig:aTGC_Extraction_f5g_f5z_contour}
  \end{subfigure}
 \caption{The observed and expected two-dimensional 95\% CL contours
    for limits in the plane of two simultaneously non-zero parameters
    for the combined  \zzlmlplmlpprimed\ and \zzllvv\ channels. Except
  for the two aTGC parameters under study, all others are set to
  zero. The horizontal and vertical lines correspond to the
  one-dimensional limits for each aTGC parameter.}  
   \label{fig:tgc2dlimbraz}
\end{figure}

Limits are set on each coupling, assuming all of the other couplings are zero (as in the SM),
and on pairs of couplings assuming the remaining two couplings are zero.
The observed and expected 95\% CL invervals for the four aTGC
parameters for the \zzlmlplmlpprimed\ and \zzllvv\ channels combined are listed
in \Cref{table:tgc-comb_2bin}. Since the energy scale at which new
physics may appear is unknown, no form factor is used when  
deriving the limits. The two-dimensional 95\% CL intervals 
are shown in \Cref{fig:tgc2dlimbraz}.

%%%%%%%%%%%%%% TGC limit 4l+2l2n
\renewcommand{\arraystretch}{1.5}
\begin{table}[htbp]
  \centering
  \begin{tabular}{ccc}
  \toprule
  \toprule
  Coupling          &    Expected~($10^{-3}$)        & Observed~($10^{-3}$)     \\
  \midrule
  
  $f_{4}^{\gamma}$  & [$-4.6$, $4.8$]   & [$-3.8$, $3.8$] \\
  $f_{4}^{Z}     $  & [$-4.0$, $4.1$]   & [$-3.3$, $3.2$] \\
  $f_{5}^{\gamma}$  & [$-4.8$, $4.8$]   & [$-3.8$, $3.8$] \\
  $f_{5}^{Z}     $  & [$-4.1$, $4.1$]   & [$-3.3$, $3.3$] \\
  
  \midrule
  \end{tabular}
  \caption{One-dimensional expected and observed 95\%\ CL limits on the aTGC parameters for both
the \zzlmlplmlpprimed\ and \zzllvv\ channels combined. The limit for
  each coupling assumes that the other couplings are fixed at 
           their SM value.}
\label{table:tgc-comb_2bin}
\end{table}

%%%%%%%%%%%%%% 2D limtis

The one-dimensional limits are more stringent than those derived from
measurements at LEP~\cite{Alcaraz:2006mx}, the Tevatron~\cite{Tevatron:d0} and
previously by ATLAS~\cite{Aad:2012awa} and are comparable to the
limits set by CMS at 8 \TeV\ \cite{CMS:2014xja}. CMS has recently
improved the limits on aTGCs by combining measurements at 7 and
8 \TeV\ \cite{Khachatryan:2015pba}.

%% file: acknowledgements/Acknowledgements.tex
% Acknowledgements for papers with collision data
% Version 6-Mar-2017

% Standard acknowledgements start here
%----------------------------------------------
We thank CERN for the very successful operation of the LHC, as well as the
support staff from our institutions without whom ATLAS could not be
operated efficiently.

We acknowledge the support of ANPCyT, Argentina; YerPhI, Armenia; ARC, Australia; BMWFW and FWF, Austria; ANAS, Azerbaijan; SSTC, Belarus; CNPq and FAPESP, Brazil; NSERC, NRC and CFI, Canada; CERN; CONICYT, Chile; CAS, MOST and NSFC, China; COLCIENCIAS, Colombia; MSMT CR, MPO CR and VSC CR, Czech Republic; DNRF and DNSRC, Denmark; IN2P3-CNRS, CEA-DSM/IRFU, France; SRNSF, Georgia; BMBF, HGF, and MPG, Germany; GSRT, Greece; RGC, Hong Kong SAR, China; ISF, I-CORE and Benoziyo Center, Israel; INFN, Italy; MEXT and JSPS, Japan; CNRST, Morocco; NWO, Netherlands; RCN, Norway; MNiSW and NCN, Poland; FCT, Portugal; MNE/IFA, Romania; MES of Russia and NRC KI, Russian Federation; JINR; MESTD, Serbia; MSSR, Slovakia; ARRS and MIZ\v{S}, Slovenia; DST/NRF, South Africa; MINECO, Spain; SRC and Wallenberg Foundation, Sweden; SERI, SNSF and Cantons of Bern and Geneva, Switzerland; MOST, Taiwan; TAEK, Turkey; STFC, United Kingdom; DOE and NSF, United States of America. In addition, individual groups and members have received support from BCKDF, the Canada Council, CANARIE, CRC, Compute Canada, FQRNT, and the Ontario Innovation Trust, Canada; EPLANET, ERC, ERDF, FP7, Horizon 2020 and Marie Sk{\l}odowska-Curie Actions, European Union; Investissements d'Avenir Labex and Idex, ANR, R{\'e}gion Auvergne and Fondation Partager le Savoir, France; DFG and AvH Foundation, Germany; Herakleitos, Thales and Aristeia programmes co-financed by EU-ESF and the Greek NSRF; BSF, GIF and Minerva, Israel; BRF, Norway; CERCA Programme Generalitat de Catalunya, Generalitat Valenciana, Spain; the Royal Society and Leverhulme Trust, United Kingdom.

The crucial computing support from all WLCG partners is acknowledged gratefully, in particular from CERN, the ATLAS Tier-1 facilities at TRIUMF (Canada), NDGF (Denmark, Norway, Sweden), CC-IN2P3 (France), KIT/GridKA (Germany), INFN-CNAF (Italy), NL-T1 (Netherlands), PIC (Spain), ASGC (Taiwan), RAL (UK) and BNL (USA), the Tier-2 facilities worldwide and large non-WLCG resource providers. Major contributors of computing resources are listed in Ref.~\cite{ATL-GEN-PUB-2016-002}.
%----------------------------------------------

%% file: atlas_authlist.tex
% ATLAS Collaboration author list
% Data extracted on 12-Aug-2016 for paper reference STDM-2014-16
%$\documentclass[11pt]{article}
%\usepackage{a4wide}\begin{document}
\begin{flushleft}
{\Large The ATLAS Collaboration}

\bigskip

M.~Aaboud$^\textrm{\scriptsize 136d}$,
G.~Aad$^\textrm{\scriptsize 87}$,
B.~Abbott$^\textrm{\scriptsize 114}$,
J.~Abdallah$^\textrm{\scriptsize 8}$,
O.~Abdinov$^\textrm{\scriptsize 12}$,
B.~Abeloos$^\textrm{\scriptsize 118}$,
R.~Aben$^\textrm{\scriptsize 108}$,
O.S.~AbouZeid$^\textrm{\scriptsize 138}$,
N.L.~Abraham$^\textrm{\scriptsize 152}$,
H.~Abramowicz$^\textrm{\scriptsize 156}$
% ...
\bigskip
\\
% ...
$^{8}$ Department of Physics, The University of Texas at Arlington, Arlington TX, United States of America\\
% ...
$^{12}$ Institute of Physics, Azerbaijan Academy of Sciences, Baku, Azerbaijan\\
% ...
$^{87}$ CPPM, Aix-Marseille Universit{\'e} and CNRS/IN2P3, Marseille, France\\
% ...
$^{108}$ Nikhef National Institute for Subatomic Physics and University of Amsterdam, Amsterdam, Netherlands\\
% ...
$^{114}$ Homer L. Dodge Department of Physics and Astronomy, University of Oklahoma, Norman OK, United States of America\\
% ...
$^{118}$ LAL, Univ. Paris-Sud, CNRS/IN2P3, Universit{\'e} Paris-Saclay, Orsay, France\\
% ...
$^{136}$ $^{(a)}$ Facult{\'e} des Sciences Ain Chock, R{\'e}seau Universitaire de Physique des Hautes Energies - Universit{\'e} Hassan II, Casablanca; $^{(b)}$ Centre National de l'Energie des Sciences Techniques Nucleaires, Rabat; $^{(c)}$ Facult{\'e} des Sciences Semlalia, Universit{\'e} Cadi Ayyad, LPHEA-Marrakech; $^{(d)}$ Facult{\'e} des Sciences, Universit{\'e} Mohamed Premier and LPTPM, Oujda; $^{(e)}$ Facult{\'e} des sciences, Universit{\'e} Mohammed V, Rabat, Morocco\\
% ...
$^{138}$ Santa Cruz Institute for Particle Physics, University of California Santa Cruz, Santa Cruz CA, United States of America\\
% ...
$^{152}$ Department of Physics and Astronomy, University of Sussex, Brighton, United Kingdom\\
% ...
$^{156}$ Raymond and Beverly Sackler School of Physics and Astronomy, Tel Aviv University, Tel Aviv, Israel\\
% ...
$^{a}$ Also at Department of Physics, King's College London, London, United Kingdom\\
$^{b}$ Also at Institute of Physics, Azerbaijan Academy of Sciences, Baku, Azerbaijan\\
$^{c}$ Also at Novosibirsk State University, Novosibirsk, Russia\\
$^{d}$ Also at TRIUMF, Vancouver BC, Canada\\
$^{e}$ Also at Department of Physics {\&} Astronomy, University of Louisville, Louisville, KY, United States of America\\
% ...
\end{flushleft}

%\end{document}
% Created with xml2latex.py